\documentclass[pre,aps,twocolumn,floats,showpacs,twoside]{revtex4}

\usepackage{amsmath}
\usepackage{amssymb}
\usepackage{latexsym}

\usepackage{graphicx}
\usepackage{epsfig}

\usepackage{epic}
\usepackage{eepic}

\newcommand{\nn}{\nonumber}

\newcommand{\rr}{\mathbf{r}}
\newcommand{\kk}{\mathbf{k}}

\newcommand{\oo}{\mathbf{0}}

\newcommand{\LL}{\left}
\newcommand{\RR}{\right}

\newcommand{\NN}{\mathcal{N}}
\newcommand{\OO}{\mathcal{O}}
\newcommand{\dd}{\delta}
\newcommand{\DD}{\Delta}
\newcommand{\DDr}{\Delta^{rnd}}

\newcommand{\al}{\alpha}

\newcommand{\eee}{e^{i \kk \rr}}
\newcommand{\eeeP}{e^{i \kk \rr'}}
\newcommand{\eeePP}{e^{i \kk \rr''}}

\begin{document}

\title{Diffusion Processes on Small-World Networks with Distance-Dependent Random-Links}

\author{Bal\'azs Kozma}
\email{kozmab@th.u-psud.fr}
\affiliation{Laboratoire de Physique
Th\'eorique (UMR du CNRS 8627), B\^atiment 210 Universit\'e
Paris-Sud - 91405 Orsay Cedex, France}

\author{Matthew B. Hastings}
\email{hastings@lanl.gov}
\affiliation{Center for Non-linear Studies and Theoretical
Division, Los Alamos National Laboratory, Los Alamos, New Mexico
87545, USA}

\author{G. Korniss}
\email{korniss@rpi.edu}
\affiliation{Department of Physics, Applied Physics, and
Astronomy, Rensselaer Polytechnic Institute, 110 8$^{th}$ Street,
Troy, NY 12180--3590, USA}

\begin{abstract}

We considered diffusion-driven processes on small-world networks with
distance-dependent random links. The study of diffusion on such networks is
motivated by transport on randomly folded polymer chains, synchronization
problems in task-completion networks, and gradient driven transport on
networks. Changing the parameters of the distance-dependence, we found a
rich phase diagram, with different transient and recurrent phases in the
context of random walks on networks (or with different smooth and rough
phases in the context of noisy task-completion landscapes). We performed
the calculations in two limiting cases: in the case of
rapid-network-update, where the rearrangement of the random links is fast,
and in the quenched case, where the link rearrangement is slow compared to
the motion of the random walker or the surface. It has been
well-established that in a large class of interacting systems, adding an
arbitrarily small density of, possibly long-range, quenched random links to
a regular lattice interaction topology, will give rise to mean-field like
behavior (i.e., the small-world-like random links can be treated in a
mean-field fashion). In some cases, however, mean-field scaling breaks down
in "low-dimensional" small-world networks, where random links are added to
a low-dimensional regular structure. Examples of such cases are the common
diffusion and the Edwards-Wilkinson process, the main subjects of this
paper.  This break-down can be understood by treating the random links
perturbatively, where the mean-field prediction appears as the lowest-order
term of a naive perturbation expansion. The asymptotic analytic results are
also confirmed numerically by employing exact numerical diagonalization of
the network Laplacian. Further, we construct a finite-size scaling
framework for the relevant observables, capturing the cross-over behaviors
in finite networks. This work provides a detailed account of the
self-consistent-perturbative and renormalization approaches briefly
introduced in Refs. \cite{KOZMA04} and \cite{KOZMA05b}.

\end{abstract}


\maketitle
\section{INTRODUCTION}
\label{sec:intro}

Our environment is pervaded by large-scale complex systems, rooted
in different backgrounds: sociology, decision-making, economy,
transportation, ecology, regulation and transport in biology,
information transfer, and so on
\cite{ALBERT02,DOROGOVTSEV02,NEWMAN03}.  The major characteristic
of these systems is that they do not have any regular structure,
and they exist over complex networks. Characterizing these
networks is a nontrivial task, and finding the most important
quantities to classify the static structure of them has been
intensely researched. Recent research on networks has shifted the
focus from the structural (topological) analysis to the study of
processes (dynamics) on these complex networks
\cite{Boccaletti_review}.

One of the most common phenomena that occur in a large number of
processes is diffusion, where the flow across the links, and
consequently, the rate of change of a generalized density at a
given node, is driven by the local density gradients between the
nodes and its neighbors. Therefore, it is important to understand
the properties of the diffusion and the related diffusion-operator
(or Laplace-operator) on these networks. In our work, besides the
results, we present a technique, based on impurity-averaged
perturbation theory, which gives a novel tool to study processes
on networks especially {\em small-world} networks.

\begin{figure}[tb]
\begin{center}
\includegraphics[width=.3\textwidth]{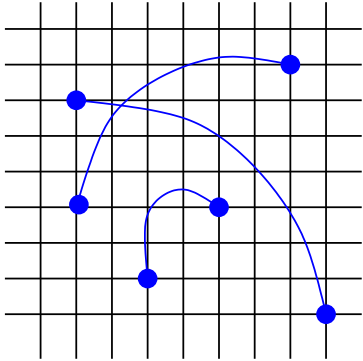}
\end{center}
\vspace*{-0.75cm}
\caption{
\label{fig:SW_2d}
Schematic construction of a small-world network}
\end{figure}

First introduced by Watts and Strogatz \cite{WATTS98,NEWMAN00}, small-world
networks were constructed to model social interactions \cite{MILGRAM67}. In
the original construction, on a regular lattice, each node is connected to $k$
nearest neighbors, then each link of a node is rewired to a randomly chosen
node with probability $p$. Changing the value of $p$, one can interpolate
between a regular ($p=0$) and a completely random, Erd\H{o}s-R\'enyi, network
($p=1$) \cite{ERDOS60, BOLLOBAS01}. The original construction of Watts and
Strogatz was difficult to treat mathematically so a new construction was
suggested \cite{NEWMAN99} combining two networks: a regular $d$-dimensional
one, and on the top of that an Erd\H{o}s-R\'enyi network (see
Fig.~\ref{fig:SW_2d}) with parameter $p$. We refer to this construction as the
``{\em plain}'' SW network in order to distinguish it from power-law SW
networks, defined later. Well known classical models (originally defined on
regular lattices) have been studied on these networks, such as the Ising model
\cite{SCALETTAR91, GITTERMAN00, BARRAT00, NOVOTNY04,SEN3}, the XY model
\cite{KIM01}, diffusion \cite{MONASSON99, BLUMEN_2000a,
  BLUMEN_2000b,ALMAAS_2002,HASTINGS04,KENKRE05,JASCH,LAHTINEN1,LAHTINEN2}, and synchronization
\cite{KORNISS02,KOLAKOWSKA03,TOROCZKAI03,KORNISS01,GUCLU04,GUCLU_PRE06}.

In this paper, we consider a class of networks, called {\em
power-law small-world} (PL-SW) networks
\cite{KLEINBERG00,SOKOLOV97,BLUMEN_2000b,SEN2}. In these networks, on
top of a regular $d$-dimensional lattice, the random long-range
links have a {\em distance-dependent power-law probability
distribution}, where the probability of two nodes being connected
is give by
\begin{equation}
\mbox{Probability ($i$ and $j$ connected)}=\frac{p}{\NN r^\al}=p f(r) \;.
\label{eq:PLSW_prob}
\end{equation}
Here, $r=|i-j|$ is the Euclidean distance between the two nodes in
the original regular lattice, $\al$ is the strength of the
distance-dependent suppression of the random long-range links, and
$p$ is the density of such long-range links (i.e., $p$ is the
average number of random links on a node.) $\NN$ is the
normalization-factor of the distribution
\begin{equation}
\NN = \sum_{i=1}^{L^d} \frac{1}{r^\al} \; ,
\end{equation}
where $L$ is the linear size of the $d$-dimensional lattice.

Both the structure of these networks \cite{MOUKARZEL02} and diffusive dynamics
on these networks \cite{BLUMEN_2000b} have been studied by others, in one
dimension. Our results provides a detailed exploration of the phase diagram in
diffusion processes as the connection topology interpolates between the
original ``plain'' SW ($\al = 0$) and the purely short-range connected ($\al =
\infty$) network. One expects that when the probability of the random links
decays rapidly enough ($\alpha \to \infty$), the large scale behavior of the
processes on them is not affected by these links.  When the probability of the
random links decays slowly ($\alpha \to 0$), the results of the original SW
network should be recovered \cite{HASTINGS03,KOZMA04}.  In between, there is a
transition which is the focus of this paper.  Our starting point is a
calculation on rapidly-updated networks which turns out to be the first-order
approximation of a naive perturbation expansion for the quenched system.  For small
$p$, this expansion breaks down and we apply a self-consistent calculation.

Earlier numerical work \cite{BLUMEN_2000b} for $d=1$, in the context of the
random-walk, suggested that a crossover from the transient to the recurrent
phase occurs at around $\alpha\approx 2$. Here, we compute the asymptotic
scaling properties of the propagator analytically (and confirm it
numerically in $d=1$), and precisely construct the full phase diagram for
all $d$. This work provides a detailed account of the self-consistent
perturbative approach, briefly introduced in Refs. \cite{KOZMA04} and
\cite{KOZMA05b}, and expanded numerical results.

In our research we will consider different diffusion-driven processes, defined
in the following sections, {\em on} these networks.  Two different cases will
be distinguished: the {\it rapidly-updated} network, when the rearrangement of the
random links on the network takes place on a timescale much faster than the
process itself, and the {\it quenched} network, when the random links can be
considered static with respect to the timescale of the process.  In the
network literature, the first case of rapidly-updated networks is often
referred to as {\it annealed} networks, and we will refer to them as annealed
in this manuscript also.  We will denote the annealed average of a quantity by
$[...]^{ann}$ and the quenched average by $[...]$. In many cases, in the
quenched system, the effect of the random long-range links can be understood
by a general criterion \cite{HASTINGS03}, based on averaging over the
long-range links in a mean-field (or annealed) fashion (see Section \ref{ch:ann}.).  For a
few models, this criterion is violated, leading to non-mean-field behavior
such as diffusion on a small-world network with a one-dimensional spatial
structure.

Such a power-law SW network can emerge in a randomly-folded polymer where,
besides the one-dimensional chain of the monomers, two distant monomers can
be adjacent to each other, facilitating random long-range links, because of
the spatial conformation of the polymer \cite{SEN1}. For example, in three dimensions,
for an ideal chain, the probability of two distant sites to be adjacent is
proportional to $r^{-3/2}$ where $r$ is the distance of the two segments,
while for a polymer chain in a good solvent, this probability is
proportional to $r^{-1.97}$ \cite{GENNES79}. Note, that in the case of
small-world networks generated by such polymer configurations, the random
links are correlated: there is a high chance to find an other long-range
link around an existing one of the same length scale. In the case of
annealed networks, these correlations are ``washed away'' by the constant
rearrangements of the links but, in the quenched case, the correlations can
change the universal, large scale, behavior of the processes on them.  In
our study, we will deal with uncorrelated random links and leave the
correlated case for further research.

 Another field where such structure can emerge is that of
synchronization problems in distributed computing
\cite{KORNISS03,GUCLU_PRE06} where synchronization is achieved by
introducing random communications between distant processors.
Choosing a power-law SW network may be preferable as it lowers the
cost associated with communications. It was also argued that
``wiring-cost'' considerations \cite{PETERMANN05} for spatially
embedded
 networks, such as cortical networks \cite{LAUGHLIN03} or on-chip logic
networks \cite{DAVIS98}, can generate such power-law suppressed link-length
distribution. In general, wiring cost in SW networks can result to the
introduction of power-law SW networks \cite{PETERMANN05}.

\subsection{Diffusion processes }
\label{sec:diff}

One case in which the diffusion equation arises on a PL-SW network
is a macromolecule randomly moving along a polymer chain jumping
over adjacent segments with nonzero probability
\cite{SOKOLOV97,BLUMEN_2000b,MARKO04,BERG76}; if the macromolecule
motion is fast compared to rearrangements of the links, then the
network may be considered as quenched, while if the macromolecule
motion is slow compared to the link rearrangements then the
network is annealed \cite{BOCKMAN03}. The equation describing
random walk processes is
\begin{equation}
  \partial_t P_{\rr'}(t)=-\sum_{\rr''} \left( \Delta^0_{\rr' \rr''}+q
    \Delta^{rnd}_{\rr' \rr''} \right) P_{\rr''}(t),
\label{eqn:diffusion}
\end{equation}
where $P_{\rr'}(t)$ is the probability of finding the walker at
site $\rr'$ at time $t$ and $q$ is the relative {\em transition
rate} through the random links. Above we have used that the
diffusion operator $\DD_{\rr,\rr'}$ on SW networks, embedded in
regular $d$-dimensional networks, can be written as
\begin{equation}
\DD_{\rr,\rr'}=\DD^0_{\rr,\rr'} + q \DDr_{\rr,\rr'},
\label{eq:SWdiff}
\end{equation}
where $\DD^0$ is the regular $d$-dimensional Laplace operator on the
underlying lattice, $\DD_{\rr,\rr'}$ is the Laplacian on the random part of the network,
and $q$ is the relative strength of the relaxation
through the random links. For example, in one dimension, $\Delta_{ij}^0=
2\delta_{i,j} -\delta_{i,j-1} - \delta_{i,j+1}$. The diffusion through the
random links is
\begin{equation}
\Delta^{rnd}_{\rr,\rr'}=
\left\{
      \begin{array}{cl}
        -A^{rnd}_{\rr,\rr'}          & \mbox{if $\rr \ne \rr'$},\\
        \sum_{\rr'' \ne \rr} A^{rnd}_{\rr,\rr''}  & \mbox{if $ \rr = \rr'$}
      \end{array}
      \right.
\end{equation}
where $A^{rnd}_{\rr,\rr'}=1$ if there is a random long-range link
connecting sites $\rr$ and $\rr'$. For PL-SW networks [see
Eq.~(\ref{eq:PLSW_prob})],
\begin{eqnarray}
\mbox{Probability ($A^{rnd}_{\rr,\rr'}=1$)}&=&\frac{p}{\NN |\rr-\rr'|^\al}=p
  f(|\rr-\rr'|) \nonumber \\ \nonumber \\
\mbox{Probability ($A^{rnd}_{\rr,\rr'}=0$)}&=&1- p f(|\rr-\rr'|).
\end{eqnarray}

For technical purposes, it is often useful to employ the {\em spectral
  decomposition} of the diffusion operator. In the orthogonal eigensystem:
\begin{equation}
  \DD\Psi^k=\lambda_k \Psi^k
\end{equation}
for all $k$-s, where $\Psi^k$ is the normalized eigenvector and
$\lambda_k$ is the corresponding eigenvalue of the operator and
\begin{equation}
\langle \Psi^k|\Psi^l \rangle=\dd_{kl}.
\end{equation}
The $i=0$ index is reserved for the uniform zero-mode of the system:
\begin{equation}
  \lambda_0=0
\end{equation}
and
\begin{equation}
  \Psi^0=\frac{1}{\sqrt{L^d}} (1,1,1,...,1).
  \label{zero_mode}
\end{equation}

The Green's function, $G_{\rr',\rr''}(t)$, of the diffusion process is the
solution of the diffusion equation with initial condition
$P_{\rr'}(t=0)=\delta_{\rr',\rr''}$.  Because the diffusion equation has
only a first-order time derivative, formally its solution can be written as
\begin{eqnarray}
G_{\rr,\rr''}(t)&=& \sum_{\rr'} \left( 
e^{-(\DD^0+q\DDr)t}\right)_{\rr,\rr'} P_{\rr'}(t=0)
\nonumber \\
&=& \left(e^{-(\DD^0+q\DDr)t}\right)_{\rr,\rr''}.
\end{eqnarray}
Using the spectral decomposition of the diffusion operator one
obtains
\begin{equation}
e^{-(\DD^0+q\DDr)t}=\sum_k e^{-\lambda_k t}|\Psi^k\rangle
\langle\Psi^k| \;,
\label{propagator}
\end{equation}
where \( (|\Psi^k\rangle \langle\Psi^k|)_{\rr,\rr'} =\Psi^k_\rr
\Psi^k_{\rr'}\) is the projector to the subspace of $\Psi^k$.  It is often
useful to introduce the Laplace transform of the Green's function,
$G_{\rr',\rr''}(\omega)= \int_0^{\infty} G_{\rr',\rr''}(t) e^{-\omega t} d
t$. Performing the integration over the time domain, one finds
\begin{equation}
G_{\rr',\rr''} (\omega)=( \DD^0+q \Delta^{rnd} + \omega)^{-1}_{\rr',\rr''}.
\end{equation}

In order to characterize the diffusion process on different networks, one
of our focuses is on the scaling properties of the {\em expected number of
returns} of the random walker by time $T$. For one node at site $\rr$, it
is defined as
\begin{eqnarray}
F_\rr(T)&=&\int_0^T {\rm d}t P_\rr(t) \nn \\
 &=& \int_0^T {\rm d}t G_{\rr,\rr}(t).
\end{eqnarray}
Averaging over all nodes in the network one then has
\begin{equation}
F(T)=\frac{1}{L^d} \sum_{\rr} \int_0^T {\rm d}t G_{\rr,\rr}(t).
\end{equation}
We are interested in its long-time behavior in the thermodynamic
limit.  A random-walk process is called {\em recurrent} if in an
infinite system, the expected number of returns is infinite as $T
\to \infty$ and $transient$ if, on average, the walker visits the
origin only finite number of times. Using the spectral
decomposition of $G_{\rr,\rr}(t)$ from Eq.~(\ref{propagator}), one
obtains
\begin{eqnarray}
  F(T)&=& \frac{1}{L^d}  \int_0^T \sum_{k} e^{-\lambda_k t}
      \sum_{\rr}  |\Psi^k_\rr|^2 dt \nonumber \\
  &=&  \frac{1}{L^d} \sum_{k} \int_0^T  e^{-\lambda_k t}  dt  \nn \\
  &=& \frac{1}{L^d} \sum_{k \ne 0} \frac{1}{\lambda_k}
 \left(1-e^{-\lambda_k T}\right) + \frac{T}{L^d} \;.
\label{F_T_exact}
\end{eqnarray}
The above form is useful for exact numerical calculations for a given
realization of the network.

To analytically extract the leading-order asymptotic scaling properties for
$F(T)$, instead of the sharp cutoff, it is sufficient to use a kernel that
suppresses the integrand for $t \gg T$. To make our calculation easier, we
chose $\int_0^{\infty} G_{\rr,\rr}(t) e^{-t/T} {\rm d}t$
\cite{Redner}. From the spectral decomposition of the Green's function, and
repeating the same steps as above, now for the exponential cutoff, one
finds
\begin{equation}
F(T)\propto \frac{1}{L^d} \sum_\rr G_{\rr,\rr} (\omega=1/T) =\frac{1}{L^d} \sum_{k \ne 0}
\frac{1}{\lambda_k+1/T} + \frac{T}{L^d}.
\label{eq:F(T) spectral}
\end{equation}
In order to make the formalism similar to that of the other phenomena
described in this section, let us introduce a modified GF,
\begin{equation}
\hat{G}_{\rr,\rr'}(\omega)=G_{\rr,\rr'}(\omega) - \frac{1}{\omega L^d}.
\label{eq:Ghat}
\end{equation}
It is basically the GF defined in the space orthogonal to the uniform zero
mode. Note that, in the next section, defining the GF in this subspace
will be necessary because the coupling matrix is non-invertible.  In terms of
this modified GF
\begin{equation}
F(T)\propto \frac{1}{L^d} \sum_\rr \hat{G}_{\rr,\rr} (\omega=1/T)
+ \frac{T}{L^d}.
\label{eq:F(T) appr.}
\end{equation}

\subsection{Synchronization in task-completion networks
and the Edwards-Wilkinson process }
\label{sec:EW}

In this section, we motivate the Edwards-Wilkinson (EW) process
\cite{EDWARDS82} on networks, which can be thought of in terms of a
synchronization paradigm in a noisy environment. Consider a distributed
task-processing network, where the processing nodes, in order to schedule
and perform new tasks, must {\em wait} for the results delivered by other
nodes. This task-dependency between nodes correspond to the links, which in
general, can be asymmetric or directed, quenched or
time-dependent. Examples include distributed-computing systems
\cite{KORNISS03}, manufacturing supply chains, or e-commerce-based services
facilitated by interconnected servers \cite{NAGURNEY05}.  Understanding the
scaling behavior of the fluctuations in our model will help us better
understand the generic features of delays and back-log formation in
large-scale networked processing systems. In particular, for certain
distributed computing schemes (parallel discrete-event simulations)
\cite{LUBACHEVSKY00,GREENBERG94,LUBACHEVSKY87,LUBACHEVSKY88,FUJIMOTO,KOLAKOWSKA_review},
it was shown \cite{KORNISS00} that on regular grids, the evolution of the
task-completion (or synchronization) landscape is governed by the
Kardar-Parisi-Zhang (KPZ) \cite{KARDAR86} equation. The resulting landscape
is ``rough", corresponding to a large spread in the locally-completed tasks
on different nodes. In general, the width provides a sensitive measure of
the average degree of de-synchronization in task-completion networks,
\begin{equation}
w^2(t)=\frac{1}{N} \sum_{i=1}^{N} (h_i(t)-\bar{h}(t))^2 \;.
\label{w2}
\end{equation}
Here, $h_i(t)$ is the local progress on node $i$, $\bar{h}=(1/N)
\sum_{i=1}^{N} h_i(t)$ is the mean progress, and $N$ is the number
of processing nodes in the system. When the network has an
underlying $d$-dimensional structure with linear size $L$, one
also has $N=L^d$.

In order to improve the uniformity of the progress of the nodes in
the above distributed-computing example, ultimately leading to
better scalable performance, one must suppress large fluctuation
in the synchronization landscape. For example, one can construct a
scalable autonomous synchronization schemes where the nodes, in
addition to their nearest neighbor on the grid, also communicate
with random (possibly distant) neighbors (with a very low
frequency); the sole purpose of communications through the random
links is to keep the spread of the task-completion landscape under
control \cite{KORNISS03,GUCLU_PRE06}. In this case, the underlying
synchronization network is a SW network (which can be quenched or
annealed, depending on the implementation).

While the precise local synchronization rules give rise to
strongly non-linear effective interactions between the nodes, one
can gain some insight by considering the linearized version of the
effective equations of motion. As it was shown for the basic
distributed-computing synchronization problem \cite{KORNISS00},
the dynamics, neglecting non-linearities, can be effectively
captured by the EW process \cite{EDWARDS82}. The generalization of
the EW process to complex networks \cite{KOZMA04,KOZMA05b}
(focusing on PL-SW networks in this paper), is given by
\begin{equation}
\partial_t h_i(t)=-\sum_{j} \Delta_{ij}h_{j}(t) + \eta_{i}(t) \;,
\label{EW_process}
\end{equation}
where $\DD_{ij}$ is the diffusion operator {\em on the network}
and $\eta_{i}(t)$ (without loss of generality) is a
delta-correlated white noise with variance $2$ :
\begin{equation}
\langle \eta_i(t) \eta_j(t') \rangle =2 \dd_{ij} \dd(t-t').
\end{equation}
In general, the symmetric relaxational couplings between the
nodes, $A_{ij}$ ($A_{ii}\equiv 0$), can be weighted. The
generalized network Laplacian then can be written as
\begin{equation}
\Delta_{ij}=\delta_{ij}\sum_{k}A_{ik}-A_{ij} \;.
\end{equation}
For unweighted networks with unit coupling strength $A_{ij}$ is
simply the adjacency matrix.

Clearly, the precise scaling behavior of synchronization landscape
(its width, in particular) is strongly affected by {\em both} the
nonlinearities and the underlying interaction topology. Our
results provide a detailed account on the network's effect on the
EW synchronization problem. Because of the linear nature of the
model, this is a much simpler problem than the actual
synchronization dynamics, but, as we shall see in the following
sections, the behavior of it is still nontrivial, particularly on
quenched SW networks.  The understanding of the EW process (on
networks) is a necessary first step before tackling more
complicated, nonlinear problems.

Using the Fokker-Planck description of the EW process, it can be
shown \cite{ALB_HES} that the steady-state probability
distribution function (PDF) of the possible surface configurations
is
\begin{equation}
P [ \vec{h} ] \propto \exp \LL(-{H[\vec{h} ]}\RR),
\label{ss_pdf}
\end{equation}
where $\vec{h}=\{h_i\}_{i=1}^{L^d}$ is the configuration-vector and
$H[\vec{h}]$ is the steady-state Hamiltonian\footnote{One can consider this
  Hamiltonian as the energy of a given configuration of the surface and the
  PDF as a finite temperature distribution function of the possible
  surfaces, in analogy with the PDF of a canonical ensemble in statistical
  physics.} of the surface
\begin{equation}
H[ \vec{h} ]=\frac12 \sum_{ij} h_i \DD_{ij} h_j =
\frac12 \sum_{ij} A_{ij} (h_i-h_j)^2 \;.
\label{eq:H[h] Aij}
\end{equation}

The steady-state Green's function (GF) or {\em propagator} of the
system is defined as the two-point correlation function. Since the
steady-state PDF is Gaussian, the two-point correlation function
is the inverse of the coupling matrix. In our case, $\DD$ is
non-invertible because of the uniform eigenmode  $\Psi^{0}$ with
zero eigenvalue [Eq.~(\ref{zero_mode})]. The singularity of this
mode can also be traced back to Eq.~(\ref{EW_process}), in that
the mean height performs a simple random walk and has a diverging
variance in the limit of $t\to\infty$ (for any system size), i.e.,
its steady-state variance does not exist. Equivalently, the
steady-state PDF Eq.~(\ref{ss_pdf}) with the Hamiltonian
Eq.~(\ref{eq:H[h] Aij}) is ill-defined,
$\langle \bar{h}^2 \rangle = \mbox{``$\infty$''}$.
This problem can be overcome by constructing observables which lie
in the space orthogonal to the zero mode of the network Laplacian.
It can be achieved by, e.g., measuring height fluctuations from
the mean [see Eq.~(\ref{w2})]. Hence, the appropriately defined GF
is
\begin{equation}
G_{ij} = \langle (h_{i}-\bar{h}) (h_{j}-\bar{h}) \rangle = \hat{\DD}^{-1}_{ij}
\label{eq:EW_Greens}
\end{equation}
where $\hat{\DD}^{-1}$ is the inverse of $\DD$ in the space orthogonal to
the uniform zero mode. Particularly useful for numerical purposes,
one can employ the spectral decomposition of the inverse of the network Laplacian in the
space orthogonal to the zero mode
\footnote{In the literature, this construction is also referred to as
the pseudo-inverse of the network Laplacian.}
\begin{equation}
G_{ij} = \sum_{k=1}^{L^d-1} \frac{1}{\lambda_k} \Psi^k_{i} \Psi^k_{j}.
\end{equation}
Note that since the GF is constructed from the height fluctuations measured
from the mean, the zero mode automatically drops out from the sum.

Employing the GF, the average width can be expressed as
\begin{equation}
\langle w^2 \rangle = \LL\langle \frac{1}{L^d} \sum_{i=1}^{L^d}
(h_i-\bar{h})^2 \RR\rangle = \frac{1}{L^d} \sum_{i=1}^{L^d} G_{ii}
\label{eq:w2GF}
\end{equation}
or, for numerical purposes,
\begin{equation}
\langle w^2 \rangle = \frac{1}{L^d} \sum_{k=1}^{L^d-1} \frac{1}{\lambda_k}.
\label{eq:w2 spectral}
\end{equation}

%
%

\subsection{Other examples and connections between transport,
resistor networks, and the EW synchronization problem}

In the network research literature, an ``abstract'' transport
efficiency can be considered in analogy with electric
conductance properties of a network
\cite{WU04,LOPEZ05,WU05,ANDRADE05,KORN_PLA_swrn,TZENG06,GK_2006}.
The average resistance of a network gives a simplified but palpable interpretation of its
transport properties. If the average network resistance is low, it
can be interpreted as a network with good transport properties; if
the resistance is high, the network is likely to be a poor choice
for transportation purposes.

Consider an arbitrary network with $N$ nodes where the conductance of the link
between nodes $i$ and $j$ is $A_{ij}$. Skipping the details (which can be
found in, e.g., Refs.~\cite{WU04,KORN_PLA_swrn,GK_2006}), using Kirchhoff's
and Ohm's laws, one can show that the effective two-point resistance of a
network can be expressed in terms of the same network propagator (GF) as
defined in the previous examples [Eq.~(\ref{eq:EW_Greens})],
\begin{equation}
R_{ij}=G_{ii}+G_{jj}-2G_{ij} \;.
\end{equation}
Comparing the above formula to those of the EW process on the same network
(and with the same set of $\{A_{ij}\}$),
connections between the two different problems
can be readily obtained \cite{KORN_PLA_swrn,GK_2006}.
Calculating the {\em height-difference} correlation function,
\begin{eqnarray}
\langle (h_i-h_j)^2 \rangle &=&
\LL\langle \LL((h_i-\bar{h})-(h_j-\bar{h})\RR)^2 \RR\rangle \nn \\
&=& G_{ii} +G_{jj} -2G_{ij}=R_{ij} \;,
\label{Rij}
\end{eqnarray}
is equivalent to measuring the two-point resistance on the same network.
The average resistance (averaged over all pairs of nodes) becomes
\begin{equation}
 \bar{R}=\frac{1}{N(N-1)} \sum_{i \ne j} R_{ij}=\frac{2N}{N-1}\langle
 w^2\rangle \simeq 2\langle w^2\rangle \;,
 \label{R_w2}
\end{equation}
which is twice the average width for sufficiently large networks
  \footnote{To obtain the above result, one exploits that the GF is defined
  in the space orthogonal to the zero mode $\Psi^0$, hence, $\sum_j G_{ij}=
  \sum_j \sum_{k=1}^{N-1} \frac{1}{\lambda_k} \Psi^k_{i} \Psi^k_{j}=
  \sqrt{N} \sum_j \sum_{k=1}^{N-1} \frac{1}{\lambda_k} \Psi^k_{i}
  \Psi^k_{j} \Psi^0_j =0 $.}. These simple relationships are useful in
  understanding intuitively how the network topology influences the
  behavior of different processes in networks. Also, note that the above
  relationships, Eqs.~(\ref{Rij}) and (\ref{R_w2}), are valid for an
  arbitrary graph.  For networks embedded in a $d$-dimensional regular
  grid, such as the one considered in this paper, $N=L^d$.  For example,
  for a one-dimensional chain with unit resistances between each adjacent
  nodes, the two point resistance is $R_{ij} = |i-j|$, therefore, without
  any calculations, the height difference correlation function is also \(
  \langle (h_i-h_j)^2 \rangle = |i-j|\).

There is also an extensive literature on the connections between
resistor networks (with link conductances $A_{ij})$ and
discrete-time random walks (RW), with {\em transition
probabilities} $P_{ij}=A_{ij}/\sum_{k}A_{ik}$ from node $i$ to
node $j$ \cite{Doyle,Lovasz,Redner,Chandra,Tetali}. Our results,
to be obtained in this paper for the GF on distance-dependent SW
networks, hence provide answers for these related problems.

For example, in transport or flow problems on networks, one often
considers the ``load" of the nodes or the links, which can be
captured by an appropriately defined betweenness measure
\cite{FREEMAN77,NEWMAN04,NEWMAN05} (for a given dynamics). This
observable captures the amount of traffic, information, or data,
the nodes or the links typically handle in the corresponding
information, social, or infrastructure network. For example, in
random routing or search problems on complex networks, an
appropriate {\em node betweenness} can defined as {\em the
expected number of visits} to node $i$ by a random walker
(starting at a source node and before reaching its randomly chosen
target), averaged over all source and target pairs
\cite{Guimera_PRL04,Danila_PRE06}. It can be shown to be given by
\cite{GK_2006}
\begin{equation}
b_{i} = \frac{\sum_{k}A_{ik}}{2}\bar{R} \;,
\end{equation}
where $\bar{R}$ is the average two-point resistance of the
associated resistor network with link conductances $A_{ij}$. The
node betweenness can be used to obtain the congestion threshold in
queue-limited network traffic
\cite{Guimera_PRL04,Danila_PRE06,GK_2006}. Similarly, one can show
that the {\em link betweenness} $b_{ij}$ for the same RW problem
(the expected number of times a walker travels across the link
between nodes $i$ and $j$, averaged over all source and target
pairs) can be written as
\begin{equation}
b_{ij} = A_{ij}\bar{R} \;.
\end{equation}
The link betweenness, capturing the load of the connections
between the nodes, can be used to extract the congestion threshold
in bandwidth-limited traffic. Further, one can also show that the
average first-passage time for the above random-walk problem
(averaged over all source and target pairs) can be written as
\cite{Chandra,Tetali}
\begin{equation}
\overline{\tau} = \frac{\sum_{i,k}A_{ik}}{2}\bar{R} \;.
\end{equation}
As all of the above examples illustrate, the average two-point
resistance of a graph $\bar{R}$ (or, equivalently, the average
steady-state width of the associated EW synchronization landscape)
plays a fundamental role in the efficiency of prototypical
transport problems.

\section{ANNEALED SMALL-WORLD NETWORKS}
\label{ch:ann}

In the literature of networks the disorder of links is considered {\em
  annealed} or {\em mean-field} (MF) when it can be approximated by its
average value.  Such a situation can emerge when the process investigated on
the network feels only the average effect of the links due to the dynamics of
the system, in other words, when the timescale of the changes in the configuration of
the links is much shorter than the characteristic timescale of the process on
it.

If $O$ is some observable depending on a random potential $V$, accounting for
the effect of the random links, then $[O]^{ann}$ is calculated by replacing
$V$ with its expected value:
\begin{equation}
[O(V)]^{ann}=O([V])
\end{equation}
Specifically, for the propagator of the diffusion operator, the focus of
the present work, one has
\begin{eqnarray}
[G_{\rr',\rr''}]^{ann}
&=& \LL[\LL(\frac{1}{\hat{\DD}^0 +
    q\hat{\DD}^{rnd}}\RR)\RR]^{ann}_{\rr',\rr''} \nn \\
&=& \LL(\frac{1}{\hat{\DD}^0 + q [\hat{\DD}^{rnd}]}\RR)_{\rr',\rr''} \;.
\end{eqnarray}
Both in the annealed and the quenched case, the ensemble is
translationally invariant, and so will be the averages of the
observables of the system. \([G_{\rr',\rr''}]^{ann}\) will only
depend on the difference $\rr'-\rr''$. Let us introduce
\begin{eqnarray}
G^{ann}(\rr)=[G_{\rr',\rr'+\rr}]^{ann}
\end{eqnarray}
where $\rr$ is the spatial separation of two sites. Since translationally
invariant operators are diagonal in Fourier space, it is useful to
calculate the Fourier (or $k$-space) representation of them.  In Fourier
space the GF can be written as
\[
 G^{ann}(\kk)=\frac{1}{k^2+ q [\DDr](\kk)}
\]
where $k^2$ and $[\DDr](\kk)$ are the Fourier transform of \(\hat{\DD}^0\)
and \([\hat{\DD}^{rnd}]\) respectively.  The behavior of
\([\hat{\DD}^{rnd}]\) for $1/L \ll k \ll 1/a$ is calculated in Appendix
\ref{app:[DDr](k)}, yielding
\begin{equation}
[\DDr](\kk)\propto
   \left\{
      \begin{array}{ll}
        p & \mbox{if $\al<d$} \\
        p k^{\al-d}  & \mbox{if $d<\al<d+2$}\\
        p k^2  & \mbox{if $d+2<\al$} \; \; ,
      \end{array}
   \right.
\label{D_annealed}
\end{equation}
with logarithmic corrections at the boundaries of these different $\al$
regimes.  In order to  keep the treatment of the annealed and the quenched
system in parallel, we also call $[\DDr]$ the {\em annealed self-energy},
\(\Sigma^{ann}(\rr)=[\DDr](\rr)\). The real-space behavior can be obtained
by inverse Fourier transformation where necessary and, specifically,
\begin{equation}
  G^{ann}(\oo) = \frac{1}{L^d}\sum_{\kk \ne 0}
    \frac{1}{k^2+q\ [\DDr](k)} \; \; .
\label{eq:G(0) ann}
\end{equation}
In the present and the next sections, the behavior of $G^{ann}(\oo)$ will be
investigated. In Section \ref{ch:app & num}, physically measurable
quantities will be associated with $G^{ann}(\oo)$, like the surface width in the EW
model and the expected number of returns of random-walk processes.

For calculating scaling properties, we take the continuum limit of these
quantities. In this limit, $\rr$ is no longer constrained to the
lattice sites, so it can point in any direction with the condition \( a \le
|\rr| \le L \), the sums  replaced by integrals, and Kronecker
$\dd_{ij}$ by Dirac $\dd(\rr-\rr')$. To calculate the scaling properties of
$G^{ann}(\oo)$, with respect to $p$, $q$, and $L$, the sum can be approximated
with its integral-form limit:
\begin{equation}
  G^{ann}(\oo) \approx \frac{S_{d}}{(2 \pi)^d} \int_{1/L}^{1/a} \frac{dk
   k^{d-1}}{k^2+q\ [\DDr](k)} \; \; .
\label{eq:G(0) ann cont}
\end{equation}
where \( S_d\) is the surface of a $d$-dimensional unit sphere.

Mention must be made that mapping out the different phases of the annealed
network is a relatively easy task, and has been investigated by others in
other contexts \cite{SOKOLOV97,HARNAU96}.

The results can be summarized as follows.

\subsection{Phase diagram in one dimension }

\begin{figure*}[t]
\centering
\includegraphics[width=.6\textwidth]{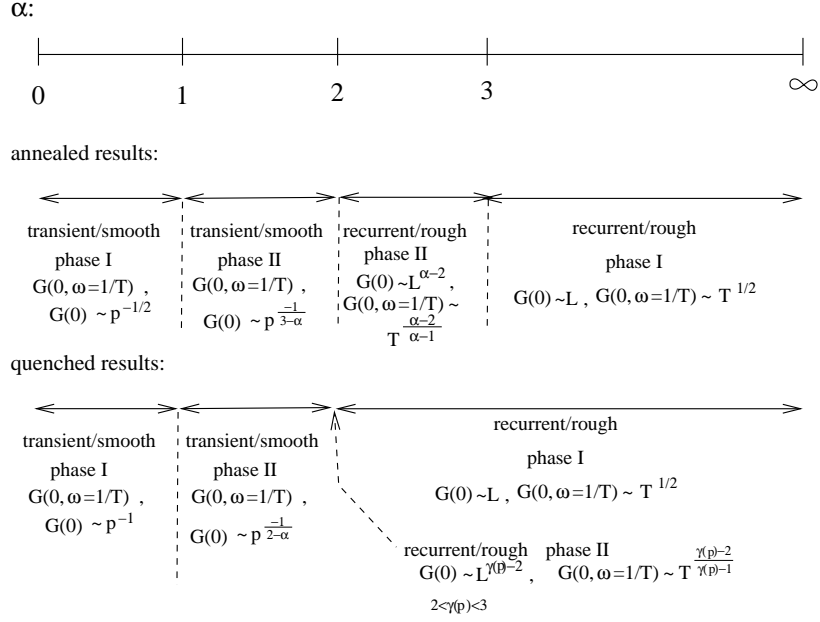}
\vspace*{-0.0cm}
\caption{
\label{fig:1d phases}
The one-dimensional phases.
}
\end{figure*}

The results for one dimension are sketched in Fig.~\ref{fig:1d phases}
and the details of the calculations can be found in Appendix
\ref{app:G(0)}.  The relationship of $G^{ann}(0)$ with measurable quantities is
explained in more detail in Section \ref{ch:app & num}.  For $\al<2$, the
integral, Eq.~(\ref{eq:G(0) ann cont}), converges at both ends, so is
system size independent for large $L$ and diverges as $p \to 0$. There are
two regimes, distinguished by their scaling behavior (reflecting on the
properties of the underlying random walk/surface) : (1) {\it
transient/smooth phase I}, where the system behaves as if the long-range
links were uniformly distributed, and $G^{ann}(0)$ is $\al$-independent; (2) {\it
transient/smooth phase II}, where $G^{ann}(0)$ is still finite in the
thermodynamic limit, but has an $\alpha$-dependent divergence as $p \to 0$.
For $\al>2$, $G^{ann}(0)$ has infrared divergence resulting in two system-size
dependent regimes: (1) {\it recurrent/rough phase I}, where the system
behaves as if there were no long-range links, and $G^{ann}(0)$ diverges linearly
with the system-size, $L$; (2) {\it recurrent/rough phase II}, with
sublinear scaling with respect to the system-size. At the boundaries of
these phases, logarithmic corrections are present.

\subsection{Phase diagram in two dimensions }

\begin{figure*}[t]
  \centering \includegraphics[width=.6\textwidth]{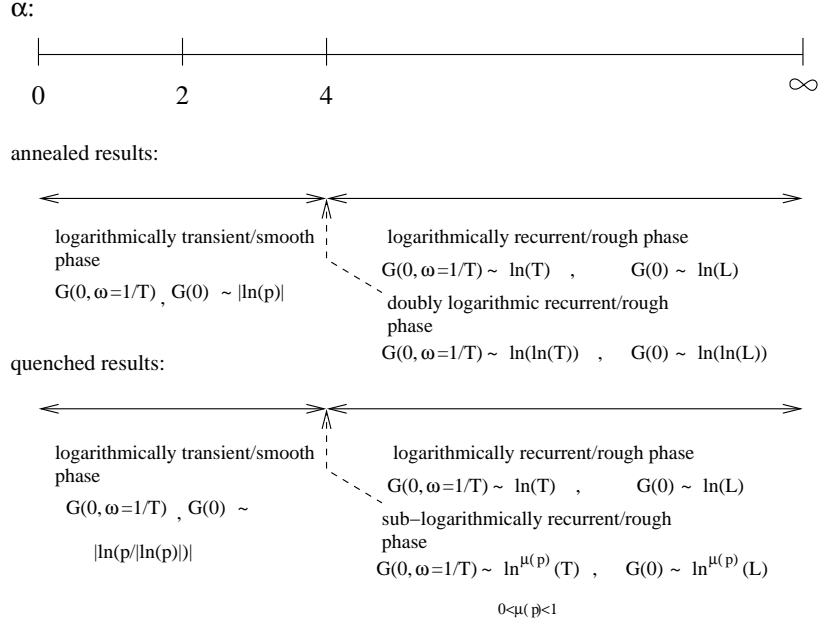}
  \vspace*{-0.0cm}
\caption{
\label{fig:2d phases}
The two-dimensional phases. (For sake of simplicity, factors of $a$, the
microscopic cutoff, were omitted from the formulas and can be calculated by
dimensional analysis.)
}
\end{figure*}

The behavior of the annealed two-dimensional system is summarized in
Fig.~\ref{fig:2d phases}. Three different phases are identified: a {\em
  logarithmically smooth/transient phase} when $\al<4$, a {\em
  doubly-logarithmic rough/recurrent phase} when $\al=4$, and a {\em
  logarithmically rough/recurrent phase} when $\al>4$. The details of the
calculations can be found in Appendix \ref{app:G(0)}.

\subsection{Higher dimensions, $d>2$}

In higher dimensions, the integral Eq.~(\ref{eq:G(0) ann cont}) diverges
for large $k$-s but has a well defined cutoff because of the lattice
spacing, $a$. As a result $G^{ann}(0)$ is always finite, which means that the
surface is always smooth and the random walker is always transient.

\section{QUENCHED SMALL-WORLD NETWORKS}
\label{ch:qu}

In this section we will develop a perturbation expansion, based on
impurity-averaged perturbation theory \cite{RAMMER91}, to calculate the
quenched average of the GF over the ensemble of random networks.  The reason
for investigating such an average is the so-called \emph{self-averaging
  hypothesis}, that for large enough networks the behavior of the GF, or in
general any global quantity, will only deviate slightly from the behavior of
its quenched average. This assumption is due to the recognition that if a
random system is large enough, the GF samples through a large amount of
randomness, which is a large number of random links in our case.  In
effect, the GF will be the same or very similar from one realization to an
other and for large enough system sizes it will converge to its quenched
value. In principle, this hypothesis can always be checked or proven after
setting up a formalism to calculate quenched averages over the randomness. To
do so, one has to obtain the quenched variance of the GF.  If the variance
disappears in the TD limit, the hypothesis is proven to be valid.

For example, the probability distribution of the random walker over the
nodes of the network for large times is expected to be self-averaging. One
can imagine that after long enough time the random walker will visit a
large number of sites, and therefore  encounter a large number of random
links.

In this section, we will investigate the limit when the density of random
links is small ($p \ll 1$). Due their small number, we will treat the
diffusion through these random links as perturbation of the $d$-dimensional
diffusion. As a first step, we will consider the ``pedagogical'' example of
a simpler problem, where there is only a single long-range link in the
system, which will help us to understand the different terms of the more
difficult problem of the perturbation expansion.  Second, we will set up a
naive perturbation expansion and show the breakdown of it. Third, we will
reformulate the expansion to obtain the leading-order behavior of the GF.
For special cases, we will obtain higher-order corrections and investigate
their behavior.

\subsection{The example of a single link}
\label{ss:sl}

In this section we will consider two  problems that have
similar mathematical structure as that of the problem of the quenched
network.  First, we will calculate the Green's function of a diffusion
process with a regular $d$-dimensional diffusion plus a ``perturbation''
which accounts for particles escaping from the origin with rate $q$.
Second, we will consider the Green's function of a process that will
consist of a $d$-dimensional diffusion plus diffusion between the sites $a$
and $b$ with rate $q$.

\begin{figure}[t]
  \centering
\includegraphics[width=.45\textwidth]{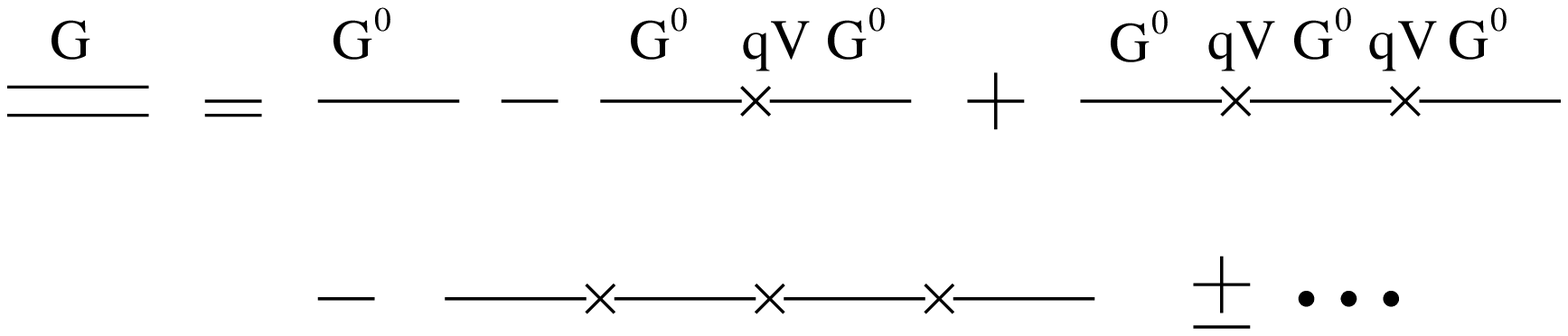}
\vspace*{-0.0cm}
\caption{
\label{fig:V}
The simple diagrams of the example of a singe link.
}
\end{figure}

For the first problem, the diffusion operator describing the process is
\begin{eqnarray}
\Gamma=\DD+qV,
\end{eqnarray}
where the perturbation to the regular diffusion is $V_{ij}=\ \dd_{i0}
\dd_{0j}$. For this process, the GF can be calculated as
\begin{eqnarray}
G&=&\frac{1}{\DD+qV}=\frac{1}{(1+qVG^0)\DD} \nn \\
 &=&G^0-G^0qVG^0+G^0 qV G^0 qV G^0 - ... \; \; ,
\label{eq:V}
\end{eqnarray}
where $G^0=(\DD^{0})^{-1}$ is the GF of the Laplacian on the regular
lattice. In Fig.~\ref{fig:V} a diagrammatic notation is introduced to
represent the above series expansion. Single lines, double lines, and
crosses denote $G^0$, $G$, and $V$ respectively.  Using the specific form
of $V$, $G$ can be calculated as
\begin{eqnarray}
G_{ij}&=&G^0_{ij} - G^0_{i0} q G^0_{0j} + G^0_{i0} q G^0_{00}  q
G^0_{0j}\nn \\
&-& G^0_{i0}  q G^0_{00}  q G^0_{00}  q G^0_{0j} + ... \nonumber \\
&=&G^0_{ij} - G^0_{i0} q \left(1-q G^0_{00} + q G^0_{00} q G^0_{00}
-... \right)G^0_{0j} \nn \\
&=&G^0_{ij} - G^0_{i0} \left( \frac{q}{1+q G^0_{00}} \right) G^0_{0j}
\label{eq:ped}
\end{eqnarray}

To have a better grip at what this result means, let us calculate $G$ for
one dimension. From Appendix \ref{app:massiveGF}, ${G^0_{ij}\approx L -
  |i-j|}$.  Substituting this to Eqn.~(\ref{eq:ped}) and noticing that
$\frac{q}{1+q G^0_{00}} = \frac{q}{((qL)^{-1}+1)qL} \approx
\frac{1}{L}(1+\frac{1}{qL})$,
\begin{eqnarray}
  G_{ij}&\approx& L-|i-j|-(L-|i|)\frac{1}{L}(1+\frac{1}{qL}) (L-|j|) \nn \\
&=&\frac{1}{q}+|i|+|j|-|i-j|.
\label{eq:single-sink 1d}
\end{eqnarray}

Before moving on to the next problem, in the case of this one-dimensional
lattice, let us have a closer look at the perturbation expansion of
Eq.~(\ref{eq:ped}) in light of the exact behavior of the GF,
Eq.~(\ref{eq:single-sink 1d}).  Comparing the first and the second order term
of the perturbation expansion,
\begin{eqnarray}
 G^0_{i0} (q) G^0_{0j} \ll G^0_{i0} (q^2 L) G^0_{0j}.
\end{eqnarray}
Even though the second order term is higher order in $q$, it diverges with
the system size. As it can be seen from the exact solution, this divergence
is not the property of the full GF, $G$, but that of the ill-defined
perturbation expansion. In the next section, a similar problem will be
encountered which will be resolved by reordering and resumming the series
expansion.

For the second problem, the perturbation is,
\begin{eqnarray}
  \label{eq:singlelink}
 q V^{(a,b)}_{ij}=q \ \dd_{ia}\dd_{aj}-
  q \dd_{ib} \dd_{aj} + q \ \dd_{ib} \dd_{bj} - q \dd_{ia} \dd_{bj} \;.
\end{eqnarray}
The meaning of the first two terms is that particles disappear at node $a$
with rate $q$ and appear at node $b$. The last two terms are responsible
for the inverse process from $b$ to $a$. In this case we have to solve the
same equation, Eq.~(\ref{eq:V}), as before only with a different perturbing
potential.

To make the calculation easier, let us rearrange the formula of $V^{(a,b)}$,
\begin{eqnarray}
q V^{(a,b)}_{ij} = q(\dd_{ia}-\dd_{ib}) (\dd_{aj}-\dd_{bj}) = q(UW)_{ij}\;,
\label{eq:V^(a,b) fact.}
\end{eqnarray}
where we factorized $V^{(a,b)}$ as a product of two matrices
$U_{ik}=(\dd_{ia}-\dd_{ib})$ and $W_{kj}=(\dd_{aj}-\dd_{bj})$. This way,
the $n^{th}$ ($n>1$) term of the series expansion of Eq.~(\ref{eq:V}) will
become
\begin{eqnarray}
G^0 q U (W G^0 q U)^{n-2} W G^0.
\end{eqnarray}
The three parts of this expression are $(G^0U)_{ik}=G^0_{ia}-G^0_{ib}$,
$(WG^0)_{lj}=G^0_{aj}-G^0_{bj}$, and $(WG^0U)_{ij}=(\dd_{ak}-\dd_{bk})
G^0_{kl} (\dd_{la}-\dd_{lb}) =G^0_{aa}-G^0_{ab}-G^0_{ba} +
G^0_{bb}=2(G^0_{aa}-G^0_{ab})$. In the last step, we used the fact that
$G^0$ is symmetric and translationally invariant. Therefore, the $n^{th}$
term in the series expansion is $G^0_{ik}(\dd_{ka}-\dd_{kb})q \left(2 q
  (G^0_{aa}-G^0_{ab})\right)^{n-2} (\dd_{al}-\dd_{bl})G^0_{lj} $. Let us
compare it to the $n^{th}$ term of the previous problem: there we had
$G^0_{ik} \dd_{k0} q \ ( q G^0_{00})^{n-2} \ \dd_{0l}G^0_{lj}$. As it can
be seen, the second problem has the same structure in its expansion as the
previous one. The closed form of the GF is
\begin{widetext}
\begin{eqnarray}
  G_{ij}= G^0_{ij} - (G^0_{ia}-G^0_{ib})
    \left( \frac{q}{1+2 q (G^0_{aa}-G^0_{ab})} \right) (G^0_{aj}-G^0_{bj}).
\end{eqnarray}
\end{widetext}

\subsection{From a naive to a self-consistent single-link perturbation expansion}

In the case of the random networks, a similar perturbation expansion can be
done as in the previous section for a given realization of the random
links.  The method, Eq.~(\ref{eq:V}), is still valid though in this case
the perturbation, $V$, is more complicated: it consist of a sum of terms
like (\ref{eq:singlelink}) corresponding to the random links in the system.
In principle, the GF could be calculated for any long-range-link
configuration. Though, such an expression is not really useful for
extracting the dependence of the GF on the parameters of the probability
distribution of the links.  Averaging over all the realizations will get
rid of the details due to the specifics of a given realization and provide
us with the dependence on the statistical properties.

\begin{figure}[t]
\centering
\includegraphics[width=0.45\textwidth]{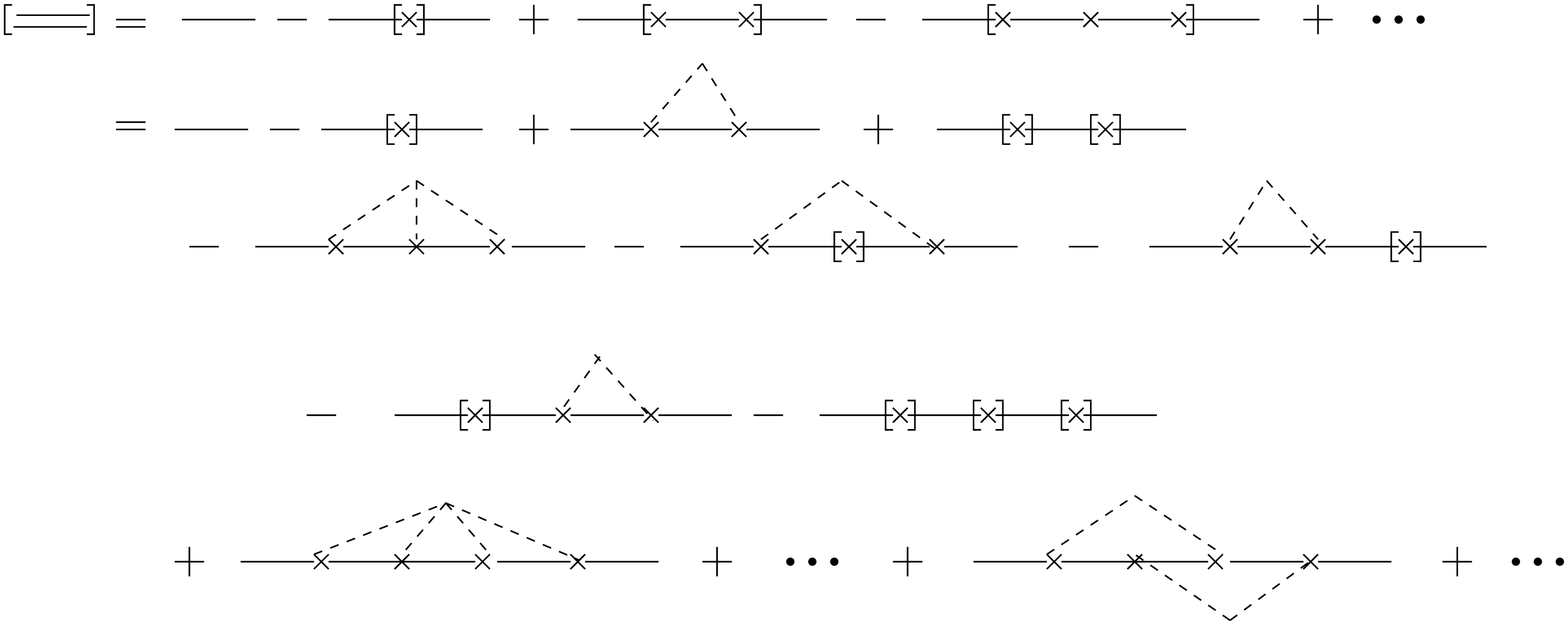}
\vspace*{-0.0cm}
\caption{
\label{fig:[GF]}
The averaged Green's function in terms of the moments and cumulants of
$\DDr$.  Single lines, double lines, and crosses denote $G^0$, $G$, and $q
\DDr$ respectively. $[ \ ]$ denote the disorder average of an expression
with respect to the random potential $\DDr$. Crosses connected by dashed
lines represent the cumulant of those random variables (see
Fig.~\ref{fig:cumulants}).  }
\end{figure}

Here, a similar diagrammatics will be used as in the previous section in
order to keep track of the different terms in the expansion. For the
definition of the symbols see Fig.~\ref{fig:[GF]}.

First, let us take the disorder average, represented by $[ \ ]$, of the
perturbation expansion of the GF (see Fig.~\ref{fig:[GF]}). In this
expression, the higher- and higher-order terms have higher- and
higher-order moments of the perturbation potential, $\DDr$. In probability
theory, moments are not really suitable for any type of expansion since
higher- and higher-order terms have larger and larger expectational values.
Cumulants are more appropriate to describe a random variable since they
decay as their order tends to infinity for well behaving PDF-s.

\begin{figure}[b]
\centering
\includegraphics[width=.45\textwidth]{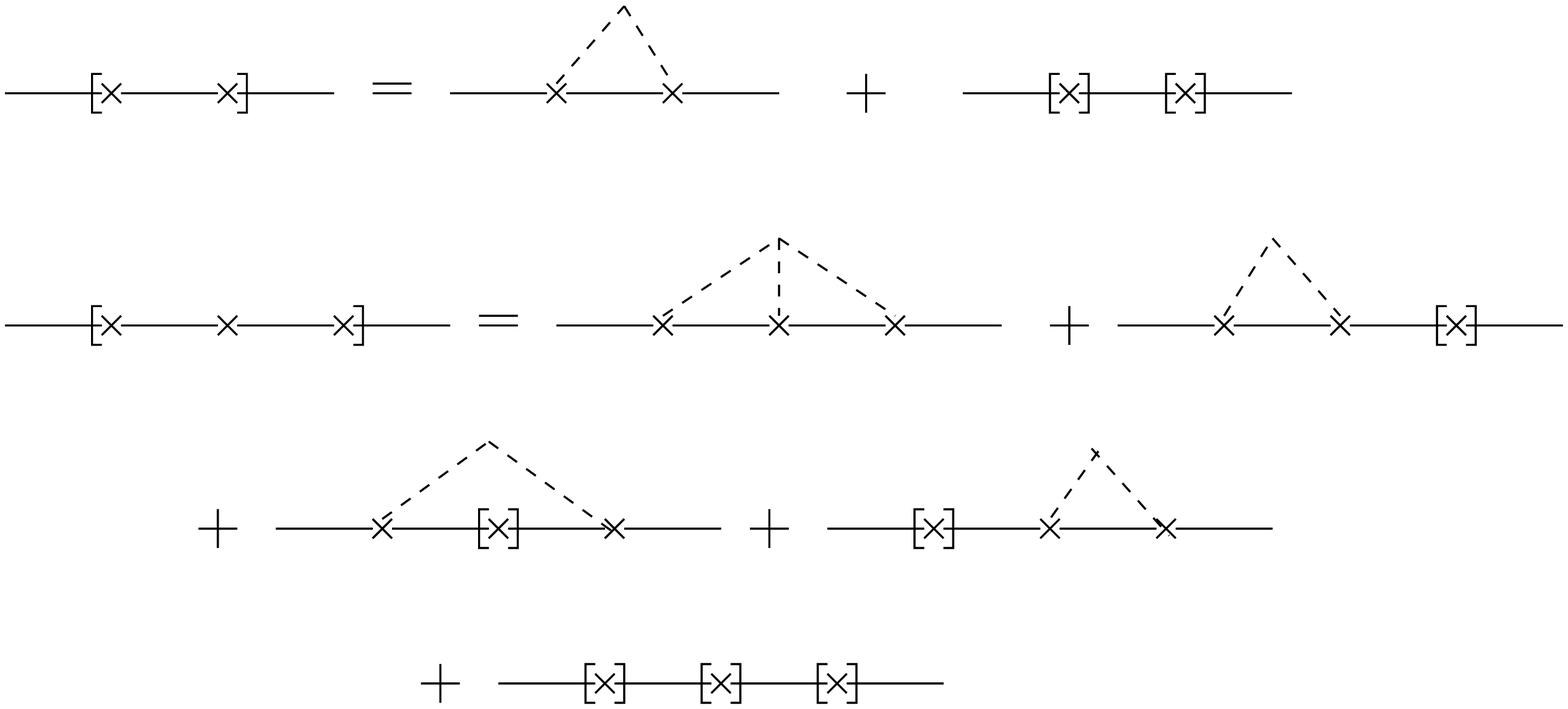}
\vspace*{-0.0cm}
\caption{
\label{fig:cumulants}
The second and the third moments in terms of cumulants. The 
notation is the same as in Fig.~\ref{fig:[GF]}.
}
\end{figure}

From a diagrammatic point of view, an important property of the cumulants
is that the moment of an $n^{th}$-order scattering is the sum of all the
possible combination of cumulants of the $n$ random potentials (see
Fig.~\ref{fig:cumulants}). One might consider this as a recursive
definition of the $n^{th}$-order cumulant in terms of the lower-order
cumulants and the $n^{th}$-order central moment. It is also used to define
the self-energy.  In the theory of disordered systems, the {\em
self-energy} is a potential added to the unperturbed Hamilton operator and
accounts for the average effect of the impurities, i.e. the random links,
in the system. The recursive property of the cumulants is also deployed in
self-consistent calculations where single lines are substituted with double
lines as a result of the recursive resummation of different order
single-line diagrams.

\subsection{Cumulants of the random perturbation potential}
\label{subsec:cumulants}

In the case of PL-SW networks, the entries of the matrix of the
perturbation potential are described by binary PDF-s: there is either a
random link between two sites or there is none. These PDF-s are
characterized by $p$: the probability of having a link between two
arbitrary sites, $i$ and $j$, is $p f(|i-j|)$ and the probability of having
none is $1-p f(|i-j|)$ (see Eq.~(\ref{eq:PLSW_prob})). The cumulants of
this distribution are shown in Table 3.1. 

\begin{table}
\centering
\begin{tabular}{p{1.8 cm}|p{3.5 cm}|p{2.5 cm}}
Order of cumulant & Binary distribution & Poisson {distribution}\\
\hline
1st & $p$ & $p$ \\
2nd & $p-p^2$ & $p$\\
3rd & $p-3p^2+2p^3$ & $p$ \\
4th & $p-7p^2+12p^3-6p^4$ & $p$\\
$\vdots$ & $\vdots$ & $\vdots$
\end{tabular}
\label{tab:cumulants}
\caption{Comparison of the cumulants of the binary and the Poisson
 distributions with expected value $p$. }
\end{table}

Observe that all the cumulants are of order $p$ to leading order. The
cumulants do not decay as the degree of them increases. Some might think
that for PDF-s with such a property an expansion in terms of the cumulants
should not work since all terms are of the same order with respect of the
expansion parameter, $p$. Later on in this section we will revisit this
question. As for now, let us assume that a cumulant expansion could work in
principle.

The random long-range links are independently distributed.  Defining
\begin{eqnarray}
  x^{(a,b)}=\left\{
      \begin{array}{ll}
        1 & \mbox{ with probability $p f(|a-b|)$}\\
        0 & \mbox{ with probability $1-p f(|a-b|)$},
      \end{array}
      \right.
\label{eq:x-bin}
\end{eqnarray}
the contribution of each random link to the perturbation potential can be
formulated as
\begin{eqnarray}
 q \DDr_{ij}=\sum_{a < b} q V^{(a,b)}_{ij} x^{(a,b)},
\end{eqnarray}
where $V^{(a,b)}$ is the {\em single-link} diffusion operator between $a$
and $b$ introduced in the previous section (see Eq.~(\ref{eq:singlelink})).
Therefore, the cumulants are:
\begin{widetext}
\begin{eqnarray}
[q \DDr_{ij}]_c &=& \sum_{a < b} q V^{(a,b)}_{ij} [x^{(a,b)}] \ ,
      \nonumber \\
\left[q \DDr_{ij} q \DDr_{kl}\right]_c &=& \sum_{a < b}
      q V^{(a,b)}_{ij} \sum_{c < d} q V^{(c,d)}_{kl} [x^{(a,b)}
      x^{(c,d)}]_c \nonumber \\
      &=& \sum_{a < b} q V^{(a,b)}_{ij} \sum_{c < d}
         q V^{(c,d)}_{kl} [(x^{(a,b)})^2]_c \dd_{ac} \dd_{bd} \\
      &=& \sum_{a< b} q^2 V^{(a,b)}_{ij} V^{(a,b)}_{kl} [(x^{(a,b)})^2]_c \ ,
      \nonumber \\
\left[q \DDr_{ij} q \DDr_{kl} q \DDr_{nm}\right]_c &=&
      \sum_{a < b} q^3 V^{(a,b)}_{ij} V^{(a,b)}_{kl} V^{(a,b)}_{nm}
      [(x^{(a,b)})^3]_c \ . \nonumber \\
&\vdots& \nn
\label{eq:cumulants}
\end{eqnarray}
\end{widetext}
The $n^{th}$ order cumulant of $\DDr$ will be the function of that of
$x^{(a,b)}$ . To make the analytics slightly easier, let us change the
statistical properties of the random networks a little in a way that does
not change the behavior of the ensemble in the $p \to 0$ limit. Instead of
binary random variables, as introduced in Eq.~(\ref{eq:x-bin}), Poisson
distributed ones will be used with
\begin{eqnarray}
\mbox{Probability}(x^{(a,b)}=n)=
    \frac{\left({pf(|a-b|)}\right)^n}{n!}e^{-pf(|a-b|)}. \nn \\
\end{eqnarray}
Using Poisson distribution, a negligible fraction of the nodes will have
multiple links between them, but this fraction is so small ($O(p^2)$) that
it will not affect the average properties of the system. From a mathematical
point of view, though, the forms of the cumulants became much simpler
\begin{eqnarray}
 [(x^{(a,b)})^n]_c=pf(|a-b|)
\end{eqnarray}
as it can be seen in Table \ref{tab:cumulants}.

Let us have a closer look at the expressions in Eq.~(\ref{eq:cumulants}) and
interpret the mathematical formulas. For example, the third order cumulant
consist of a sum of terms like \(V^{(a,b)}_{ij} V^{(a,b)}_{kl}
V^{(a,b)}_{nm}\).  Comparing it to the terms of the single-link problem of
the previous section, one can see that it corresponds to the $3^{rd}$ order
scattering off a single link between $a$ and $b$. In general, the $n^{th}$
order cumulant is a sum of $n^{th}$ order single-link scatterings weighted
with the expected number of those links, $pf(|a-b|)$.

Using the specific form of $pf(|a-b|)$ and $V^{(a,b)}$
\footnote{See Eqs.~(\ref{eq:PLSW_prob}) and (\ref{eq:singlelink})}, let us
  calculate the exact form of the different cumulants:
\begin{eqnarray}
q [\DDr_{ij}]_c&=&\sum_{a < b} (\dd_{ia}-\dd_{ib}) (\dd_{aj}-\dd_{bj})
\frac{p}{\NN |a-b|^\al} \nn \\
 &=&
    \left\{
      \begin{array}{cl}
        q p                    & \mbox{if $i = j$} \\
        - \frac{q p}{\NN |i-j|^\al} & \mbox{if $i \ne j$} \; ,
      \end{array}
      \right.
\end{eqnarray}
i.e. the first cumulant is the annealed ``superdiffusion'' operator. Note
that
\begin{eqnarray}
[\DDr_{ii}]_c=-\sum_{i\ne j} [\DDr_{ij}]_c \; ,
\label{eq:conservative}
\end{eqnarray}
namely, the diffusion is ``{\em conservative}''.

For the higher-order cumulants, in order to make it easier to interpret the
results, let us have $G_{ij}$ in between the random perturbation operators,
as they will appear in the formulas of the following sections (see Section
\ref{sec:sc}),
\begin{eqnarray}
[q\DDr_{ik}G_{kl}q\DDr_{lj}]_c&=&\sum_{a < b} (\dd_{ia}-\dd_{ib})
(\dd_{ak}-\dd_{bk}) \nn \\ &\times& G_{kl} (\dd_{la}-\dd_{lb})
(\dd_{aj}-\dd_{bj}) \frac{q^2 p}{\NN |a-b|^\al} \nonumber \\ &=&\sum_{a <
b} (\dd_{ia}-\dd_{ib}) (\dd_{aj}-\dd_{bj}) \nn \\ &\times& \frac{ 2
(G_{aa}-G_{ab}) q^2 p }{\NN |a-b|^\al}.
\label{eq:2nd cumulant}
\end{eqnarray}
This is also a superdiffusion operator as in the previous case but the
weight of the matrix-entries changed to \( \frac{ 2 (G_{aa}-G_{ab}) q^2 p
}{\NN |a-b|^\al} \). One can easily generalize to the $n$th order cumulant:
If there are $n$ crosses connected by a dashed line, we call these {\em
single-link diagrams}, and when double lines are running in between them,
the corresponding matrix is
\begin{eqnarray}
\sum_{a < b} (\dd_{ia}-\dd_{ib}) (\dd_{aj}-\dd_{bj}) \frac{ (2 q
  (G_{aa}-G_{ab}))^{n-1} q p }{\NN |a-b|^\al}
\label{eq:nth scsl}
\end{eqnarray}
resulting from the aforementioned properties of multiple scatterings from a
single link operator, $V^{(a,b)}$. The generalization to cases when there
are different operators in between the cumulants of the perturbation
potential is straightforward.

\subsection{Introduction of the self-energy, self-consistent resummation
  of the diagrams}
\label{sec:sc}

\begin{figure}[t]
\centering
\includegraphics[width=.45\textwidth]{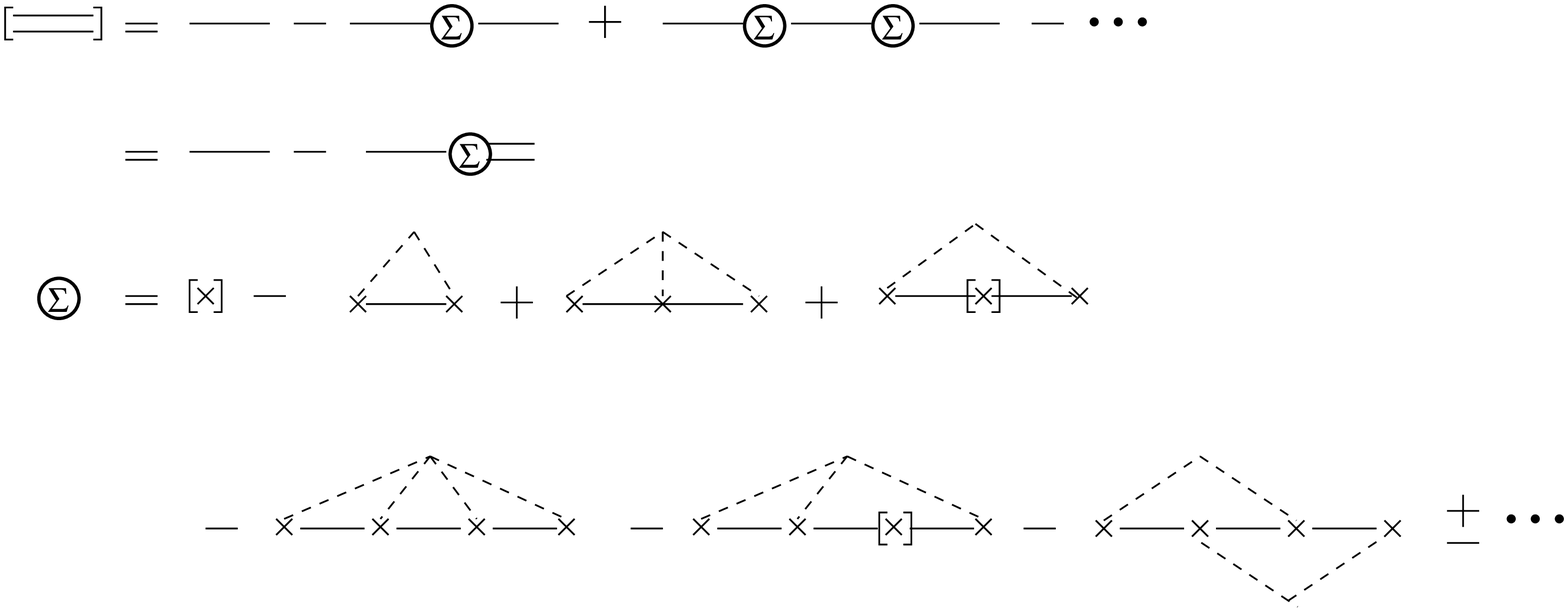}
\vspace*{-0.0cm}
\caption{
\label{fig:self-energy}
The introduction of the self-energy, $\Sigma$. The 
notation is the same as in Fig.~\ref{fig:[GF]}.
}
\end{figure}

As in quantum field theory, the diagrams of Fig.~\ref{fig:[GF]} can be
rearranged so that the self-energy is defined. The {\em self-energy }
consist of diagrams that are {\em irreducible}, i.e. they cannot be cut into
two pieces by cutting one of the propagators, $G^0$, in them (see Fig
\ref{fig:self-energy}). In terms of mathematical equations:
\begin{eqnarray}
G=G^0-G^0 \Sigma G = (\Delta+\Sigma)^{-1}.
\label{eq:DE}
\end{eqnarray}

\begin{figure}[t]
\centering
\includegraphics[width=.45\textwidth]{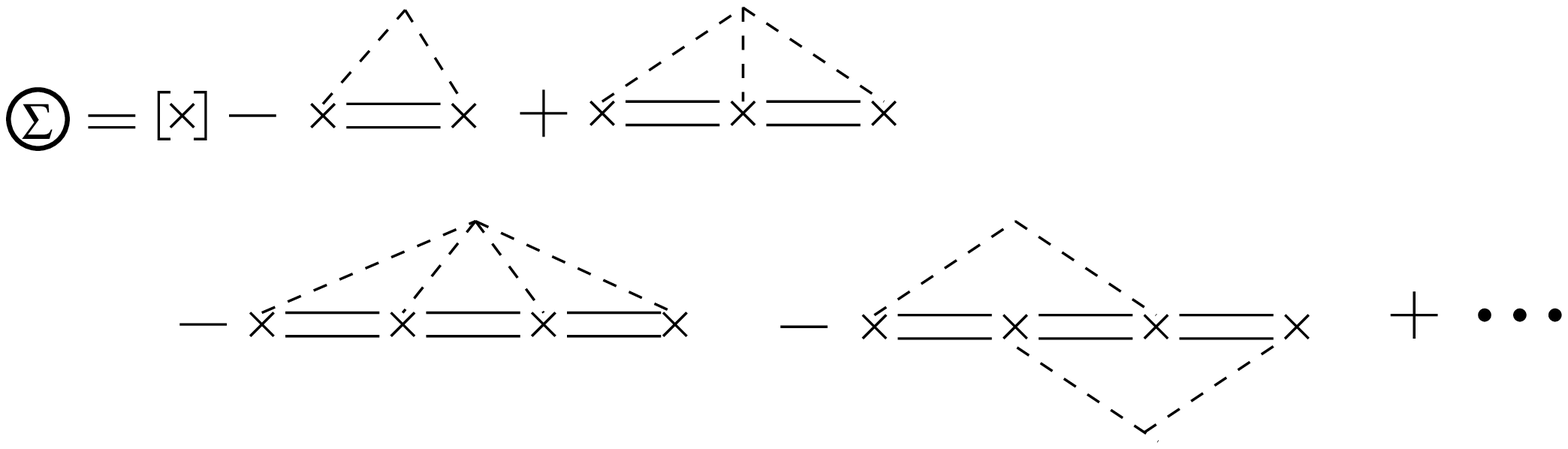}
\vspace*{-0.0cm}
\caption{
\label{fig:Sigma^sc}
The self-energy self-consistently. The 
notation is the same as in Fig.~\ref{fig:[GF]}.
}
\end{figure}

At this point, let us compare some higher-order diagrams to the first-order
one in the self-energy. Let us consider diagrams with $n$ crosses connected
by a dashed line and single lines running in between them in one dimension.
In this case, $(G^0_{aa}-G^0_{ab}) \propto r$, where $r=|a-b|$. The
contribution of such a diagram to the self-energy is
\begin{eqnarray}
\Sigma^{sl}_n(r) \approx - q^n \frac{p \ (2r)^{n-1}}{\NN r^\al}
\end{eqnarray}
and
\begin{eqnarray}
\Sigma^{sl}_n(0) &=& \sum_{r\ne 0} \Sigma^{sl}_n(r) \approx q^n p \int_a^L
\frac{(2r)^{n-1}}{\NN r^\al} dr \nn \\ &\propto& q^{n-1} p L^{n-1-\al}
\stackrel{L \to \infty}{\longrightarrow} \infty,
\end{eqnarray}
for some large enough $n$. Just like in the example
of a single link, the diagrams diverge with the system size.  These
divergences can be avoided by including enough intervening scattering
events of the propagator between two crosses \cite{RAMMER91}. In terms of
diagrams, it means that a set of diagrams are resummed in between two
crosses. If there is a subset of processes with intervening scatterings
that will avoid the system-size dependent divergence of the higher-order
terms, then resumming over all the possible scattering events between two
crosses should also work since it includes those processes too. In terms of
diagrams, it means that single lines are replaced by double lines (see
Fig.~\ref{fig:Sigma^sc}).  To avoid the overcounting of the diagrams, one
must note that, expressing $\Sigma$ in terms of double lines, some diagrams
of $\Sigma$ with single lines will become ``obsolete.'' For example, in
Fig.~\ref{fig:Sigma^sc}, there is no double line counterpart of the fourth
diagram of $\Sigma$ in Fig.~\ref{fig:self-energy}. This is so because such
diagrams are already included in the second diagram of Fig.~\ref{fig:Sigma^sc}.
In general, to avoid overcounting, the disallowed
diagrams in this expansion of $\Sigma$ are the ones that can be cut into
two by cutting two double lines in them.

In order to have a better idea as to how this resummation happens, let us
find all the single-link diagrams that contribute to the second-order
self-consistent diagram of $\Sigma$ in Fig \ref{fig:Sigma^sc}. Note that
this diagram comes with sign $(-1)$. The double line between the cumulant
of the two scatterings is built of single lines and cumulants of the
scatterings between them.  Expanding the double line in terms of single
lines, diagrams with $m$ scatterings have sign $(-1)^{m}$ in $G$. Therefore
their contribution to the investigated diagram has sign $(-1)(-1)^m$.

In the single-line expansion of $\Sigma$ lets consider a subset of diagrams
within the $(m+2)^{nd}$ order scatterings. Let this specific subset consist of
diagrams which have the second-order cumulant of the first and the last
scattering (i.e. the crosses representing them are connected to each other
by a dashed line but not to any other crosses) and all the possible
combination of cumulants of the intermediate $m$ scattering potentials.
Note that these are all parts of the self-energy since they cannot be cut
to two by cutting one of the single lines in them.  The sign of these
diagrams is $(-1)^{(m+1)}$ which agrees with the sign of the $m$-th order
term of $G$ above. Therefore, we found all the single line diagrams that
contribute to the $m$th order term of $G$ in the self-consistent diagram
investigated.

The above procedure of replacing single lines with double lines is called
{\em self-consistent perturbation expansion} in the literature of
field-theory. Assuming that the expansion works, to obtain the $n^{th}$
order correction to the self-energy one has to calculate the Green's
function using the $(n-1)^{st}$ order approximation of $\Sigma$ then
substitute this GF into the $n$th order diagram of $\Sigma$. In other
words, $\Sigma$ has to be calculated using {\em successive approximation}.

As done in the single-line case, let us compare the first- and second-order
diagrams of the self-energy, denoted by $\Sigma^1$ and $\Sigma^2$, to check
the validity of this new self-consistent expansion.
As long as the higher-order diagrams are higher order in the expansion parameter ($p$ or $q$) the
perturbation expansion works.

Without loss of generality, let us investigate the long-distance behavior of
two diagrams: the first-order diagram, the annealed diffusion operator,
\begin{eqnarray}
  \Sigma^{(1)}(r) = \frac{qp}{\NN r^\al};
\end{eqnarray}
and the second-order diagram, from Eq.~(\ref{eq:2nd cumulant}),
\begin{eqnarray}
 \Sigma^{(2)}(r) =  \frac{2 (G(0)-G(r)) q^2 p}{\NN r^\al}.
\end{eqnarray}
Note that, in $\Sigma^2(r)$, $(G(0)-G(r))$ is approximated to first order,
i.e. it is calculated with only the first diagram in the self-energy. The
details of its calculations are given in Appendix \ref{app:G(0)-G(r)}.

In one dimension, there are four different cases:\\
1) $\al<d$: \((G(0)-G(r))\propto (pq)^{-\frac{1}{2}}\) and
\begin{eqnarray}
 \Sigma^2(r)\propto \frac{1}{r^\al} (qp)^{-\frac{1}{2}} pq^2.
\end{eqnarray}
\(\Sigma^2(r) \ll \Sigma^1(r)\) if \( q \ll p\). The perturbation expansion
can be applied in the weak interaction limit, \(q \to 0\), keeping the
number of links fixed, but breaks down in a network with a few but strong
long-range links, as \(p \to 0\).\\
2) $d<\al<d+1$: \((G(0)-G(r))\propto
(pq)^{\frac{-1}{3-\al}}\) and
\begin{eqnarray}
\Sigma^2(r) \propto \frac{1}{r^\al} (qp)^{\frac{-1}{3-\al}} pq^2.
\end{eqnarray}
\(\Sigma^2(r) \ll \Sigma^1(r)\) if \( q^{2-\al} << p\). A similar condition
as in case 1) though, here, the strength of the interaction must have a
smaller value for the expansion to be applicable.\\
3) $d+1<\al<d+2$: \((G(0)-G(r)) \propto \frac{1}{qp}r^{\al-2}\) and
\begin{eqnarray}
\Sigma^2(r) \propto \frac{q}{r^2}.
\end{eqnarray}
The perturbation expansion breaks down for all parameter values.\\ 4)
$d+2<\al$: The perturbation expansion breaks down. Though, unlike in the
cases above, in this regime the effect of the random links is negligible at
large distances to that of the regular diffusion on the chain in both the
annealed and the quenched system.

In two dimensions there are two cases:\\
1) $\al<d+2$: \((G(0)-G(r))\propto - \ln (pq a^2)\). Therefore, the
condition for the validity of the perturbation expansion is \(
q^{-1}e^{-1/q} \ll p\).  The perturbation expansion works when $q \to 0$.
In the limit of a small number of strong links, the situation is better than
in the one-dimensional cases since the expansion is valid for a large
range of $p$ values, but still breaks down in the $p \to 0$ limit. \\
2)$d+2<\al$: Similar to case 4) in one dimension, the perturbation
expansion breaks down, though the diffusion through the random links are
irrelevant at large distances compared to the regular lattice diffusion.

In three and higher dimensions \((G(0)-G(r))\propto a^{d-2}\) and
\begin{eqnarray}
\Sigma^2(r)\propto \frac{pq^2 a^{d-2}}{r^\al}.
\end{eqnarray}
The expansion can be applied in the weak interaction limit. In the case of
vanishingly small number of links, unlike in the lower-dimensional case,
the higher-order terms do not dominate the behavior of the self-energy. In
fact, all the higher-order terms are the same or higher order in $p$ as
$\Sigma^1$. The appearance of higher-order diagrams with same
$p$-dependence is due to the fact that the cumulants of $\DDr$ are all the
same order in $p$. Though, diagrams with multiple cumulants
(see Fig.~\ref{fig:2cumulants}) are of order $p^2$. As it will be shown later, the
approximation of the first order (annealed) diagram is valid in this case,
the perturbation expansion provides the right scaling of the self-energy.

In lower dimensions, $d \le 2$, the failure of the self-consistent
perturbation expansion in terms of the expansion parameter $p$ is due to
the fact that diagrams with multiple scatterings off a single link (i.e.
diagrams with only one cumulant in them) contribute with the same weight,
$p$, to the self-energy. These scattering processes also contain the GF
propagating between two scattering events. The GF is determined from lower
order approximations of the self-energy, and therefore is a function of $p$
itself.  In the $p \to 0$ limit, the GF should converge to the original
pure GF, $G^0$, that is divergent as a function of the system size in these
dimensions.  Therefore, $G$ is divergent as well when $p \to 0$. Following
this argument, one can understand that the higher-order single-link
(or one-cumulant) diagrams dominate the small-$p$ limit behavior of the
self-energy since the first-order, single-cross, diagram does not depend on
the GF, while all the other single-link diagrams do.

As it was done in Subsection \ref{ss:sl}, the infinite set of single-link
diagrams\footnote{These are the diagrams where all the crosses are
connected together by dashed lines. Diagrams with crosses connected to
multiple sets by dashed lines are not single-link diagrams, like those of
Fig.~\ref{fig:2cumulants}.} can be summed.  Using Eq.~(\ref{eq:nth scsl}),
\begin{eqnarray}
\Sigma^{sl}_{ij} &=& \sum_{n=1}^{\infty} \sum_{a < b} (\dd_{ia}-\dd_{ib})
(\dd_{aj}-\dd_{bj}) \nn \\
&\times& \frac{ (2 q (G_{aa}-G_{ab}))^{n-1} q p }{\NN |a-b|^\al} \nonumber \\
 &=& \sum_{a < b} (\dd_{ia}-\dd_{ib})
(\dd_{aj}-\dd_{bj}) \nn \\
&\times& \frac{p}{\NN |a-b|^\al} \frac{q}{1+2q(G_{aa}-G_{ab})}.
\label{eq:slsc discrete}
\end{eqnarray}
Since the operator is translationally invariant and the continuum limit is
meaningful one can write
\begin{eqnarray}
  \Sigma^{sl}(\rr) = \frac{-p}{\NN r^\al} \frac{q}{1+2q(G(\oo)-G(\rr))}
 \label{eq:SCF}
\end{eqnarray}
Since the self-energy is a conservative (diffusion) operator,
\begin{eqnarray}
\Sigma^{sl}(\oo)= - \sum_{\rr \ne \oo} \Sigma^{sl}(\rr).
\end{eqnarray}
This property can also be derived from Eq.~(\ref{eq:slsc discrete}). The
above pair of equations is referred to as the {\em Self-Consistent Formula}
(SCF).

In order to solve the self-consistent formula, Eq.~(\ref{eq:SCF}),
the following steps are to be taken: \\
First, an {\em ansatz} has to be made
about the large-distance behavior of the self-energy. In most cases, we can
make a very good guess by assuming that the long-distance spatial behavior
of the self-energy will be the same as that of the self-energy of the
annealed system,
\begin{eqnarray}
\Sigma^{sl}(\rr)=\frac{s}{r^\al} \propto \Sigma^{an}(\rr) = [\DDr](\rr)
\ \ \  \mbox{if $|\rr| \gg a$},
\label{eq:ansatz}
\end{eqnarray}
and the strength of the interaction, $s(p,q)$, will be some function of $p$
and $q$ determined by the self-consistent formula. \\
Second, $(G(\oo)-G(\rr))$ has to be calculated using the above ansatz, \(
G(\rr) = (\DD + \Sigma^{sl})^{-1} (\rr)\). As done before, in order to
calculate $G(\rr)$, the Fourier transforms of the operators are used
because of their diagonal form in $k$-space:
\begin{eqnarray}
G(\rr)=\int_{1/L}^{1/a}\frac{e^{i \kk \rr}}{k^2+\Sigma(k)} \frac{d^dk}{(2
\pi)^d}
\end{eqnarray}
and
\begin{eqnarray}
(G(\oo)-G(\rr))=\int_{1/L}^{1/a} \frac{1-e^{i \kk \rr}}{k^2+\Sigma(k)}
  \frac{d^dk}{(2 \pi)^d}
\label{eq:G(0)-G(r) 1d}
\end{eqnarray}
The calculation of the scaling of the latter quantity can be found in
Appendix \ref{app:G(0)-G(r)} for various forms of $\Sigma(k)$.\\
Third, the asymptotic form of $(G(\oo)-G(\rr))$ is substituted into the
self-consistent formula \footnote{ This procedure can also be considered as
  an {\em iteration process} since, if the initial ansatz was incorrect,
  one can take the result of the SCF as a new ansatz of $\Sigma$ and apply
  the same steps over and over again until $\Sigma$ converges to a function
  that does not change in the iteration process.}. If our ansatz about
$\Sigma(\rr)$ was correct in the first step, the self-consistent formula has
to be satisfiable by choosing $s$ appropriately; if it was not correct, a
new ansatz has to be made.\\
Fourth, having $\Sigma^{sl}$, $G(0)$ is calculated to obtain physically
measurable quantities of the system. The details of the calculations are as
follows:

\subsection{The one-dimensional phases}
\label{ss:1d quenched}

The leading-order results for one dimension are sketched in Fig.~\ref{fig:1d
  phases}.  Similar phases appear as were predicted by the annealed argument.
But, in the transient/smooth phases, we have a different scaling property - a
faster divergence as $p$ vanishes.  Furthermore, the recurrent/rough phases
have different phase-boundaries. Recurrent/rough phase I, where the effect
of the long-range links is negligible, spans a wider interval of the
$\alpha$-axis, and recurrent/rough phase II is collapsed to one point on
this axis. In this phase, the sublinear behavior of $G(0)$ is no longer
determined by the distance-distribution of the random links but rather by
their density, $p$, through Eq.~(\ref{eq:SCeq}).  The form of this equation is
approximate, but the fact that the exponent $\mu$ depends continuously on $p$
is likely to be exact, in light of the scaling arguments in Section
\ref{sec:ren}.

For $0<\al<1$, let us go through the process step by step.
The ansatz, Eq.~(\ref{eq:ansatz}), is $\Sigma(r)=\frac{s}{r^\al}$, and its Fourier
transform is $\Sigma(k) \approx s$. In the second step, from Appendix
\ref{app:G(0)-G(r)},
\begin{eqnarray}
(G(0)-G(r))\propto \frac{1}{\sqrt{s}}
\end{eqnarray}
for large $r$-s.  Substituting this asymptotic form into the SCF, Eq.~(\ref{eq:SCF}),
\begin{eqnarray}
  \frac{s}{r^\al}=\frac{p}{\NN r^\al} \frac{q}{1+\frac{2q}{\sqrt{s}}}
  \stackrel{s \to 0}{\longrightarrow} \frac{p \sqrt{s}}{\NN r^\al}.
\end{eqnarray}
Therefore
\begin{eqnarray}
  s \propto p^2.
  \label{eq:s_1d_plain}
\end{eqnarray}
In the last step we assumed that $s \to 0$ as $p \to 0$. This assumption,
which is indeed satisfied, can be checked for consistency after obtaining
the functional form $s(p)$. In the final step, from Appendix
\ref{app:G(0)},
\begin{eqnarray}
 G(0)=\int_{1/L}^{1/a}\frac{dk}{k^2+p^2}\propto p^{-1}.
\end{eqnarray}
For $3<\al$, the Fourier transform of the ansatz is $\Sigma(k) \propto
k^2$. Therefore,
\begin{eqnarray}
(G(0)-G(r)) \propto r.
\end{eqnarray}
Substituting this in the SCF,
\begin{eqnarray}
\frac{s}{\NN r^\al} \ne \frac{p}{\NN r^\al} \frac{1}{r}.
\end{eqnarray}
The spatial behavior of the two sides of the equation are different, our
ansatz does not work. This problem can be fixed easily by setting
\(\Sigma(r)=\frac{s}{\NN r^{\alpha+1}}\). The Fourier transform of any self
energy that decays faster than \(r^{-(d+2)}\) is \(\Sigma(k) \propto k^2\),
therefore \((G(\oo)-G(\rr)) \propto r \) is still valid for the new ansatz.
Using this new form of $\Sigma(r)$, the self-consistent equation is
satisfiable if $s \propto p$.  The scaling behavior of the GF is
$G(\oo)\propto L$. \\
For $1<\al<2$, the Fourier transform of the ansatz is $\Sigma(k) \propto s
k^{\al-1}$. Therefore,
\begin{eqnarray}
(G(0)-G(r)) \propto  s^{\frac{-1}{3-\al}} \; .
\end{eqnarray}
Substituting this asymptotic form into the self-consistent formula,
\(
 s \propto p^{\frac{3-\al}{2-\al}}
\) and
\begin{eqnarray}
 G(0) \propto p^{\frac{-1}{2-\al}}.
\end{eqnarray}
For $2<\al<3$, the Fourier transform of the ansatz is $\Sigma(k) \propto s
k^{\al-1}$, which is the same as in the previous case but $G(r)$ has a
different long-distance property:
\begin{eqnarray}
(G(0)-G(r)) \propto \frac{1}{s} r^{\al-2}.
\end{eqnarray}
Using this result in the self-consistent formula,
\begin{eqnarray}
\frac{s}{\NN r^\al} \ne \frac{p}{\NN r^\al} \frac{s}{r^{\al-2}},
\end{eqnarray}
A better choice is $\Sigma(r) \propto \frac{s}{r^{\al+1}}$. Since
$\al+1>3$, its Fourier transform is $\Sigma(k) \propto s k^2$, and
therefore \((G(0)-G(r)) \propto r\). The SCF is satisfiable if $s \propto
p$, resulting in
\begin{eqnarray}
  G(0) \propto L.
\end{eqnarray}
For $\al=2$, the Fourier transform of the ansatz is $\Sigma(k)=k$. Therefore,
\begin{eqnarray}
(G(0)-G(r)) \propto \frac{1}{s}\ln (rs).
\end{eqnarray}
Substituting into the SCF,
\begin{eqnarray}
\frac{s}{\NN r} \ne \frac{p}{\NN r} \frac{s}{\ln (rs)},
\end{eqnarray}
the two sides have a different $r$ dependence. One iteration step generated
a faster decaying function. This tells us that the fixed point of the SCF
has to be a function with a faster decay than that of out initial ansatz at
long distances. Let us modify the ansatz of $\Sigma$, to $\Sigma(r)\propto
\frac{s}{n r^{\gamma}}$, where $2<\gamma<3$ and $n$ is a constant factor.
For self-energies with such an exponent, \((G(0)-G(r))\) has already been
calculated in the $2<\al<3$ case above. As it will be shown below, in this
phase the ($\gamma$ dependent) constants in $\Sigma(r)$ will have a crucial
role in determining the exponent $\gamma$.

For this reason one has to calculate the scaling properties of the ansatz
of the self-energy with more scrutiny: from Appendix \ref{app:[DDr](k)} the
result of the Fourier transformation is
\begin{eqnarray}
\Sigma(k)=\frac{s}{n_1}k^{\gamma-1} \; ,
\end{eqnarray}
where \(n_1=n \LL( 2 \sin \left(\frac{\pi}{2}\gamma\right) \Gamma(1-\gamma)
\RR)^{-1}\). Also, from Appendix \ref{app:G(0)-G(r)}, using the above form
of the self-energy,
\begin{eqnarray}
(G(0)-G(r)) \approx \frac{n_2}{s} r^{\gamma-2}
\end{eqnarray}
for large distances, where
\begin{eqnarray}
n_2=\frac{n}{2 \pi} \frac{\cos \left(\frac{\pi}{2}\gamma\right)
  \Gamma(2-\gamma)} {\sin
  \left(\frac{\pi}{2}\gamma\right)\Gamma(1-\gamma)} = \frac{n}{2 \pi} \cot
  \left(\frac{\pi}{2}\gamma\right) (1-\gamma) \; .\nn \\
\end{eqnarray}
Substituting this into the SCF,
\begin{eqnarray}
\frac{s}{n r^\gamma} = \frac{p}{\NN r^2}\  \frac{s}{n_2
  r^{\gamma-2}}=\frac{p\ s}{\NN n_2}\ \frac{1}{ r^\gamma} \; .
\end{eqnarray}
The two sides of the equation have the same functional form. Though $s$ is
undetermined by the single link approximation, we will see later that some
information is still known. The only remaining tunable parameter is
$\gamma$ (in $n_2$), to equate the two sides of the formula.  Noting
that \( \NN=2 \sum_{i=1}^{\infty} \frac{1}{i^2}=\frac{\pi^2}{3}\), the
equation determining the value of $\gamma$ as a function of $p$,
\begin{eqnarray}
 1=p \ \frac{3}{2 \pi} \frac{ \ \tan(\frac{\pi}{2}\gamma)}{\gamma-1}, \ \
 \mbox{where $2<\gamma<3$}
\label{eq:SCeq}
\end{eqnarray}
has to be solved for different $p$-s numerically. Therefore,
\begin{eqnarray}
  G(0) \propto \frac{1}{s} L^{\gamma(p)-2} \;.
\label{G_L_degenerate}
\end{eqnarray}

Note that even though the single-link approximation does not give
information about $s(p,q)$, one can still argue about its $q$ dependence:
The dimension of $s$ has to be that of $(length)^{\gamma-3}$ since $k^2$
and $sk^{\gamma-1}$ should have the same dimensions in $G(k)$. Since $p$ is
dimensionless for $d=1$ and $\al=2$, $q$ with dimension $(length)^{-1}$ is
the only parameter of the problem that can restore the dimensionality of
$\Sigma(k)$.  Therefore one may expect that $s(p,q) \propto
q^{3-\gamma}f(p)$ where $f(p)$ is undetermined.

\subsection{The two-dimensional phases}
\label{sec:2dquenched}

The results for the quenched two-dimensional system are summarized in
Fig.~\ref{fig:2d phases}. The overall difference between the annealed and
the quenched cases is that the quenched self-energy acquires logarithmic
corrections to that of the annealed system. As a result the quenched phases
have stronger (or equal) logarithmic divergences than their annealed
counterpart.  Three different phases are identified: a {\em logarithmically
  smooth/transient phase} when $\al<4$, a {\em sub-logarithmically
  rough/recurrent phase} when $\al=4$, and a {\em logarithmically
  rough/recurrent phase} when $\al>4$. To map the phase space, the
single-link self-consistent formula [Eqn.~(\ref{eq:SCF})] is used to obtain
the leading order behavior of the GF.

The steps of solving the self-consistent formula are the same as in the
one-dimensional case. The details are as follows:\\
For $\al<2$ the Fourier transform of the ansatz is \(\Sigma(k) \propto s\).
Therefore,
\begin{eqnarray}
(G(\oo)-G(\rr)) \propto \ln (s a^2)
\end{eqnarray}
and the result of the self-consistent equation as $s \to 0$ is
\begin{eqnarray}
  \frac{p}{\ln s}=s
\end{eqnarray}
so
\begin{eqnarray}
s \approx \frac{p}{\ln p} + ...
\label{eq:s_2d_plain}
\end{eqnarray}
and
\begin{eqnarray}
G(\oo) \propto \ln \left(
  \frac{p}{\ln p}\right).
\label{eq:G(0)_2d_plain}
\end{eqnarray}
For \(2<\al<4\) the Fourier transform of the ansatz is \(\Sigma(k) \propto
s k^{\al-2}\). Therefore,
\begin{eqnarray}
(G(\oo)-G(\rr)) \propto \ln (sa^{4-\al})
\end{eqnarray}
and the result of the self-consistent equation is the same as in the
previous case: $s \approx \frac{p}{\ln p} + ...$ and \(G(\oo) \propto \ln
\left( \frac{p}{\ln p}\right)\).\\
For $\al>4$ the Fourier transform of the ansatz is \( \Sigma(k) \propto
sk^2\). Therefore,
\begin{eqnarray}
(G(\oo)-G(\rr))&=&\int_{1/L}^{1/a}\frac{1-J_0(kr)}{(1+s)k^2} \frac{k dk}{(2
\pi)^2} \nn \\ &\propto& \int_{1/r}^{1/a}\frac{dk}{k}=\ln (r/a).
\end{eqnarray}
Substituting this to the self-consistent formula,
\begin{eqnarray}
  \frac{s}{\NN r^\al} \ne \frac{p}{\NN r^\al}\frac{1}{\ln (r/a)}.
\end{eqnarray}
Since the spatial behavior of the two sides of the equation are different,
our ansatz does not work. As in the one-dimensional problem, this problem
can be fixed by setting \(\Sigma(r)=\frac{s}{\NN r^\alpha \ln (r/a)}\). The
Fourier transform of any self-energy that decays faster than \(r^{-(d+2)}\)
is \(\Sigma(k) \propto k^2\), and therefore \((G(\oo)-G(\rr)) \propto \ln
(r/a)\) is still valid for the new ansatz. Using this new form of
$\Sigma(r)$, the self-consistent equation is satisfied if $s \approx p$.
The scaling behavior of the GF is
\begin{eqnarray}
G(\oo)\propto \ln L.
\end{eqnarray}

For $\al=4$, let us look into the details. In real space, \( \Sigma(\rr) =
\frac{s}{\NN r^4}\), and its Fourier transform is \( \Sigma(k)=s k^2 \ln
(ka) \) (see the calculations of Appendix \ref{app:[DDr](k)}).  Therefore,
from Appendix \ref{app:G(0)-G(r)},
\begin{eqnarray}
(G(\oo)-G(\rr))\propto \frac{1}{s} \ln (s \ln (r/a)).
\end{eqnarray}
Substituting the above result into the self-consistent equation,
\begin{eqnarray}
\frac{s}{r^4} \ne \frac{p}{\NN r^4} \frac{s}{\ln (s \ln (r/a))}.
\end{eqnarray}
The two sides of the equation do not match because the iteration process
generated a double-logarithmically stronger decay of the self-energy. If
one tried another ansatz, $\Sigma(r) \propto \frac{1}{r^{4+\epsilon}}$,
$\epsilon \ll 1$, the iteration step would weaken the decay making it
$\propto \frac{1}{r^4 \ln (r/a)}$. Therefore the fixed point of the
self-consistent equation for $\al=4$ should be a function with a decay
between $\frac{1}{r^4}$ and $\frac{1}{r^{4+\epsilon}}$. After some guess
work, a better ansatz is
\begin{eqnarray}
 \Sigma(\rr)= \frac{s}{n \  r^4 \ln^\mu (r/a)},
\end{eqnarray}
where $0>\mu>1$ and $n$ is a constant.  Its Fourier space behavior from
Appendix \ref{app:[DDr](k)} for small $k$-s is
\begin{eqnarray}
\Sigma(k) \propto \frac{s c_1(\mu)}{n} k^2 \ln^{1-\mu}(ka) +
   \mathcal{O}\left( \frac{k^2}{\ln^\mu (ka)} \right) \; ,
\end{eqnarray}
where $c_1(\mu)$ is a constant, the result of the Fourier transformation.
From Appendix \ref{app:G(0)-G(r)}
\begin{eqnarray}
G(\oo)-G(\rr) \propto \frac{n \ c_2(\mu)}{s \ c_1(\mu)} \ln^\mu (r/a) +\mathcal{O}
  \left(\ln^{\mu-1}(r/a) \right) \; , \nn \\
\end{eqnarray}
where $c_2(\mu)$ is a constant, the result of the inverse-Fourier
transformation.  Substituting this result to the self-consistent formula
when $r \to \infty$,
\begin{eqnarray}
\frac{s}{n \ r^4 \ln^\mu (r/a)} = \frac{p}{\NN r^4} \times
\frac{n \ c_2(\mu)}{s \ c_1(\mu)}
\times \frac{1}{\ln^{\mu} (r/a)}
\end{eqnarray}
yielding
\begin{eqnarray}
1=\frac{p}{\NN} \times \frac{c_1(\mu)}{c_2(\mu)} \; .
\label{eq:2d al 4 SCF}
\end{eqnarray}
where $\NN=\sum_\rr \frac{1}{r^4}$ is the normalization constant of the
probability distribution of the random links, and $c_1(\mu)$ and $c_2(\mu)$
are constants defined by the Fourier transformation of specific functions.
As in the one-dimensional case, $\mu$ became a continuously varying
function of $p$, defined by Eq.~(\ref{eq:2d al 4 SCF}). Though, in this
calculation, $c_1$ and $c_2$ are not determined, in future research one may
try to perform the Fourier transformations more accurately or determine
their value numerically by performing discrete-Fourier transformation.\\
Finally, one finds
\begin{eqnarray}
G(0) &\approx& \int_{1/L}^{1/a} \frac{k dk}{k^2+s k^2 \ln^{1-\mu} (ka)} \\
&\propto& \frac{1}{s} \ln^{\mu(p)}(L), \ \ \ 0<\mu(p)<1,
\end{eqnarray}
a ``{\em sub-logarithmic}'' divergence with the system-size.

From numerical and experimental point of view, one must note that the above
logarithmic differences are difficult to measure.

\subsection{Higher dimensions, $d \ge 3$}

In the naive perturbation expansion it was seen that higher-order
single-link diagrams give the same order contribution in $p$ as the first
order one. From the SCF, one can verify this result since the GF is finite
and determined by its behavior at the lattice scale (see Appendix
\ref{app:G(0)-G(r)}),
\begin{equation}
(G(\oo)-G(\rr)) \approx g \propto a^{d-2} \;,
\end{equation}
where $g$ is some constant
and the exact value of it is determined by the microscopic details of the
lattice.  Substituting this into the SCF,
\begin{equation}
 \Sigma^{sl}(r)=\frac{p}{\NN r^\al}\frac{q}{1+2qg}
\end{equation}
and $G(0)$ is independent of $p$ and $q$ to leading order, $G(0) \approx g
+ ...$.

\subsection{Higher-order corrections}
\label{ssec:HOrdCorr}

\begin{figure}[t]
\centering
\includegraphics[width=.45\textwidth]{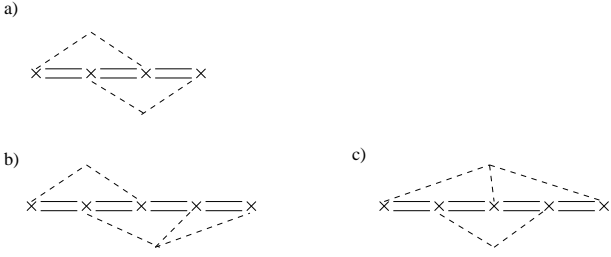}
\vspace*{-0.0cm}
\caption{
\label{fig:2cumulants}
Two cumulant diagrams. The 
notation is the same as in Fig.~\ref{fig:[GF]}. Moreover, square brackets, $[\ ]$ were
dropped from around double lines for the sake of clarity. 
}
\end{figure}

In the previous sections the self-energy was approximated with the infinite
sum of single-link diagrams. Here, the contribution of higher-order
diagrams is investigated in the case of the plain SW network ($\al=0$) and
the other cases, $\al \ne 0$, are left to future research. The first such
diagram is that of Fig.~\ref{fig:2cumulants}(a), the lowest-order
two-cumulant diagram. It is called a two-cumulant diagram since it has two
sets of crosses: in each set, crosses are connected to each other by dashed
line.  According to the definition of the diagrams, these sets correspond
to separate cumulants of the perturbation potential.

\begin{figure}[t]
\centering
\includegraphics[width=.4\textwidth]{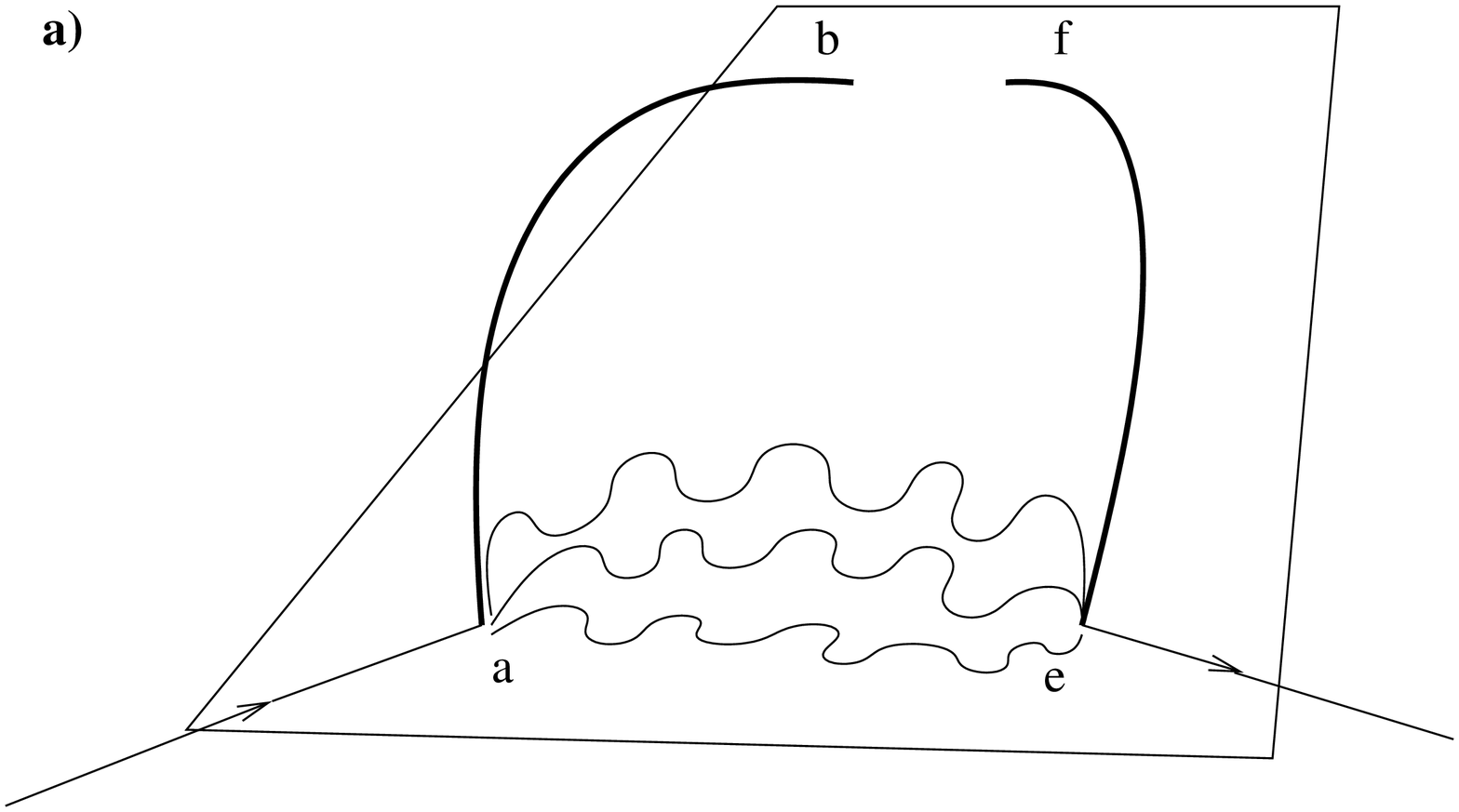}
\includegraphics[width=.4\textwidth]{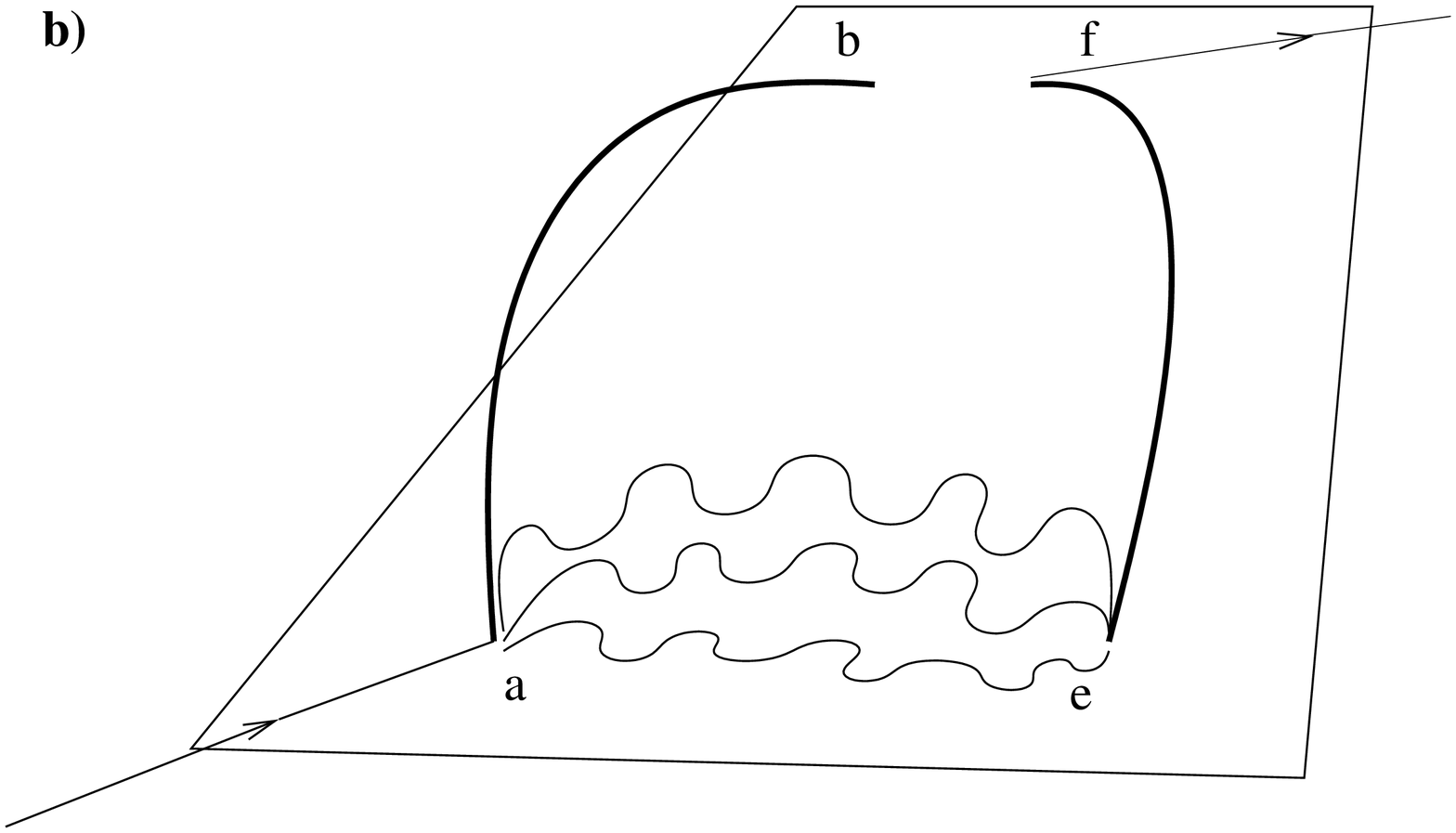}
\vspace*{-0.0cm}
\caption{
\label{fig:propagators1}
Sketch of the leading order processes in the two-cumulant diagram of
Fig.~\ref{fig:2cumulants}(a). Links are represented by fat lines, GF-s are
represented by wavy lines, and the underlying $d$-dimensional lattice is
represented by a flat sheet. The two indices of the corresponding matrix
are represented by straight lines with an arrow. a) the leading order
process at short distances; (b) the leading order process at long
distances.}
\end{figure}

In terms of matrices the corresponding scattering potential is
\begin{eqnarray}
\Sigma^{2c} &=& q^4 \sum_{a,b,e,f} V^{(a,b)} G V^{(e,f)} G V^{(a,b)} G
V^{(e,f)} \nn \\
&\times& [(x^{(a,b)})^2] [(x^{(e,f)})^2] \nonumber \\
&=& q^4 \frac{p^2}{L^{2d}}\sum_{a,b,e,f} V^{(a,b)} G V^{(e,f)} G V^{(a,b)} G
V^{(e,f)} \; , \nn \\
\end{eqnarray}
where the indices of the matrices were dropped for the sake of simplicity
and $[(x^{(e,f)})^n]=p/L^d$ for plain SW networks.  Note that each operator
in the sum is conservative [see Eq.~(\ref{eq:conservative})] hence the whole
diagram is so too.  Let us investigate a single term of this sum.  Using
Eq.~(\ref{eq:V^(a,b) fact.}), the factorized form of $V^{(a,b)}$ and
$V^{(e,f)}$, one can see that such a term consist of three scatterings of
the GF back and forth between two links in the system: a link between $a$
and $b$ and an other one between $e$ and $f$. All the possible combinations
of scatterings between the links are present. It can be shown and we will
see it in an example that the most important terms in the sum are such
where all the three GF-s in the diagram scatter between the same legs of
the links, let us say between $a$ and $e$.  At small distances, the most
important process of them is when the process starts with scattering off
$a$ then the GF scatters between $a$ and $e$ three times and finally it
``scatters out'' at $e$ (see Fig.~\ref{fig:propagators1}(a)).  The matrix
corresponding to this process is
\begin{eqnarray}
\Sigma^{2c,A}_{ij}&=& q^4 \frac{p^2}{L^{2d}} \sum_{a,e} \sum_{b,f} \dd_{ia}
(G_{ae})^3 \dd_{ej} \nonumber\\
&=& q^4 \frac{p^2}{L^{2d}} L^d L^d \sum_{a,e} \dd_{ia}
(G_{ae})^3 \dd_{ej} \nn \\
&=& (G_{ij})^3 q^4 p^2 .
\end{eqnarray}
At large distances the main contribution comes from a process depicted in
Fig.~\ref{fig:propagators1}(b) \footnote{Note that the other two processes
which have three GF-s between two endpoints cancel each other}. It starts
off exactly the same as in the previous case but in the final step the
particle jumps over the long-range link between $e$ and $f$ and scatters
out at $f$. The corresponding matrix is
\begin{eqnarray}
\Sigma^{2c,B}_{ij} &=&- q^4 \frac{p^2}{L^{2d}} \sum_{a,f} \sum_{b,e}
\dd_{ia} (G_{ae})^3 \dd_{fj} \nonumber \\
&=& - \sum_{a,f} \dd_{ia} \dd_{fj} q^4 \frac{p^2}{L^{2d}}
\sum_b \sum_e (G_{ae})^3 \nn \\
&=& - q^4 \frac{p^2}{L^{d}} \sum_e (G_{ie})^3 \; .
\end{eqnarray}
Note that these two processes are not conservative, yet the sum of them is,
which is required for the self-energy. In summary, in the continuum limit,
\begin{eqnarray}
 \Sigma^{2c,A+B}(r)=q^4 p^2 \left( G(r)^3-\frac{1}{L^d}
  \int G(r')^3 \mathrm{d}^d r' \right) \; . \nn \\
\label{eq:Sigma^2c,A+B}
\end{eqnarray}
Note that for long distances
\begin{eqnarray}
\Sigma^{2c}(r) \approx - q^4 \frac{p^2}{L^d} \int G(r')^3 \mathrm{d}^d r'
\; ,
\end{eqnarray}
the diagram is distance independent and uniform, and the contribution of it
is a mean-field interaction just like the contribution of the leading order
terms but with a different strength.

\begin{figure}[t]
\centering
\includegraphics[width=.4\textwidth]{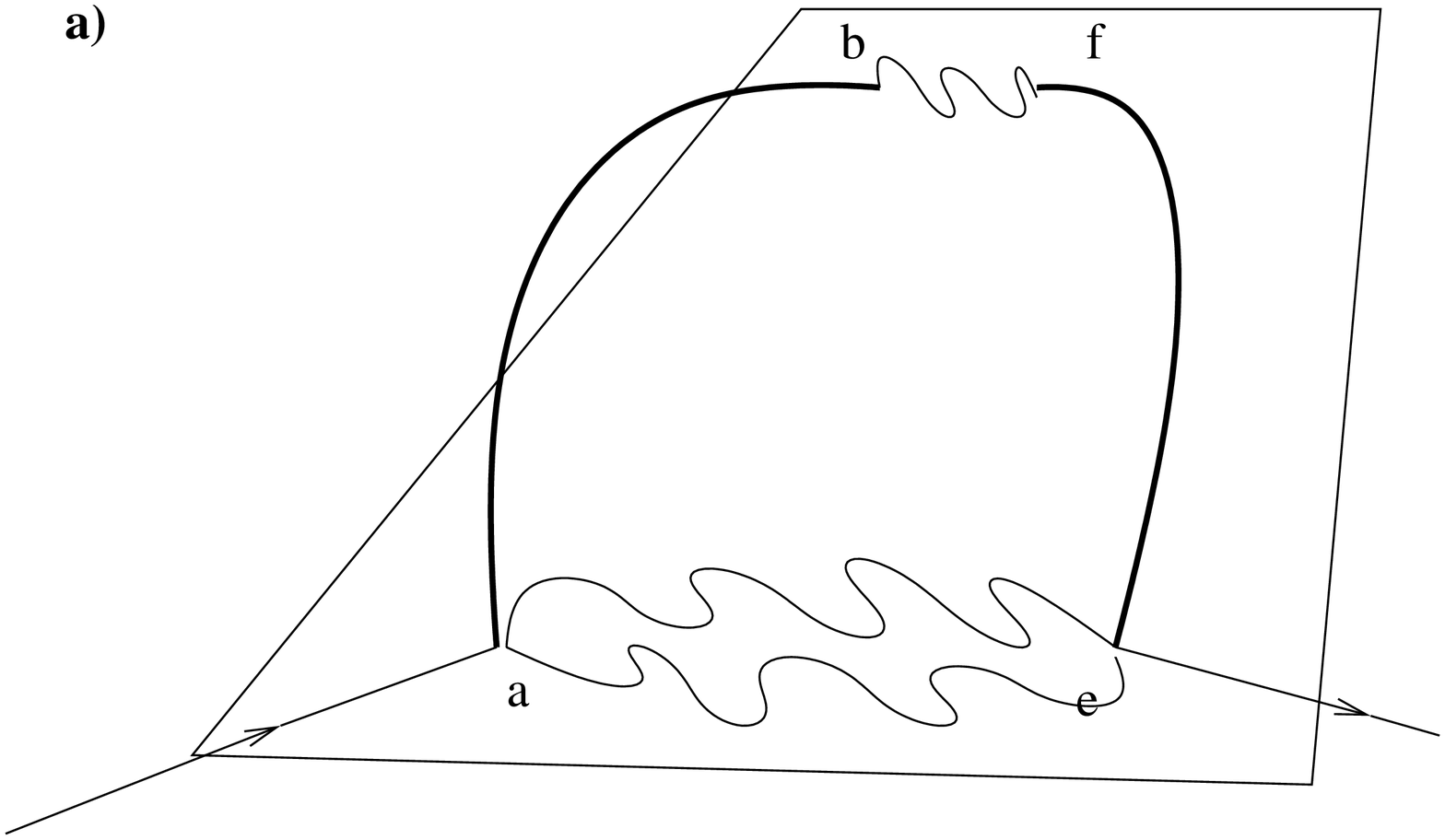}
\includegraphics[width=.4\textwidth]{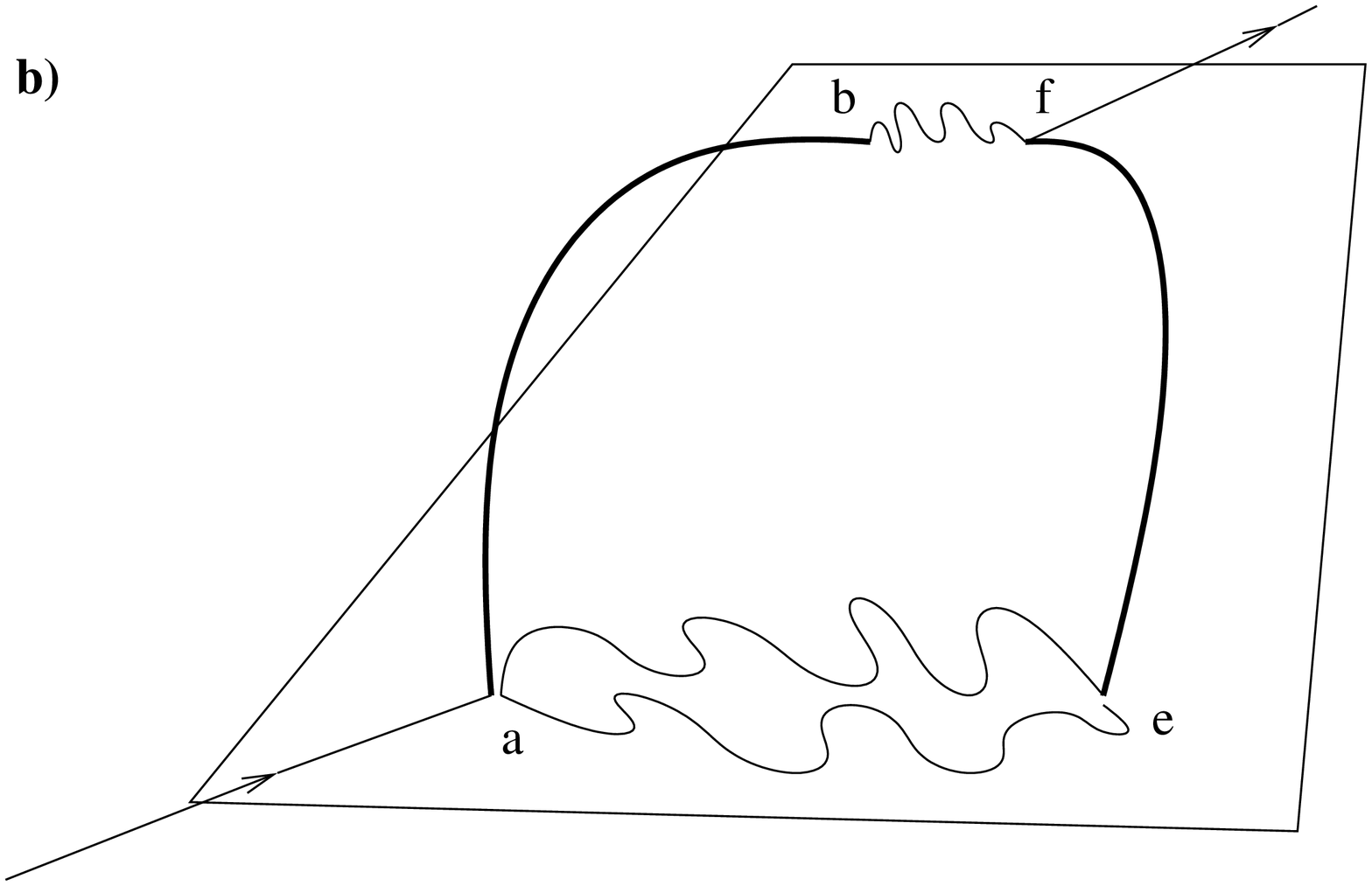}
\vspace*{-0.0cm}
\caption{
\label{fig:propagators2}
Sketch of a process the contribution of which vanishes in the
thermodynamic limit compared to the leading order one, depicted in
Fig.~\ref{fig:propagators1}. }
\end{figure}

As an example, let us investigate another term in the diagram, where not
all the GF scatter between the same legs of the two links, and see that the
contribution of such a diagram is vanishing in the large system size limit.
This process starts with scattering off $a$. The GF then propagates back
and forth between $a$ and $e$ twice before it jumps over the long-range
link between $a$ and $b$, where it then propagates from $b$ to $f$, and
finally jumps from $f$ to $e$ (see Fig.~\ref{fig:propagators2}(a)). The
corresponding matrix, up to combinatorial factors, is
\begin{eqnarray}
\Sigma^{2c,C}_{ij}&=& q^4 \frac{p^2}{L^{2d}} \sum_{a,e} \dd_{ia} \dd_{ej}
(G_{ae})^2 \sum_{b} \sum_{f} G_{bf} \nn \\
&=& (G_{ij})^2 q^4 \frac{p^2 }{L^{d}}
\sum_{f} G_{bf}
\end{eqnarray}
and the long-distance counter-term of it is [see Fig.~\ref{fig:propagators2}(b)]
\begin{eqnarray}
\Sigma^{2c,D}_{ij}=- q^4 \frac{p^2}{L^{2d}} \sum_f (G_{af})^2 \sum_{b}
G_{be} \; .
\end{eqnarray}
In the continuum limit
\begin{eqnarray}
\Sigma^{2c,C+D}(r) &=& q^4 \frac{p^2}{L^d} \left( G(r)^2 \int G(r')
\mathrm{d}^d r' \right. \nn \\
  &-& \left. \frac{1}{L^d} \int G(r')^2 \mathrm{d}^d r' \int G(r'') \mathrm{d}^d r''
  \right) \; . \nn \\
\end{eqnarray}
Compared to the leading order term, $\Sigma^{2c,A+B}(r)$, this matrix
vanishes in the thermodynamic limit.

Qualitatively, the above result can be understood from the following
argument: For the process depicted in Fig.~\ref{fig:propagators1}, only two
endpoints, $a$ and $e$, of the two links have to be within a distance
proportional to the correlation length of $G$ of each other so the GF
can propagate with finite weight between them. For the process of
Fig.~\ref{fig:propagators2}, both endpoints of both links have to be within a
distance of the correlation length from each other, $a$ close to $e$ and
$b$ close to $f$. Since the two links are randomly spread on the lattice in
a uniform fashion, having both endpoints of both links in the vicinity of
each other is much less likely than having only one endpoint of one link
close to one endpoint of the other link.

In low dimensions, some higher-order two-link diagrams have the same
problem as the higher-order single-link diagrams. For single-link diagrams,
it was demonstrated that the second-order diagram was more dominant than
the first-order one in the $p \to 0$ limit. 
There are two-link (or two-cumulant) diagrams that involve GF-s propagating
between the endpoints of the same link, just like in the case of
single-link diagrams. For example, the last double line of
Fig~\ref{fig:2cumulants}(b) represents such an event (it can be checked by
expressing the diagram in terms of matrices).  Such a GF will have a
contribution to the diagram which is divergent as $p \to 0$. Since this
divergent contribution is the only difference between this diagram and the
diagram in Fig.~\ref{fig:2cumulants}(a), the higher-order diagrams will
dominate the small $p$ behavior. As it was done before, this problem can be
avoided by resumming over all the single-link scatterings of such diagrams.
As a result, when there is a scattering off a link, the scattering is
substituted by all the single link scatterings. In terms of matrices the
substitution
\begin{eqnarray}
q V^{(ab)} &\to& q V^{(ab)} - q V^{(ab)} G  q V^{(ab)} \nn \\
 &+& qV^{(ab)} G qV^{(ab)} G
qV^{(ab)} - ...
\end{eqnarray}
is made for all $\times$-es in the diagrams. It is known from the
single-link calculations that the above infinite sum yields
\begin{eqnarray}
V^{(ab)}\frac{q}{1+2q(G_{aa}-G_{ab})}.
\end{eqnarray}
One can say that the result of the resummation is that the strength of the
interaction gets {\em renormalized} from $q$ to
$\frac{q}{1+2q(G_{aa}-G_{ab})}$.

\begin{figure}[t]
\centering
\includegraphics[width=.45\textwidth]{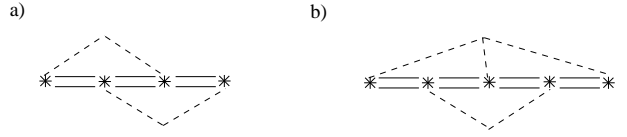}
\vspace*{-0.0cm}
\caption{
\label{fig:2cumulants_resummed}
Resummed two-cumulant diagrams.  Stars represent the resummed scattering
potentials introduced in Sect.~\ref{ssec:HOrdCorr}.  Otherwise, the
notation is the same as in Fig.~\ref{fig:[GF]}. Square brackets, $[\ ]$
were dropped from around double lines for the sake of clarity.  }
\end{figure}

\begin{figure}[t]
  \centering
  \includegraphics[width=.45\textwidth]{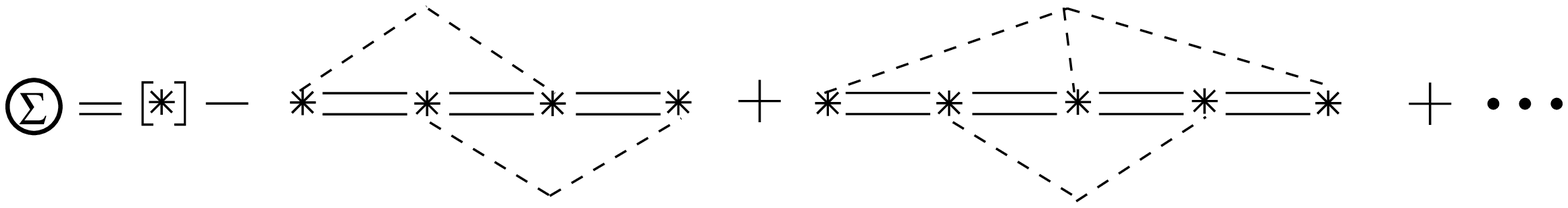}
  \vspace*{-0.0cm}
\caption{
\label{fig:Sigma_star}
The self-energy in terms of the resummed potentials.  The notation is the
same as in the previous figure.}
\end{figure}

In a typical situation the separation of $a$ and $b$ diverges with the
system size. Therefore, in such situations, $G_{ab}$ is vanishingly small
because the GF has a finite correlation length. As a result, the resummed
scattering potential can be well approximated by
\begin{eqnarray}
 V^{(ab)}\frac{q}{1+2qG(0)}
\end{eqnarray}
in the continuum limit. In diagrammatic notation, the above resummation is
represented by replacing the crosses with stars as shown in
Fig.~\ref{fig:2cumulants_resummed}. The expansion of the self-energy in
terms of the resummed potentials is shown in Fig.~\ref{fig:Sigma_star}.

Using these renormalized single-link scattering potential, let us calculate
the leading order behavior of the diagram in
Fig.~\ref{fig:2cumulants_resummed}(a).  The calculation is the same as was
done for the diagram of Fig.~\ref{fig:2cumulants}(a) except that $q$ is
replaced by $\frac{q}{1+2qG(0)}$ in Eq.~(\ref{eq:Sigma^2c,A+B}).

In one dimension, $G(0) \propto p^{-1}$. In the limit $p \to 0$,
\begin{eqnarray}
\frac{q}{1+2qG(0)} \to p.
\end{eqnarray}
Therefore, the second-order term in the resummed self-energy scales as
\begin{eqnarray}
\Sigma^{2c}(r) \propto - \frac{p^4 p^2}{L} \int G(r')^3
\mathrm{d}r'
\end{eqnarray}
at large distances. The integral in the expression can be approximated
using the results of Appendix~\ref{app:massiveGF},
\begin{eqnarray}
  G(r) \propto
   \left\{
      \begin{array}{ll}
        s^{-1/2} & \mbox{if $r \ll \xi$} \\ \\
        0 & \mbox{if $r \gg \xi$}
      \end{array}
   \right.
\end{eqnarray}
where $s \propto p^2$, obtained from the single-link approximation
[see Eq.~(\ref{eq:s_1d_plain})], and $\xi=s^{-1/2}$ is the correlation length as
before. Therefore
\begin{eqnarray}
\int G(r')^3 \mathrm{d} r' \propto p^{-4}
\end{eqnarray}
and
\begin{eqnarray}
  \Sigma^{2c}(r) \propto \frac{p^2}{L}.
\end{eqnarray}
The Fourier transform of this operator for small wavenumbers
$\Sigma^{2c}(k)\propto p^2$ has the same $p$ dependence as the leading
order one, Eq.~(\ref{eq:s_1d_plain}). Higher-order diagrams, like the one
in Fig.~\ref{fig:2cumulants_resummed}(b), yield contributions of the same
order, $p^2$. As a result, it can be concluded that the self-energy scales as
$p^2$ for small $p$-s, though neither the coefficient of it nor higher-order
corrections are known.

In two dimensions, $G(0) \propto \ln p$ from Eq.~(\ref{eq:G(0)_2d_plain}),
therefore $q$ is renormalized to $1/\ln p$. The second-order correction to
the self-energy at large distances
\begin{eqnarray}
\Sigma^{2c}(r) \propto - \frac{p^2}{(\ln p)^4 L^2} \int G(r')^3
\mathrm{d}^2 r'.
\end{eqnarray}
From Appendix \ref{app:massiveGF}, the integral can be approximated using
\begin{eqnarray}
  G(r) \propto
   \left\{
      \begin{array}{ll}
        const. + \ln (r/\xi) & \mbox{if $r \ll \xi$} \\ \\
        0 & \mbox{if $r \gg \xi$}
      \end{array}
   \right.
\end{eqnarray}
where \(\xi=s^{-1/2}\propto \sqrt{\ln p/p}\) from Eq.~(\ref{eq:s_2d_plain}). Therefore,
\begin{eqnarray}
\int G(r')^3 r' \mathrm{d} r' \propto \frac{\ln p}{p}
\end{eqnarray}
and
\begin{eqnarray}
\Sigma^{2c}(r) \propto \frac{p}{(\ln p)^3 L^2}.
\end{eqnarray}
In Fourier space, as $p \to 0$,
\(\Sigma^{2c}(k)\propto p/(\ln p)^3 \ll \Sigma^{sl}(k) \propto p/\ln p\).

In conclusion, in two dimensions, the single-link perturbation expansion of
Fig.~\ref{fig:Sigma_star} works. The mean-field model acquires logarithmic
corrections from sample-to-sample fluctuations. The fact that the
perturbation expansion works in two dimensions raises the possibility of an
$\epsilon$-expansion for $d=2-\epsilon$ which was studied in
\cite{HASTINGS04}. For small $\epsilon$-s, it was shown that the
higher-order diagrams of the perturbation expansion produce higher-order
terms in $\epsilon$. Though $\epsilon=1$ for $d=1$, the $\epsilon$
expansion still defines an ordering of the diagrams making it possible to
obtain more accurate analytic predictions for the behavior of the system in
the $p \to 0$ limit.

\section{Scaling arguments}
\label{sec:ren}

After the more careful perturbative treatment of the previous section, let
us revisit the effect of long range links and estimate the behavior of the
quenched system using scaling and renormalization arguments.

\subsection{Estimating the validity of the annealed approximation}

In many statistical systems introducing random long-range links to
the underlying lattice results in mean-field (MF) or, in our
terminology, annealed behavior \cite{HASTINGS03}. In the
following, we will study the self-consistency of the annealed
approximation by estimating the sample-to-sample fluctuations of
the quenched systems around it\cite{GOLDENFELD93} \footnote{Note that the
fact that the MF approximation is self-consistent grants only that
the universal behavior of the quenched and the annealed system
will be the same (e.g. the scaling with respect to $p$) though
other features, like constants in the formulae, can exhibit
non-universal behavior.}.

First, we consider the plain SW network ($\al=0$). For this
estimation, we utilize the EW Hamiltonian [Eq.~(\ref{eq:H[h] Aij})]
expressed in terms of the random component of the adjacency
matrix, $A_{\rr,\rr'}$, and the difference of the fields at site
$\rr$ and $\rr'$:
\begin{eqnarray}
H[\vec{h}] &=& H_0[\vec{h}] + q \sum_{\rr,\rr'} A_{\rr,\rr'}
(h_\rr-h_{\rr'})^2 \nn \\
&=& H_0[\vec{h}] + q \sum_{\rr,\rr'} [A_{\rr,\rr'}] (h_\rr-h_{\rr'})^2 \nn
\\
&+& q\sum_{\rr,\rr'} (A_{\rr,\rr'}-[A_{\rr,\rr'}]) (h_\rr-h_{\rr'})^2 \nn \\
&=& H_0[\vec{h}]+ [H^{rnd}][\vec{h}] +H^{rnd}_{fluct.}[\vec{h}],
\end{eqnarray}
where $H_0$ is the Hamiltonian of the EW model on the regular, unperturbed,
lattice and $H^{rnd}$ is the ``energy'' due to the random part of the
network.

According to MF theory, the last term in $H$,
\begin{equation}
q\sum_{\rr,\rr'} (A_{\rr,\rr'}-[A_{\rr,\rr'}]) (h_\rr-h_{\rr'})^2
\label{eq:H_fluct}
\end{equation}
is negligible compared to the second one because the sample-to-sample
fluctuations of $A_{\rr,\rr'}$ are suppressed due to averaging over a large
number of sites. For a link between two sites, the relative size of these
fluctuations can be quantified by
\begin{equation}
 \frac{\sqrt{[(A_{\rr,\rr'})^2]-[A_{\rr,\rr'}]^2}}{[A_{\rr,\rr'}]},
\end{equation}
the ratio between the standard deviation and the mean of the random
potential.

For the whole system, one has to average over many sites to estimate the
magnitude of the fluctuations in the energy, $H[\vec{h}]$. Averaging over
the entire system would suppress the fluctuations, though doing so would
underestimate their strength. This can be understood in the following
argument:

If the separation of $\rr$ and $\rr'$ is less than the correlation length,
$\xi$, of the field $h_\rr$, the values of the field at those two sites are
close to each other, and \((h_\rr -h_{\rr'})^2\) is small. Therefore, most
of the contribution to the energy comes from sites where $\rr'$ is outside
of the correlation volume of the field at site $\rr$.

On the other hand, the fluctuations in the energy will only amplify the
effect of each other where the values of the field are correlated, i.e.
over volumes $\propto \xi^d$. Therefore, the condition for the MF or annealed
approximation to be valid is
\begin{equation}
\frac{\sqrt{\int\limits_{|\rr|<\xi} d^d r \int\limits_{\xi<|\rr-\rr'|}
    d^d r' ([(A_{\rr,\rr'})^2]-[A_{\rr,\rr'}]^2)}} {
  \int\limits_{|\rr|<\xi} d^d r \int\limits_{\xi<|\rr-\rr'|} d^d r'
  [A_{\rr,\rr'}]} \ll 1.
\label{eq:MFcond}
\end{equation}
Here, we used the fact that the standard deviation of a sum of random
variables is the square root of the sum of their variances. From Subsection
\ref{subsec:cumulants}, \([A_{\rr,\rr'}]= ([(A_{\rr,\rr'})^2]
-[A_{\rr,\rr'}]^2)=p/L^d\) and the correlation length is \(\xi \propto
{(pq)}^{-1/2}\). The above condition can be expressed as
\begin{equation}
  p^{-1+d/2}q^{d/2} \ll 1.
\end{equation}
The assumptions of the MF approximation are always valid when $q \to 0$.
In one dimension, if $p \ll q$, the sample-to-sample fluctuations destroy
the MF behavior otherwise MF approximation can still be applied. In two
dimensions, if $p \to 0$ the MF predictions acquire logarithmic corrections
from the sample-to-sample fluctuations (see Section~\ref{sec:2dquenched}).
In three dimensions and above, the MF criterion is always satisfied.

Second, when $\al<d$, $[H^{rnd}]$ generates a mass (or finite correlation
length) just like in the case of a plain SW network and \(\xi \propto
{(pq)}^{-1/2}\) as well. The same argument can be repeated as in the
previous case. The condition of the MF behavior, Eq.~(\ref{eq:MFcond}), is
\begin{equation}
  p^{-1+d/2}q^{d/2} \ll 1.
\end{equation}

Third, when $d<\al<d+2$, strictly speaking, there is no correlation length
in the system, though the typical size of correlations is proportional to
$l_\times$, defined by the lengthscale above which the long-range random
links dominate the behavior of the system. In Fourier space,
\begin{equation}
k_\times^{2}=pqk_\times^{\al-d},
\end{equation}
where \(k_\times=l_\times^{-1}\). Thus, for the lengthscale $l_\times$ one
finds \(l_\times =(pq)^{-\frac{1} {2+d-\al}}\). Repeating the same argument
as above, replacing the role of $\xi$ by $l_\times$, and using
\([A_{\rr,\rr'}]= ([(A_{\rr,\rr'})^2] -[A_{\rr,\rr'}]^2)=p/(\NN r^\al)\),
one arrives at the condition
\begin{equation}
 p^{d-2}q^{2d-\al} \ll 1
\label{eq:MFbreakdown}
\end{equation}
for the MF approximation to be valid. Observe that in one dimension, if
$2<\al<3$, the MF approximation breaks down for all $p$-s and $q$-s. In all
the other cases the condition is similar to those of the plain SW
network, but due to the distance-dependent cutoff of the links, a higher
density of long-range links is required for the MF approximation to be
valid (i.e. $p$ has to be larger for observing MF behavior) .

Fourth, when $d+2<\al$, the typical size of the correlations is that of the
system size, so one cannot apply the above argument. Though, one can
introduce a ``virtual'' correlation length to the system, $\xi_{\mbox{v}}$,
check the condition of the MF approximation, Eq.~(\ref{eq:MFcond}), as a
function of $\xi_{\mbox{v}}$, and in the final step take the limit when the
correlation length goes to infinity (or, at least, to the system-size). The
result is
\begin{equation}
  p^{-1} \xi_{\mbox{v}}^{(\al-2d)} \ll 1 \; \;.
\end{equation}
In conclusion: For $d+2<\al<2d$, the condition for the MF approximation is
satisfied since $\xi_{\mbox{v}}^{(\al-2d)} \to 0$ as $\xi_{\mbox{v}} \to
\infty$. For $2d<\al$, the condition of for the MF condition is not satisfied
since $\xi_{\mbox{v}}^{(\al-2d)} \to \infty$ as $\xi_{\mbox{v}} \to \infty$.

\subsection{Renormalization}

\begin{figure}[t]
  \centering \includegraphics[width=.4\textwidth]{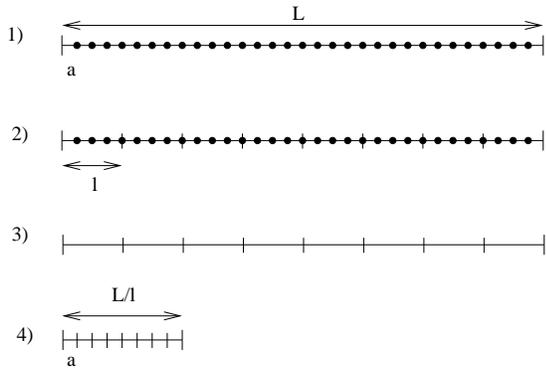}
  \vspace*{-0.0cm}
\caption{
\label{fig:renorm}
The steps of the renormalization procedure: 1) Take the original system.
2) Divide it into boxes of size $l$.  3) Average over the behavior within
each $l$-sized box.  4) Rescale the system so that $l \to a$.  }
\end{figure}

As it was shown above, MF theory is insufficient to describe diffusive
processes on quenched SW networks for certain parameter regimes. Though,
renormalization approach can capture the universal behavior of these
systems: Assuming that the large scale properties of the system are not
influenced by the microscopic details, one can average over the small
lengthscale behavior according to the steps of Fig.~\ref{fig:renorm}.

In the following, we will investigate how the diffusion operator changes
doing the steps of renormalization. Diffusion on PL-SW networks is
described only by two parameters: $p$ and $q$ (besides $d$, the dimensionality
of the system). First, the change of the density of random links under
a change in scale is
considered. Second, we will look at the renormalization of $q$ for only a
single link in the system. Third, we will combine the results at hand to
calculate the scaling of the GF on a PL-SW network; since such a network
includes many links, we will have to include interaction between links
when the density of links becomes of order unity at the given scale.

\subsection{Renormalization of $p$}

As defined in the Introduction, $p$ is the probability of a site
having a random link emanating out of that site irrespective of where the
link ends. The first two steps of the renormalization procedure are
straightforward. In the third step, after averaging over a cell of size
$(al)^d$ the probability of a node within the cell to have a random link
pointing outside of that cell is \( \int_{al}^L p \frac{r^{d-1} dr}{\NN
r^\al} \).  For all the nodes within the cell, this probability is
\begin{equation}
\tilde{p}=l^d \int_{al}^L p \frac{r^{d-1} dr}{\NN r^\al}
\propto l^d p \times
   \left\{
      \begin{array}{ll}
        \frac{L^{d-\al}}{\NN} & \mbox{if $\al<d$} \\ \\
        \frac{(al)^{d-\al}}{\NN} & \mbox{if $d<\al$}
      \end{array}
   \right.
\end{equation}
where the normalization constant is
\begin{equation}
\NN = \sum_{\rr} \frac{1}{r^\al} \propto \int_a^L \frac{r^{d-1}dr}{r^\al }
\approx
   \left\{
      \begin{array}{ll}
        L^{d-\al} & \mbox{if $\al<d$} \\ \\
        a^{d-\al} & \mbox{if $d<\al$}
      \end{array}
   \right.
\end{equation}
Therefore,
\begin{equation}
  \tilde{p}\propto p \times
   \left\{
      \begin{array}{ll}
        l^{d} & \mbox{if $\al<d$} \\ \\
        l^{2d-\al} & \mbox{if $d<\al$} .
      \end{array}
   \right.
\end{equation}
In the fourth step, the $(al)^d$ sized cell is rescaled to size $a^d$ and
the probability of one such node to have a random link will be that of the
original cell, $(al)^d$. In summary, the flow equation of the parameter of
the probability distribution of the random links is
\begin{equation}
\label{pflow}
  p \to \tilde{p} \propto p \times  \left\{
                       \begin{array}{ll}
                       l^{d} & \mbox{if $\al<d$} \\ \\
                       l^{2d-\al} & \mbox{if $d<\al$} .
                       \end{array}
                     \right.
\end{equation}

\subsection{Renormalization of $q$ for a single sink and a single link} 

In renormalization group theory, the procedure to obtain the universal
long-wavelength properties of the system is averaging over short wavelength
modes. Due to interactions which mix long and short wavelength modes, the
Hamiltonian that describes the statistical properties of the system will
change: these mixing interactions will generate modified or new
interactions between long wavelength modes after averaging.

Let us follow through this procedure in the case of when there is only a
single sink in the system.  The Hamilton operator of the system is
$H=H^0+qV$.The unperturbed Hamiltonian is the diffusion operator,
\(H^0_{kk'}=\dd_{kk'}k^2\) in Fourier space. The perturbation potential is
\(q V_{ij}=q \dd_{i0}\dd_{0j}\) in real space and
\begin{equation}
q V_{kk'}=\frac{q}{L^d}
\label{eq:qFT}
\end{equation}
a ``uniform'' matrix in Fourier space.  We will distinguish between long and
short wavelength modes
\begin{equation}
\Phi_k=
   \left\{
      \begin{array}{ll}
        \Phi'_k & \ \mbox{if \ $k<1/l$} \\ \\
        \Phi''_k & \ \mbox{if \ $k>1/l$}.
       \end{array}
   \right.
\end{equation}
Making this distinction, let us introduce the following division of $V$ in
Fourier basis:
\begin{equation}
  V= \left( \begin{array}{cc}
      V_1 & V_A \\
      V_B & V_2
      \end{array} \right),
\end{equation}
where $V_1$ and $V_2$ act on the long and short wavelength modes and $V_A$
and $V_B$ are mixing between them. The Laplace operator is divided
likewise:
\begin{equation}
  H^0= \left( \begin{array}{cc}
      H_1 & 0 \\
      0   & H_2
      \end{array} \right),
\end{equation}
The usual method of doing the averaging is by considering the {\em
  partition function}:
\begin{eqnarray}
\mathcal{Z}=\int \mathcal{D}\Phi e^{\Phi (H+q V) \Phi}  = \int \mathcal{D}
\Phi' \ e^{\Phi' (H_1+q V_1) \Phi'} \nn \\
\times \int \mathcal{D} \Phi'' \
e^{\Phi'' (H_2+q V_2) \Phi'' + \Phi'( q V_B) \Phi'' +
  \Phi'' (q V_A) \Phi' }.
\end{eqnarray}
After averaging over the short wavelength modes by performing the integral
\( \int \mathcal{D} \Phi''(...)\), a $\Phi'$ dependent
expression will remain which is the generated interaction due to the short
wavelength ``mixing''. The result is a new {\em renormalized}
perturbation potential, $q_{ren} V^{ren}$, defined by.
\begin{equation}
\mathcal{Z}=\int \mathcal{D}\Phi e^{\Phi (H+q V) \Phi}=
\int \mathcal{D}\Phi' e^{\Phi' (H_1+q_{ren} V^{ren}) \Phi'}.
\end{equation}
We will see that for the single sink problem $q_{ren} V^{ren}$ has the same
form as $V_1$ only the interaction strength gets renormalized $q \to q_{ren}$.
The above mentioned procedure of performing the short-wavelength integral
could be carried out but it was merely introduced to recapitulate the basics
of renormalization group (RG) transformation. Instead, let us calculate
$q_{ren} V^{ren}$ using GF-s and the diagrammatic technique introduced in
Section \ref{ch:qu}.

\begin{figure}[t]
  \centering \includegraphics[width=.4\textwidth]{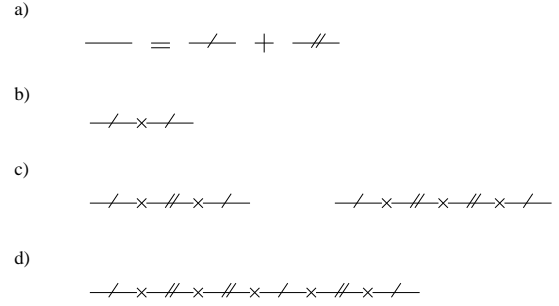}
  \vspace*{-0.0cm}
\caption{
Introduction of the short and long wavelength diagrams. a) Decomposition of
the GF to short and long-wavelength modes. b) Simple scattering of the
long-wavelength GF. c) Scatterings of the long-wavelength GF mediated by
short-wavelength GF-s. d) A typical scattering with multiple scatterings
between the long and short wavelength modes.}
\label{fig:rG1}
\end{figure}

\begin{figure}[t]
  \centering \includegraphics[width=.4\textwidth]{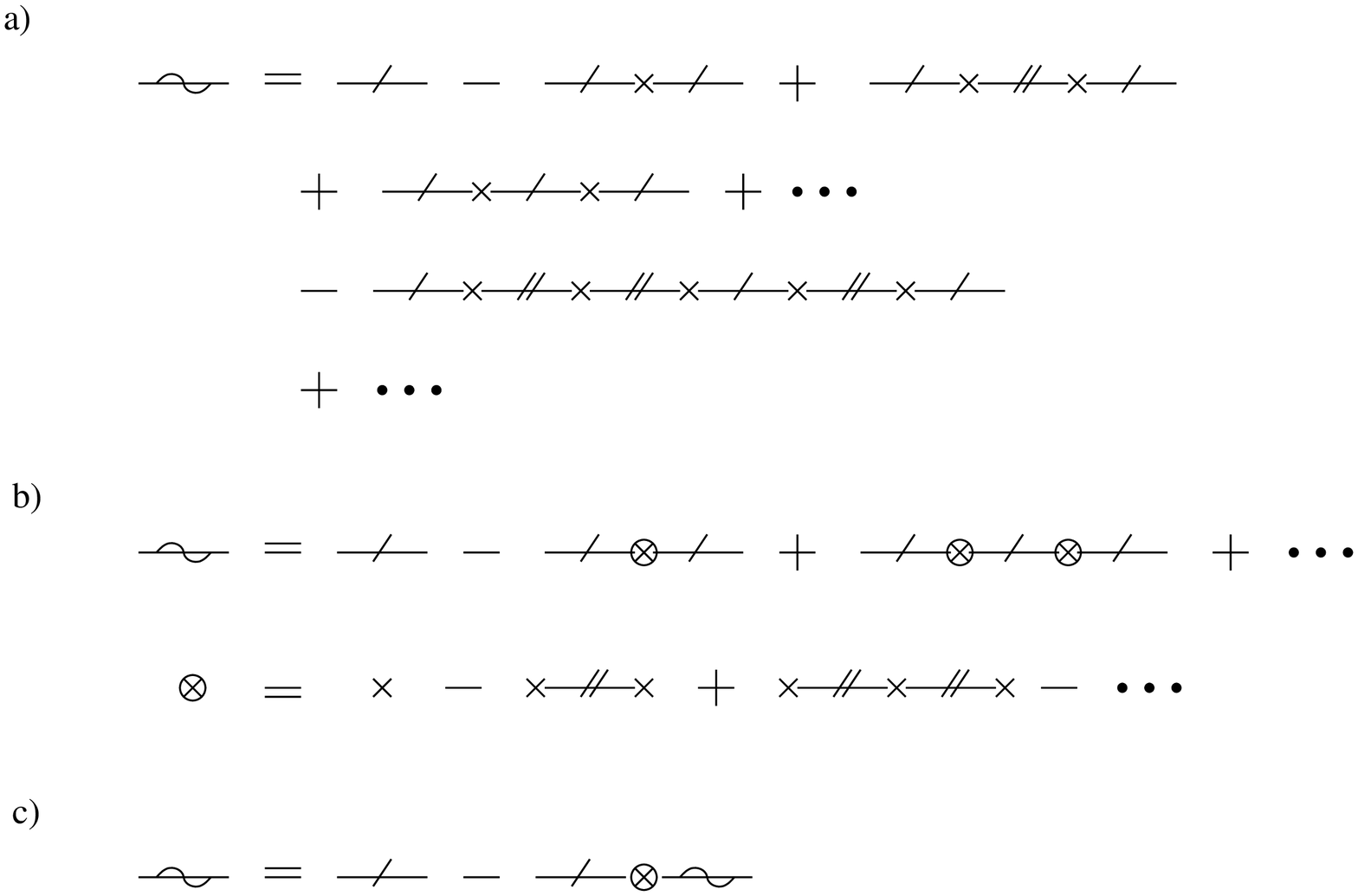}
  \vspace*{-0.0cm}
\caption{
\label{fig:rG2}
Obtaining the renormalized GF for the single sink and link problem.  a) The
perturbation expansion of the renormalized GF, represented by a line with a
wave. b) Introduction of the renormalized scattering operator, $\otimes$.
c) The Dyson equation of the renormalized GF.  }
\end{figure}

Single lines represent $G^0$ divided into long-, ${G^0}'$,
and short-wavelength, ${G^0}''$, propagators, represented by single lines
with a slash and single lines with double slash. Single lines with a wave
represent the renormalized GF, \(G^{ren}=(H_1+q_{ren}V^{ren})^{-1}\).
Crosses represents scatterings off $qV$ and a cross in a circle
represents $q_{ren}V^{ren}$ (see Figs. \ref{fig:rG1} and \ref{fig:rG2}).

$G^{ren}$ which is only defined for long wavelengths can be calculated
using the original scattering potential, $V$. Scatterings of the long
wavelength propagators can happen in two ways: {\em i)} a long wavelength
propagator can scatter off $V$ to a long wavelength propagator
[Fig.~\ref{fig:rG1}(b)], and {\em ii)} a long wavelength propagator can scatter
off $V$ to short wavelength propagators and then scatter ``back'' to a long
wavelength propagator [Fig.~\ref{fig:rG1}(c)], i.e., the scatterings of
the long wavelength propagators are mediated by short wavelength
propagators.

In Fig.~\ref{fig:rG2}(a), all the possible scatterings of the long
wavelength propagators are represented. In Fig.~\ref{fig:rG2}(b) the
scatterings are reordered introducing a new symbol, a cross in a circle
that represents all the possible ${G^0}''$ mediated scatterings of
${G^0}'$-s. Figure~\ref{fig:rG2}(c) is the reordering of Fig.\ref{fig:rG2}(b)
to the similar form of a Dyson equation where one can identify $\otimes$
as $q_{ren}V^{ren}$.

The sum in the formula of $\otimes$ can be solved. Notice that this is
exactly the same problem as that of Section~\ref{ch:qu} except that in that
problem the intervening propagator between scatterings was $G^0$ while
here ${G^0}''$. Therefore
\begin{equation}
\otimes= \frac{q}{1+q{G^0}''(\rr=\oo)} V_1=q_{ren} V^{ren},
\end{equation}
where
\begin{eqnarray}
{G^0}''(\rr=\oo)&=&\int_{1/l}^{1/a}\frac{d^dk}{k^2} \nn \\
&\propto&
 \left\{
      \begin{array}{ll}
        l  & \mbox{if $d=1$} \\
        \ln l & \mbox{if $d=2$}\\
        a^{2-d} & \mbox{if $2<d$} \; .
      \end{array}
   \right.
\end{eqnarray}
The dimensionality of $q$ is $(length)^{d-2}$ \footnote{This property can
  be obtained by noting that the two constituents of the Hamiltonian, $H^0$
  and $qV$, must have the same dimension. Also, we know that $H^0 \sim k^2
  $ and $qV \sim q/L^d$ in Fourier space. Since $k$ has dimension
  $(length)^{-1}$, the dimensionality of $q$ is $(length)^{d-2}$.}
therefore after the fourth step of renormalization:
\begin{eqnarray}
q \to \tilde{q} = q_{ren} l^{(2-d)} \; .
\end{eqnarray}
In one dimension, when $l \to \infty$ ,
\begin{eqnarray}
\label{qflow1}
\tilde{q} \propto \frac{q}{1+ql} l  \to const.
\end{eqnarray}
In two dimensions,
\begin{eqnarray}
\label{qflow2}
\tilde{q} \propto \frac{q}{1+q \ln l} = \OO(\frac{1}{\ln l}) \to 0   .
\end{eqnarray}
For $d>2$,
\begin{eqnarray}
\label{qflow3}
\tilde{q} \propto \frac{q}{1 + q a^{2-d}} l^{2-d} = \OO(l^{2-d}) \to 0   .
\end{eqnarray}

In conclusion, $q$ is relevant in one dimension, marginally irrelevant in
two dimensions, and irrelevant in higher dimensions.

For a single-link, there are two cases to be differentiated: i) When the
separation of the two endpoints ($\lambda$) is much larger than the
lengthscale over which the renormalization is carried out ($l \ll
\lambda$). In this case, the vicinity of these endpoints can be considered
as independent $d$-dimensional spaces for the short-wavelength GF to exist
in, connected by a link at the origin with strength $q$. As
in the case of a single sink, the renormalized interaction strength can be
calculated using the diagrams of Fig. \ref{fig:rG2}. As a result,
\begin{equation}
q_{ren} = \frac{q}{1+2q{G^0}''(\rr=\oo)} 
\end{equation}
and the RG equation for the interaction strength, $\tilde q$, is the same
as for the single-sink problem: (Eqs.~(\ref{qflow1}, \ref{qflow2},
\ref{qflow3})). ii) When $l \gg \lambda$, links and sinks get treated very
differently: links can simply be ignored ($\tilde{q}=0$) since the
perturbation potential caused by a link has vanishing strength at these
lengthscales while the effect of sinks can be still strong (see
Eqs.~(\ref{qflow1}, \ref{qflow2}, \ref{qflow3})).

\subsection{The renormalization of $q$ including interactions between links
and the scaling of the Green's function}

\begin{figure*}[t]
\centering 
\includegraphics[width=.8\textwidth]{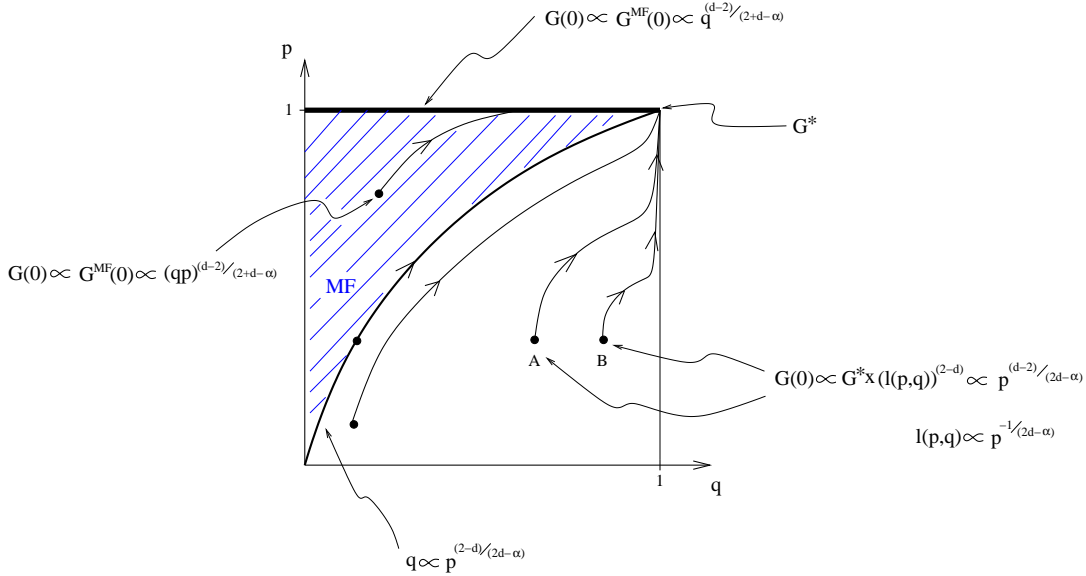}
\vspace*{-0.0cm}
\caption{
\label{fig:RGflow}
The schematic representation of the RG flow when $d<2$ and $d<\al<2d$ for
the quenched system. The parameter space is divided into two parts: i)
where the behavior of the system is MF and ii) where the large-scale
behavior is only determined by $p$. We also indicated the scaling of the GF
in the two regimes. In part ii), consider two points in the phase space $A$
and $B$ with different $q$ but the same $p$ values: Renormalization
transformed both systems to an identical one in the parameter space, where
the GF is $G^*$. Furthermore, it took the same number of renormalization
steps, $l$, to get to $G^*$ from both points because $l$ is only determined
by $p$. As a consequence, considering that the GF can be well approximated
by the free GF at small lengthscales (until $\tilde p$ becomes of order
unity): $G_A(0) \propto G_B(0) \propto G^* l^{(2-d)} \propto
p^{\frac{2-d}{2d-\alpha}}$. In other words, $G(0)$ have the same scaling
behavior in the two points.
}
\end{figure*}

We now combine the results for the RG flow of $\tilde{p}$ in
Eq.~(\ref{pflow}) with the results for $\tilde{q}$ to describe the scaling
of the Green's function on a small-world network.  When $\tilde{p}$ is
small, we can use the single-link results for the scaling of $\tilde{q}$ in
Eqs.~(\ref{qflow1},\ref{qflow2},\ref{qflow3}).  When $\tilde{p}$ becomes of
order unity, however, these results will need to be modified.

Consider a network with $p$ and $q$ both much smaller than $1$.  If
$\alpha<2d$, then $\tilde{p}$ is increasing under the flow and will eventually
become of order unity.  We will start by investigating this case.  We restrict
ourselves to the study of the case when $d< \al < 2d$, which is the most
complicated one, for the sake of brevity. In the omitted case, $\alpha < d$,
the derivation of the scaling properties can be obtained similarly using the
results of the two previous sections.

We begin by studying the case when $d < \alpha<2d$ and $d<2$ (see
 Fig.~\ref{fig:RGflow}).  As long as both $\tilde{p}$ and $\tilde{q}$ are much
 smaller than unity, $\tilde{q}$ obeys the naive scaling $\tilde{q}\propto
 l^{2-d}$.  This naive scaling for $\tilde{q}$ will break down when either
 $\tilde{q}$ becomes of order unity (due to multiple scattering off of a
 single link) or when $\tilde{p}$ becomes of order unity (due to scattering
 off multiple links).  So, we must determine whether it is $\tilde{p}$ or
 $\tilde{q}$ that becomes of order unity first.  The link density,
 $\tilde{p}$ becomes of order 1, from the naive scaling, at $l \sim
 (1/p)^{1/(2d-\alpha)}$, while $\tilde{q}$ becomes of order unity, from the
 naive scaling, at $l\sim (1/q)^{1/(2-d)}$.  So, if
 $(1/q)^{1/(2-d)}\ll(1/p)^{1/(2d-\alpha)}$ then $\tilde{q}$ becomes of
 order unity first.  That is, if $q \gg p^{(2-d)/(2d-\alpha)}$ then
 $\tilde{q}$ becomes of order unity while $\tilde{p}$ is still much smaller
 than unity.  In this case, we can continue to use the single link results
 (\ref{qflow1}) for the RG flow of $\tilde{q}$ and so $\tilde{q}$
 approaches a constant of order unity as the length scale continues to
 increase, while $p$ still increases following Eq.~(\ref{pflow}).  Thus,
 beyond this length scale, everything is set by the scaling of $\tilde{p}$
 and the original value of $q$ becomes unimportant.  Therefore, by
 dimensional analysis, using the scaling of $G(0)$, we find that $G(0)\sim
 p^{-(2-d)/(2d-\alpha)}$.  On the other hand, if $q \ll
 p^{(2-d)/(2d-\alpha)}$, then $\tilde{p}$ becomes of order unity while
 $\tilde{q}$ is still much less than one.  In this case, mean-field theory
 becomes accurate \footnote{This is the same criterion as that obtained in
 Eq.~(\ref{eq:MFbreakdown}).}, and we find that $G(0)\sim
 (pq)^{(d-2)/(2+d-\alpha)}$.  Thus,
\begin{eqnarray}
q\gg p^{(2-d)/(2d-\alpha)} & \rightarrow &
G(0)\sim p^{-(2-d)/(2d-\alpha)}\\ \nonumber
q\ll p^{(2-d)/(2d-\alpha)} & \rightarrow &
G(0)\sim (pq)^{(d-2)/(2+d-\alpha)}.
\end{eqnarray}
As a consistency check, note that the two expressions for $G(0)$
agree when
$q\sim p^{(2-d)/(2d-\alpha)}$.

Next, consider the case that $d < \alpha<2d$ and $d\geq 2$.  For $d=2$, we find
that $\tilde{q}$ is marginally irrelevant under the RG flow (\ref{qflow2}),
decreasing inversely with the logarithm of the length scale.  For
$d < \alpha<2d$, $\tilde{p}$ will become of order unity at a scale $l \sim
(1/p)^{1/(2d-\alpha)}$.  The scale $l$ is a power of $p$, hence the
logarithm of the length scale is proportional to a logarithm of $p$.  Thus,
we can apply mean-field theory at this scale $l$ with a $\tilde{q}\sim 1/
\ln (1/p)$.  Therefore, there are logarithmic corrections to mean-field
theory for $d=2$ and $G(0) \sim \ln (p/\ln p)$.  In the case when
$d < \alpha<2d$ and $d>2$, $\tilde{q}$ is irrelevant and we can apply naive
scaling until the scale $l \sim (1/p)^{1/(2d-\alpha)}$ at which point we
can apply mean-field theory.  Therefore, for $d>2$, mean-field is always
accurate in the limit of small $p,q$.

Finally, there is the case $\alpha=2d$, when $\tilde{p}$ is unchanging
under the flow.  For $d>2$, $\tilde{q}$ flows to zero, and then the effect
of the links becomes negligible.  However, for $d<2$, $\tilde{q}$ flows to
a constant of order unity.  Then we have a nontrivial fixed line of
the RG flow with $\tilde{p}$ arbitrary and $\tilde{q}$ of order
unity.  This leads to a $G(0)$ which varies as an anomalous power of the
system size, where the power depends continuously on the value of $p$.

Note that, when $\alpha > 2d$, $\tilde{p}$ always flows to zero therefore the 
long-distance scaling of the GF becomes insensitive to the presence of the 
random links.

\section{APPLICATIONS AND NUMERICS}
\label{ch:app & num}

\subsection{Calculating physical properties from the Green's function}

In Section \ref{ch:ann} and \ref{ch:qu}, an impurity averaged perturbation
expansion was set up to calculate the propagator of the diffusion operator
for different configurations of the network by changing $\al$, $p$, and
$q$. Here we interpret the results for the GF in terms of the relevant
observables of surface growth (Section \ref{sec:EW}) and random-walk
processes (Section \ref{sec:diff}).

For EW processes, from Eq.~(\ref{eq:w2GF}),
\begin{eqnarray}
[\langle w^2 \rangle] = \frac{1}{L^d} \sum_{i=1}^{L^d} [G_{ii}]
= \frac{1}{L^d} \sum_{i=1}^{L^d} G(0)= G(0).
\end{eqnarray}
Similarly in the annealed case \( [\langle w^2 \rangle]^{ann}=G^{ann}(0)\).
Therefore, all the results derived for $G(0)$ can be translated in terms of
the average width of the EW process, represented by the naming of the
different phases of Figures \ref{fig:1d phases} and \ref{fig:2d phases}

Though the perturbative calculations were done for $G(0)$, they can be
repeated for $\hat{G}(0,\omega)$, defined in Eq.~(\ref{eq:Ghat}). The
difference one has to consider is that, for the EW processes, the
unperturbed GF is \(G^{0}=(\DD^0)^{-1}\); for random-walks, it is
\(\hat{G}^0(\omega)=(\DD^0+\omega)^{-1}\). This substitution also results in
a slight change in the Fourier space calculations
\begin{eqnarray}
 G(k)=\frac{1}{k^2+\Sigma(k)} \ \ \ \to \ \
   \hat{G}(k,\omega)=\frac{1}{k^2+\Sigma(k)+\omega}, \nn \\
\end{eqnarray}
as can be seen in Fig.~\ref{fig:k_scaling}, where $\Sigma(k)$ is the
self-energy generated by the random links as calculated in
Sections~\ref{ch:ann} and \ref{ch:qu}. In order to obtain the scaling
properties of \(\hat{G}(\rr,\omega)\) and, specifically, of
\(\hat{G}(0,\omega)\), the approximations of Appendix~\ref{app:G(0)} are
used. As a general rule, in the case of \(\hat{G}(k,\omega)\), the
small-$k$ divergence is cut off by the $\omega$ term in the denominator.
Using the approximation of Eq.~(\ref{eq:F(T) appr.}), the scaling of the
expected number of returns can be related to $\hat{G}(0,\omega=1/T)$ by
\begin{eqnarray}
[F(T)]&\propto& \frac{1}{L^d} \sum_\rr [\hat{G}_{\rr,\rr} (\omega=1/T)] +
\frac{T}{L^d} \nonumber \\
&=& \frac{1}{L^d} \sum_\rr \hat{G}(0,\omega=1/T) +\frac{T}{L^d} \nonumber \\
&=& \hat{G}(0,\omega=1/T) +\frac{T}{L^d}.
\label{eq:[F(T)]}
\end{eqnarray}
Initially, we are only interested in the long-time behavior of the expected
number of returns in the infinite system-size limit, therefore, $ [F(T)]
\propto \hat{G}(0,\omega=1/T)$ in the thermodynamic limit. As it was done
for the width of the EW process, the interpretation of the scaling of
\(\hat{G}(0,\omega=1/T)\) as the scaling of the expected number of returns
is represented in the naming of the different phases of Figures \ref{fig:1d
phases} and \ref{fig:2d phases}.  Later in this section, we will also see
how the finite size of the system changes the behavior of $F(T)$.

\subsection{Numerical approach}

Numerical results were obtained in one-dimension ($N=L$). In order
to check the analytic predictions of Section \ref{ch:qu}, we
numerically calculated different physical properties of the
processes introduced in Section \ref{sec:intro}. For the quenched
networks, numerical results were obtained by: (i) generating a
large number of realizations (from $100$ to $1000$) of PL-SW
networks, then (ii) generating the diffusion operator, Eq.~(\ref{eq:SWdiff}),
on each realization, then (iii) using exact
numerical diagonalization \cite{NUMREC} to obtain the eigensystem
of the diffusion operator on each realization, then (iv) using the
spectral decomposition of the different physical quantities,
Eqs.~(\ref{eq:w2 spectral}) and (\ref{F_T_exact}), to obtain their
value for each realization; then (v) averaging these quantities
over a large number of realizations of the networks.

For the annealed networks, we numerically integrated the
corresponding time-discretized stochastic EW process in a
dynamically annealed network: at every time step the {\em random
links}, with density $p$, are reassigned. The EW equation was
numerically integrated using the most basic time-discretization
scheme
\begin{eqnarray}
h_i(t+\Delta t) &=& h_i(t) + \Delta t\left[ h_{i+1}(t) +h_{i-1}(t)
 -2h_i(t)\right] \nn \\
 &+& \Delta t \; q A_{i,r(i)}(t)\left[ h_{r(i)}(t) - h_i(t)\right]  \nn \\
 &+& \eta_i(t)\sqrt{2\Delta t} \;,
\label{EW_numint}
\end{eqnarray}
where unit lattice spacing for the underlying regular lattice is
assumed, $\eta_i(t)$-s are independent and identically distributed
random variables for all $i$ and $t$ with Gaussian distribution of
zero mean and unit variance, and $r(i)$ is the random neighbor of
node $i$ at time $t$. In this annealed construction, at every time
step, each node has a random neighbor with probability $p$;
$A_{i,r(i)}$$=$$1$ if node $i$ has a random neighbor,
$A_{i,r(i)}$$=$$0$ otherwise. At every time step the random links
are independently reassigned, a new PL-SW-network configuration is
generated. For the time-discretization scheme to be convergent for
the EW process on any {\em fixed} network, one can show that
\begin{equation}
  |\Delta t \lambda_{\max} - 1|<1 \;,
\end{equation}
i.e., $\Delta t<2/\lambda_{\max}$ is required, where
$\lambda_{\max}$ is the largest eigenvalue of the network
Laplacian. While for the annealed network, where the network
changes at every time step, we do not have a similar rigorous
requirement, as a guiding estimate, one can consider the largest
eigenvalue of the {\em average} (annealed) Laplacian. For example,
for $\alpha<1$ PL-SW networks, using our results for the annealed
random Laplacian [Eq.~(\ref{D_annealed})] (embedded in discrete
lattices), one has
\begin{eqnarray}
\lambda^{ann}_{\max} \sim \max_{k}\{2[1-\cos(k)] + [\DD^{rnd}](k)\} \nn \\
 = \max_{k}\{2[1-\cos(k)] + C qp\} = 4 + C qp ,
\end{eqnarray}
where $C$ is a constant which does {\em not} depend on $p$, $q$,
and the system size $N$. The resulting time-discretization scheme
then requires
\begin{equation}
\Delta t \stackrel{\sim}{<}\frac{2}{4 + C pq} \;.
\end{equation}
Similarly, for $1<\alpha<3$, one finds
\begin{equation}
\Delta t \stackrel{\sim}{<}\frac{2}{4 + C pq 4^{\alpha-1}} \;.
\end{equation}
The important consequence of the above order-of magnitude
estimates for the time discretization is that for {\em annealed}
plain or PL-SW networks, $\Delta t$ does not depend on the system
size. Thus, if a sufficiently small $\Delta t$ is tested and
confirmed to lead to a converging scheme for a small system size,
it will do so for all system sizes \footnote{The scenario is
drastically different for quenched {\em strongly heterogeneous}
(scale-free) networks, where the largest eigenvalue of the network
Laplacian is of the order of the largest degree in the network,
which scales with a power of the system size $N$. Thus, the
required $\Delta t$ becomes progressively smaller for larger
system sizes.}. In our numerical integrations, we used $\Delta
t=0.10$, which turned out to be sufficient to achieve numerical
convergence for the values $q=1$ and $p\leq1$ considered here.

\subsection{Roughness scaling on plain SW networks}

\begin{figure}[t]
  \centering
  \includegraphics[width=.45\textwidth]{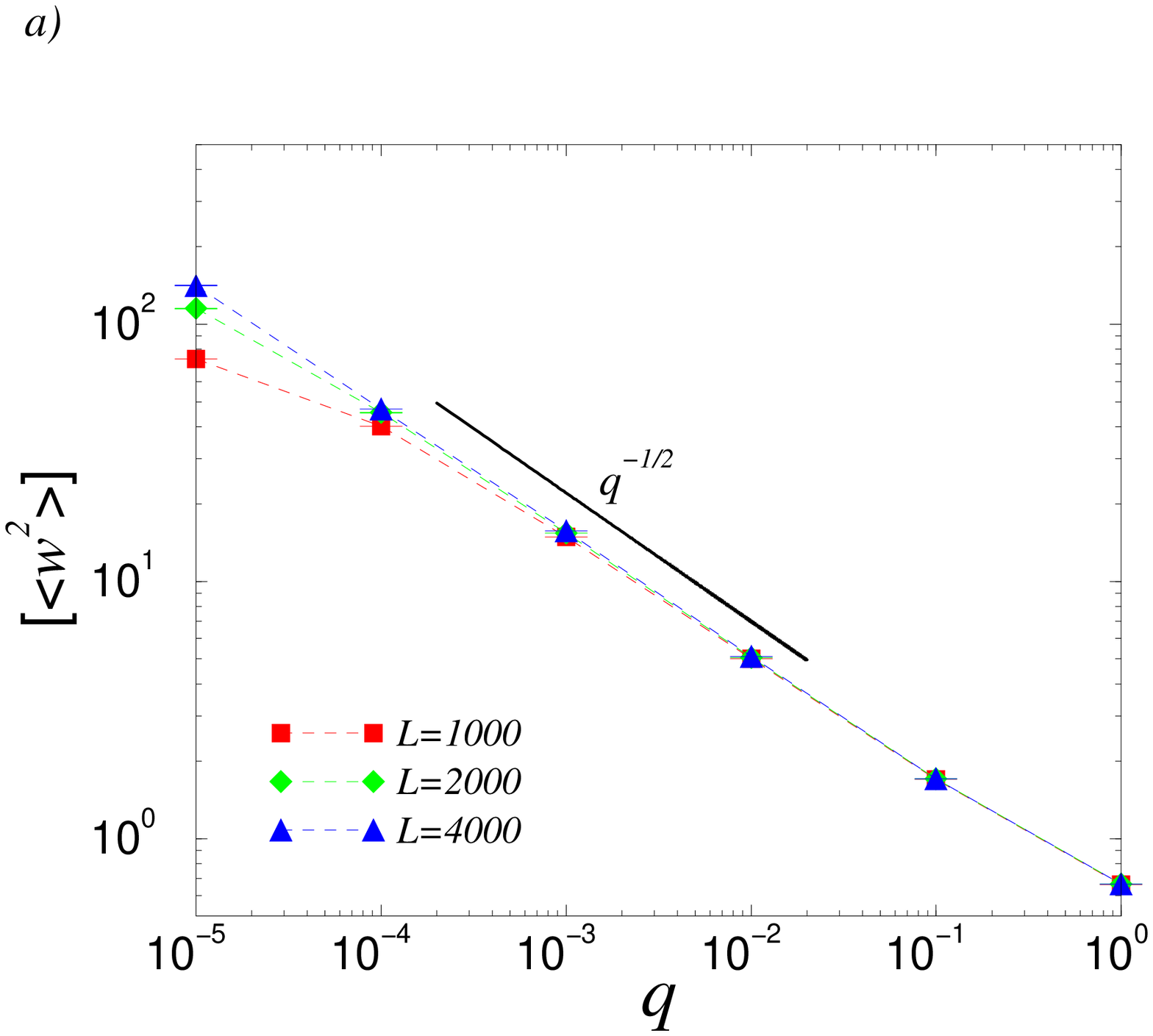}
  \includegraphics[width=.45\textwidth]{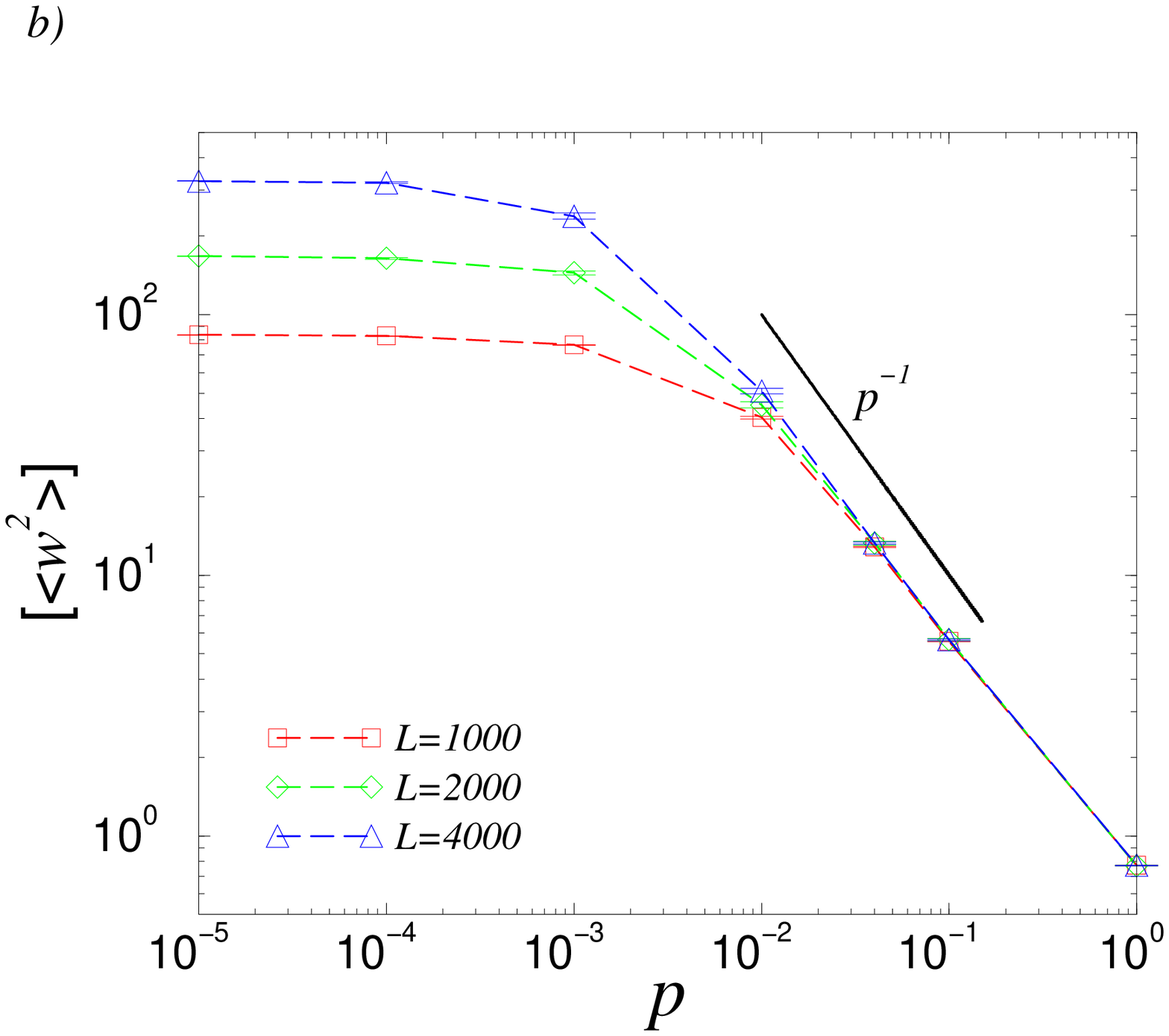}
  \vspace*{-0.0cm}
\caption{
\label{fig:plainSW}
Disorder averaged width obtained by exact numerical
diagonalization of the diffusion operator (a) changing the
strength of the long-range links while keeping their density fixed
($p=1$); and (b) changing the density of the long-range links
while keeping the strength constant ($q=1$) for system sizes
indicated in the figure. The two slopes indicate the analytic
predictions in the two cases.  Note that the MF argument would
predict the same behavior for both cases.}
\end{figure}

In Section~\ref{ch:qu}, for plain SW networks ($\al=0$), it was shown that
the MF approximation is valid in the high density of weak links limit ($q
\to 0$) and $[\langle w^2 \rangle ] \propto q^{-1/2} $, but breaks down
when the number of random links is small ($p \to 0$). In the later case,
the self-consistent perturbation expansion yielded $[\langle w^2 \rangle ]
\propto p^{-1}$. In Fig.~\ref{fig:plainSW}, we compared our analytic
predictions to numerical results and found a good agreement between the
numerics and the analytics for an intermediate interval of $q$ and $p$.
For small values finite-size effects are observed, addressed in
Section~\ref{sec:scaling}.

\subsection{Roughness scaling in transient/smooth phase II and in
recurrent/rough phases for annealed and quenched networks}

\begin{figure}[t]
  \centering
  \includegraphics[width=.45\textwidth]{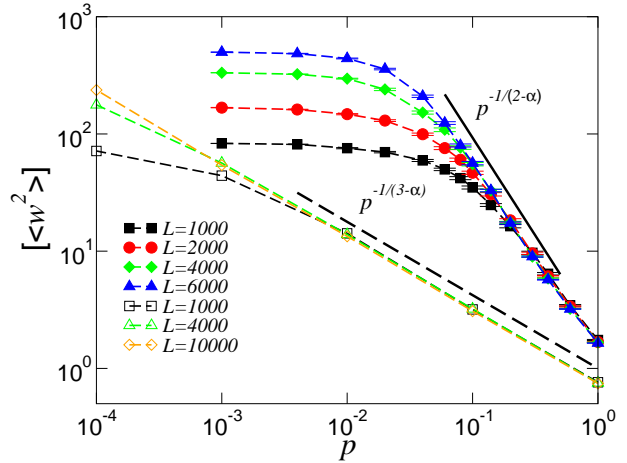}
  \vspace*{-0.0cm}
\caption{
\label{fig:a_1.4}
Numerically obtained finite-size propagators $G(0)$ vs.\ $p$ for
$\alpha$$=$$1.4$, $q$$=$$1$ (transient/smooth phase II) for
various system sizes.  Solid symbols represent data obtained by
exact numerical diagonalization averaged over 100 realizations for
the quenched network. The solid straight line represents the slope
obtained from the perturbative analytic calculations in the
$L$$\to$$\infty$ limit.  Open symbols correspond to the
steady-state averaged propagator on the annealed network with the
straight dashed line being the analytic result for the annealed
system in the $L$$\to$$\infty$ limit.  }
\end{figure}

In Fig.~\ref{fig:a_1.4}, the different scaling behavior of the
annealed and quenched system is shown in transient/smooth phase II
($\alpha$$=$$1.4$). The quenched results were obtained by exact
numerical diagonalization techniques, and the annealed
results were produced by integrating the stochastic EW process in a
dynamically annealed network, as described above. As can be seen in
Fig.~\ref{fig:a_1.4}, the numerical results and the annealed
analytic prediction are in good agreement.

\begin{figure}
\centering
\includegraphics[width=.45\textwidth]{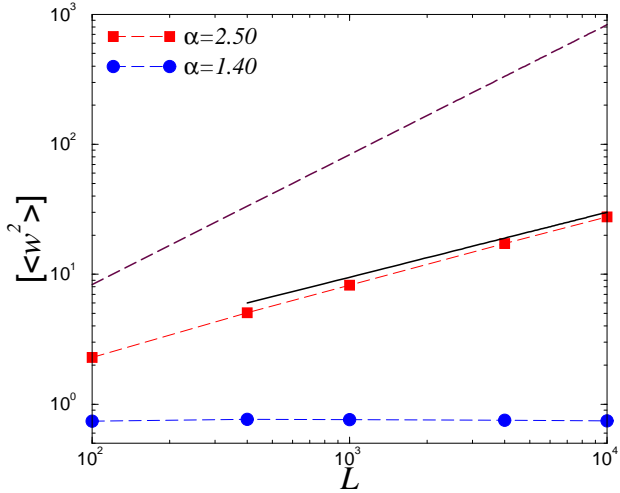}
\vspace*{-0.0cm}
\caption{
\label{w2_L_annealed}
System-size dependence of the width obtained by
numerically integrating the EW process on annealed PL-SW networks [Eq.~(\ref{EW_numint})]
(for $q$$=$$1$ and $p$$=$$1$) at a representative point in the
transient/smooth phase II ($\alpha$$=$$1.4$) and in the
recurrent/rough phase II ($\alpha$$=$$2.5$). The solid line
corresponds to the analytically-predicted asymptotic scaling in the
recurrent/rough phase II, $G(0)\sim L^{\alpha-2}$.
The dashed line indicates the exact scaling behavior in a network
with no long-range links in one dimension, $G(0)\sim L$, as a reference.
}
\end{figure}

For the annealed case, Fig.~\ref{w2_L_annealed} shows the
system-size dependence of the width (or $G(0)$) in
transient/smooth phase II ($\alpha$$=$$1.4$) and in
recurrent/rough phase II ($\alpha$$=$$2.5$). The agreement between
the analytic predictions (summarized in Fig.~\ref{fig:1d phases}) and the
simulation of the EW process in an annealed random network is
good in the asymptotic large system-size limit.

\begin{figure}[t]
  \centering
  \includegraphics[width=.45\textwidth]{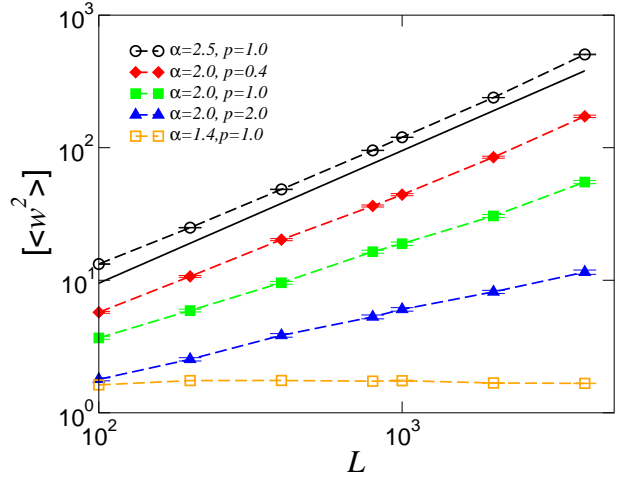}
  \vspace*{-0.0cm}
\caption{
\label{fig:a_2.0}
Comparison of system-size dependence of the width calculated by exact
numerical diagonalization (all data for $q$$=$$1$) for representative
$\al$-s from different phases of the quenched network as predicted in
Fig.~\ref{fig:2d phases} Data in the recurrent/rough phase I ({\large$\circ$}
symbols) indicate the linear scaling with $L$ (data is shifted up for
clarity).  Filled symbols represent the scaling behavior in the
recurrent/rough phase II ($\alpha$$=$$2$) for various values of $p$,
indicating the $p$-dependence of the exponent. The solid line represents the
linear system-size dependence to guide the eye.  Data in the
transient/smooth phase II ($\Box$ symbols), indicating a finite $G(0)$ in
the $L$$\to$$\infty$ limit, are also shown for comparison.  }
\end{figure}

Figure~\ref{fig:a_2.0} shows numerical results (from numerical
diagonalization) for both rough phases in quenched networks. For
the quenched case, the recurrent/rough phase II is somewhat
``degenerate'' in that it collapses onto a single point,
$\alpha$$=$$2$ (Fig.~\ref{fig:1d phases}). Here, the asymptotic
analytic results indicated that the exponent of the divergence
depends continuously on $p$, $G(0)\sim L^{\gamma(p)-2}$
[Eq.~(\ref{G_L_degenerate})]. Although the self-consistent formula
predicts this feature {\em qualitatively} (not shown on the plot
for the sake of clarity), the actual exponent appears to be
strongly affected by higher-order corrections. In the
recurrent/rough phase I, the width diverges as $G(0)\sim L$,
just as in a network without long-range links.

\subsection{Short- and long-time behavior of the expected number of returns
  of a random walker on PL-SW networks}

In experimental and numerical setups, it is important to understand the
behavior of the observed quantities for finite times and finite system
sizes in order to interpret the results. Because of finite system sizes,
one can observe different crossovers in the behavior of the expected number
of returns of a random walker. One of the crossover times, $T_\times$, can
be predicted by studying the integral representation of
$\hat{G}(0,\omega=1/T)$ in Eq.~(\ref{eq:[F(T)]}): determining whether, at
timescale $T$, the behavior of the integral is dictated by the $k^2$ term
or by $\Sigma(k)$, the self-energy generated by the random links. The other
crossover happens at times, $T_{\times\times}$, when $T_{\times\times}/L^d
\approx \hat{G}(0,\omega=1/T_{\times\times})$.  For $T \gg
T_{\times\times}$, the scaling of $F(T)$ is dominated by finite-size
effects.

As an example, let us have a closer look at transient phase II.  Before
$T_\times \propto p^{\frac{-2}{2-\alpha}}$, the expected number of returns
scales as that of a regular one-dimensional walker; for later times,
typically, the walker escapes from the vicinity of the origin through some
long-range link; though, after $T_{\times \times} \propto L p^{\frac{-1}
  {2-\alpha}}$, the walker starts to return ``from the perimeter'' of the
network due to the finite size of the system.

\begin{figure}[t]
\centering
\includegraphics[width=.45\textwidth]{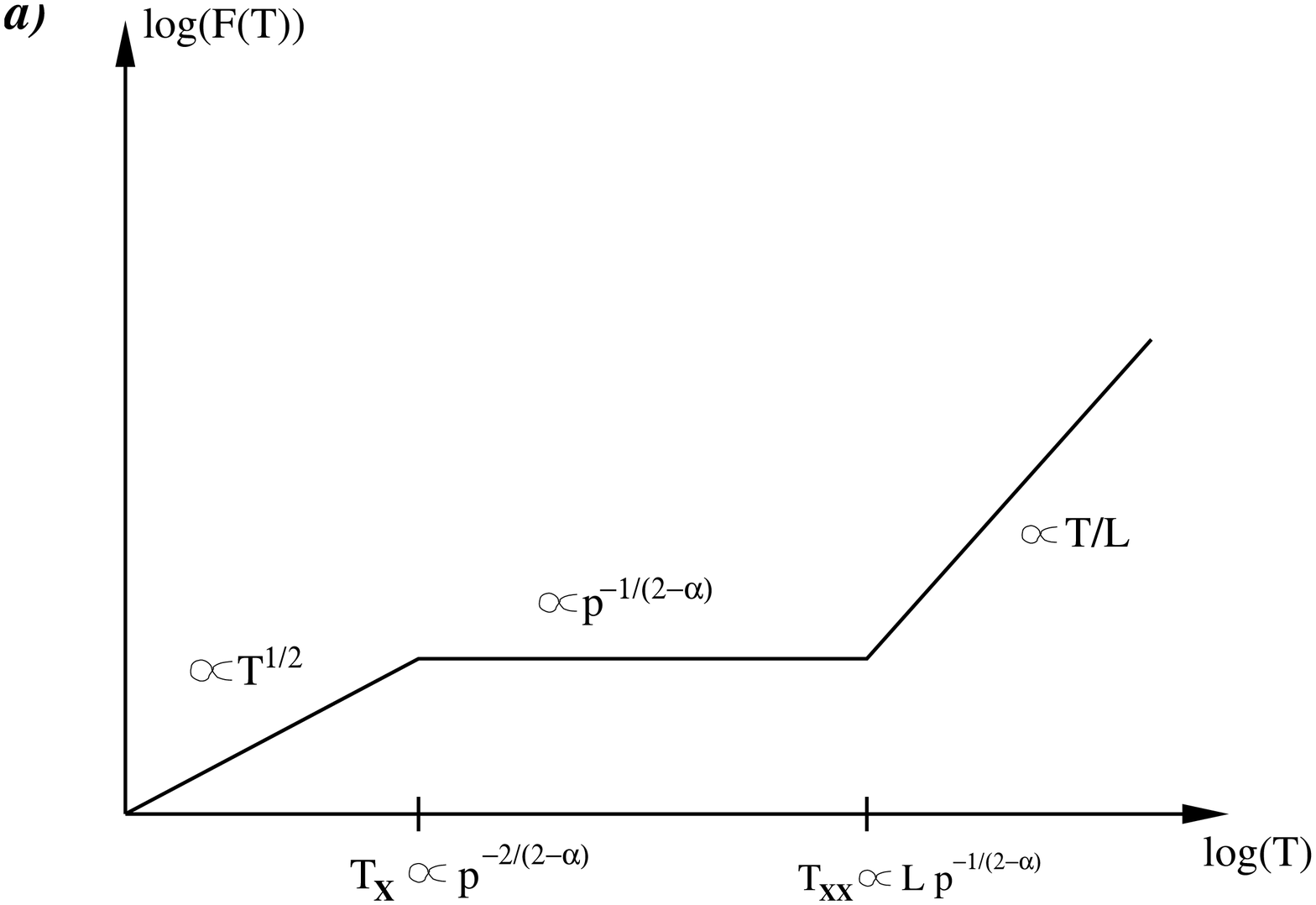}
\includegraphics[width=.45\textwidth]{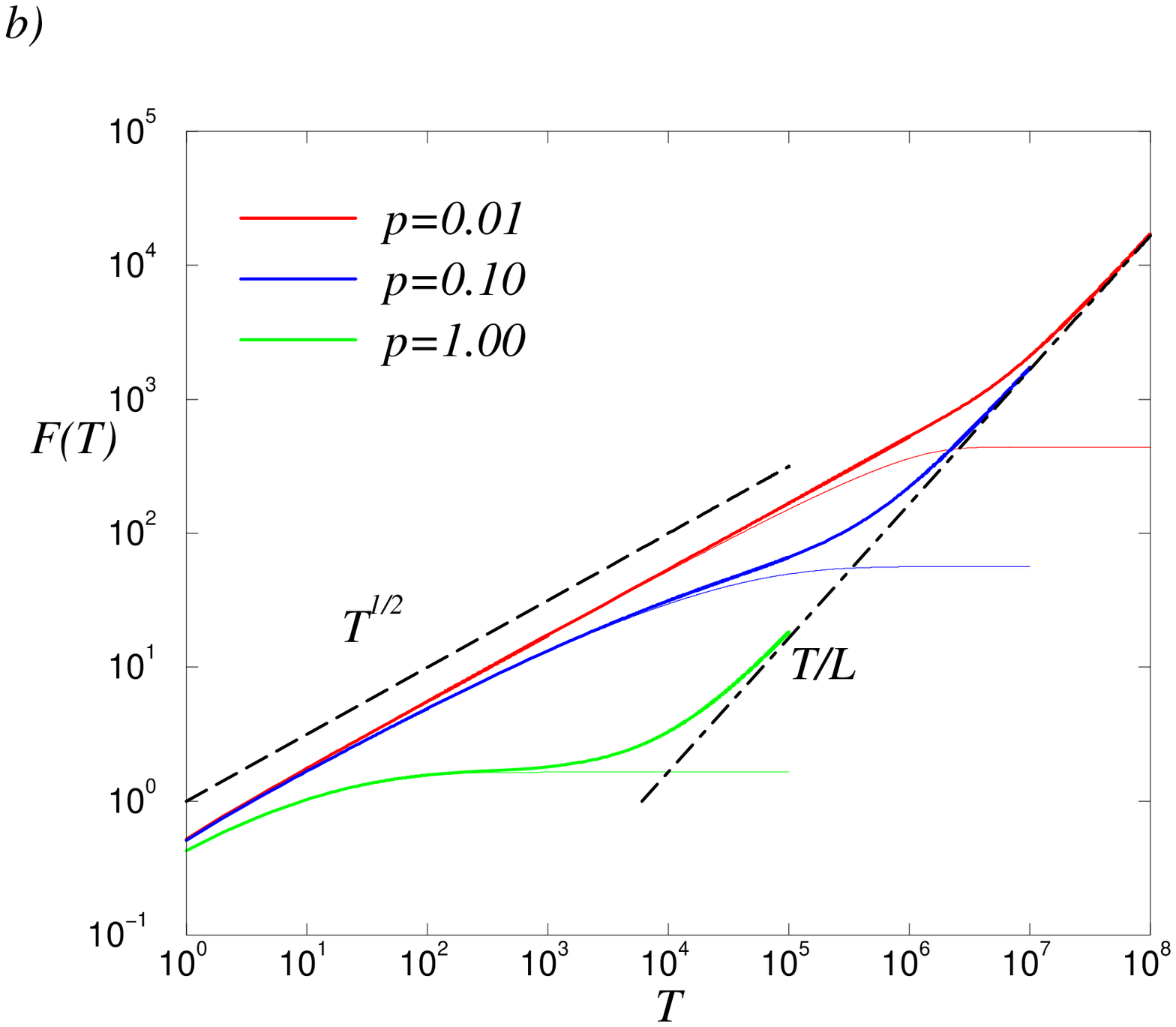}
\vspace*{-0.0cm}
\caption{
\label{fig:F(T)}
(a) The analytic prediction for the expected number of returns and
(b) its numerical verification (fat lines) in transient phase II
on a quenched PL-SW network with $\al=1.4$, $L=6000$, and $q=1$.
Thin lines represent numerical results obtained by excluding the
effect of the uniform mode in Eq.~(\ref{F_T_exact}) responsible
for the ``finite-size'' behavior for large $T$ values.}
\end{figure}

In Fig.~\ref{fig:F(T)}(a), we show the sketch of the scaling
predictions for $F(T)$ as discussed above. In Fig.~\ref{fig:F(T)}(b),
we show the disorder-averaged F(T), obtained by
employing Eq.~(\ref{F_T_exact}) with the eigenvalues from exact
numerical diagonalization, for each realization of the network,
and averaging over $100$ realizations. The two plots compare very
well.

\subsection{Finite-size behavior and scaling functions}
\label{sec:scaling}

\begin{figure}[tb]
\centering
\includegraphics[width=.45\textwidth]{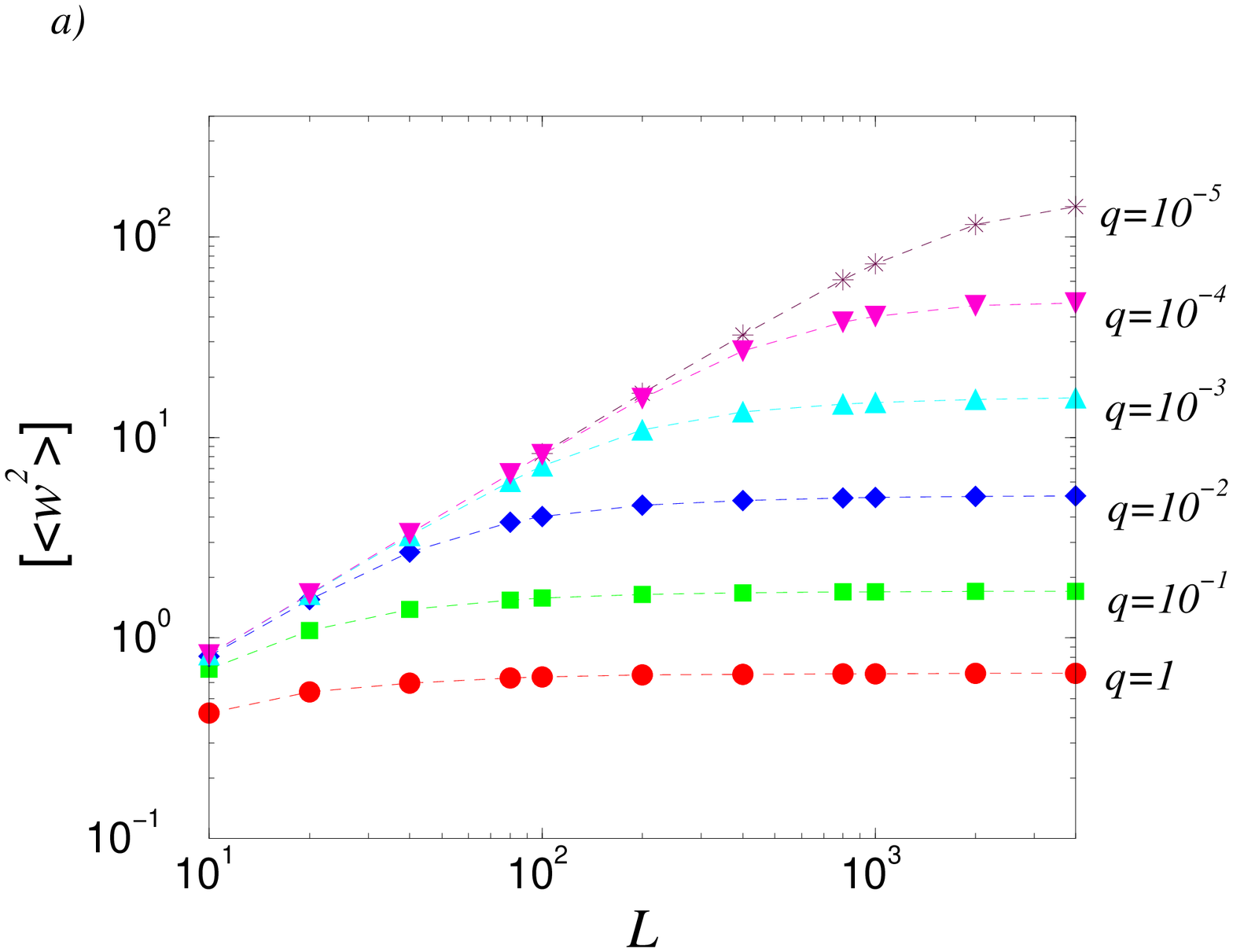}
\includegraphics[width=.40\textwidth]{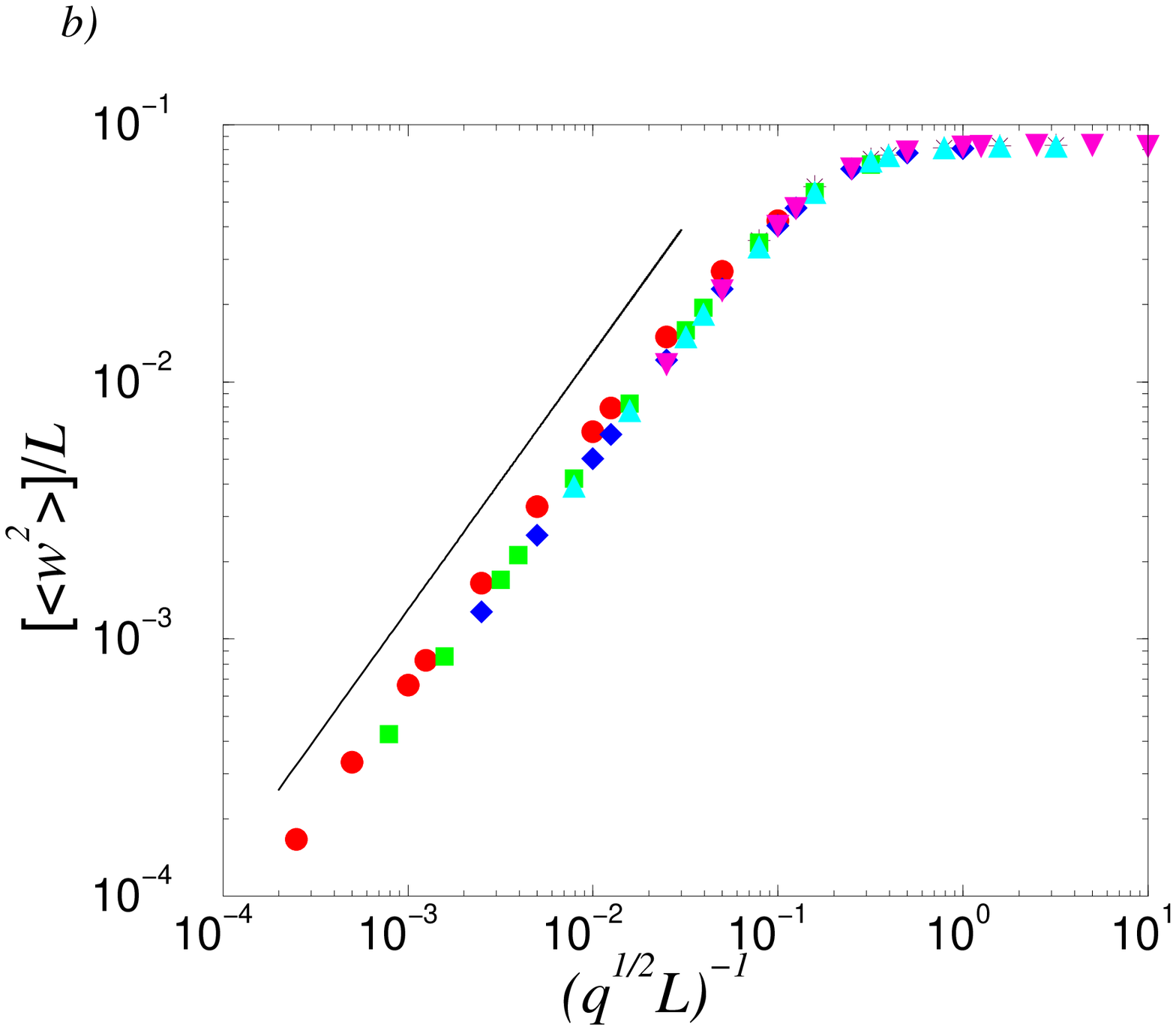}
\vspace*{-0.50cm}
\caption{
\label{fig:hardSWscaling}
(a)The system-size dependence of the width for different random connection
strengths ($q$-s) on hard SW networks. (b) The data of (a) collapsed to a
single curve approximating $\Sigma$ to first order in $q$ and rescaling the
axes according to Eq.~(\ref{w2_fss}).
Solid-line segments correspond to the
asymptotic small-$x$ behavior of the scaling function $f(x)$ given in
Eq.~(\ref{f_fss}).}
\end{figure}

Up to this point, in the analytic forms, we mostly considered the infinite
system-size behavior of the width, although it is clear from the plots,
Figs.~\ref{fig:plainSW} and \ref{fig:a_1.4}, that in the limit of weak
long-range interactions the divergence of the width is cut off by system
size effects. This observation is plausible since in the case $p \to 0$ (or
$q \to 0$ obviously) the width should not have a stronger divergence than
the pure (in this case one-dimensional) lattice.  Therefore, one can
conclude that for $p=0$ (the limit of a regular one-dimensional network)
$[\langle w^2 \rangle ]\sim L$, while for $p \neq 0$, in the infinite
system-size limit, it approaches a constant, $[\langle w^2 \rangle ] \simeq
1/\sqrt{\Sigma} =\xi$ [see Fig.~\ref{fig:hardSWscaling}(a)]. Thus, the
finite-size behavior of the average width can be expressed as
\begin{eqnarray}
[\langle w^2 \rangle ] = L f(\xi/L)
 \;,
\label{w2_fss}
\end{eqnarray}
where $f(x)$ is a scaling function such that
\begin{eqnarray}
f(x) \sim \left\{ \begin{array}{ll}
x & \mbox{if $x\ll 1$} \\
{\rm const.} & \mbox{if $x\gg 1$} \;.
\end{array} \right.
\label{f_fss}
\end{eqnarray}

For the hard SW network, $\Sigma=q+...$ hence $\xi=1/\sqrt{q}$.  The scaled
numerical data, $[\langle w^2 \rangle ]/L$ vs $q^{-1/2}/L$ in
Fig.~\ref{fig:hardSWscaling}(b), shows good collapse, as suggested by
Eq.~(\ref{w2_fss}).

\begin{figure}[tb]
\centering
\includegraphics[width=.53\textwidth]{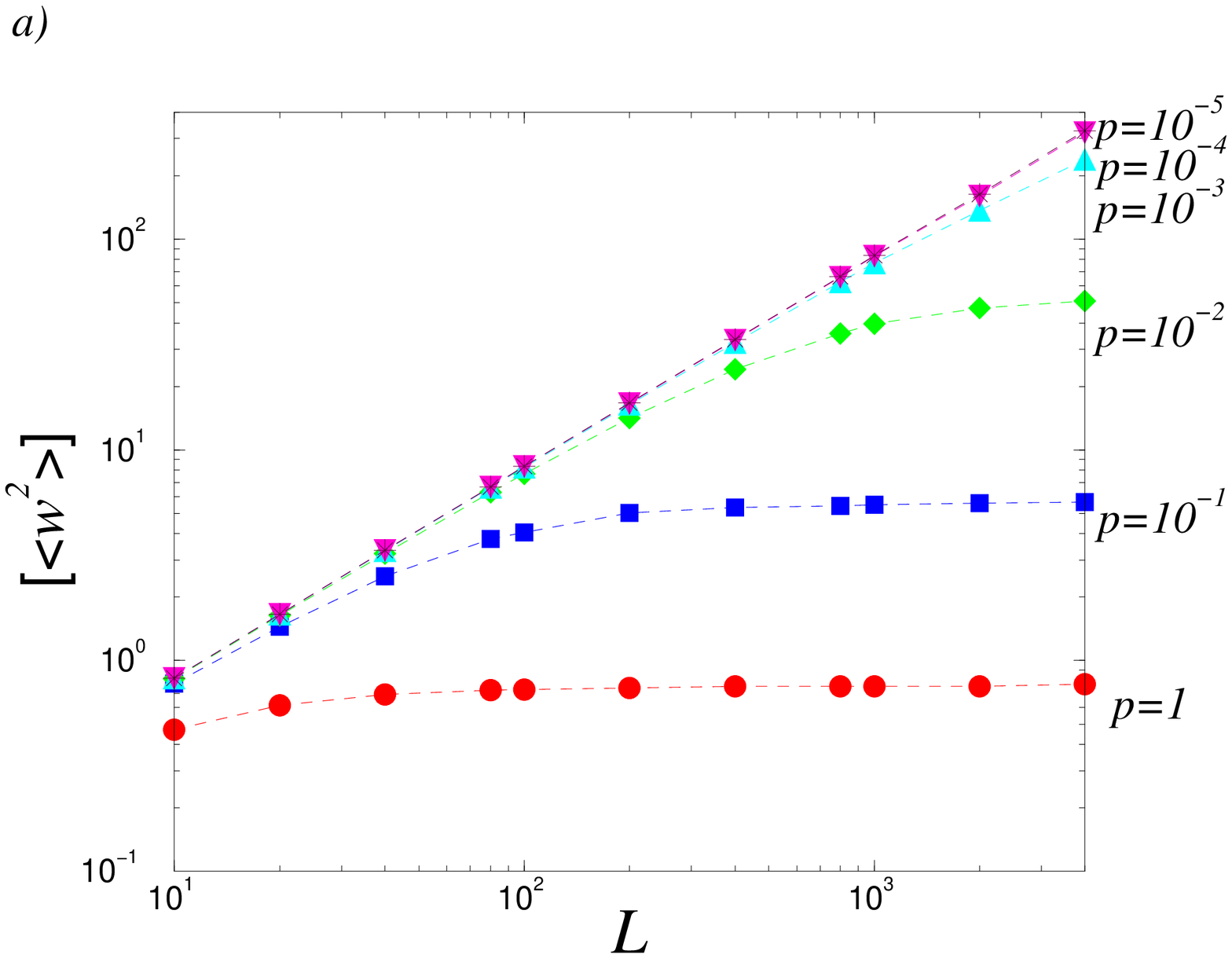}
\includegraphics[width=.45\textwidth]{./soft_plainSW_L.scaled.eps}
\vspace*{-0.20cm}
\caption{
\label{fig:softSWscaling}
(a) Width as a function of the system size with different crossovers for
different $p$-s for quenched plain SW networks. (b) Collapse of the same
data to yield the scaling function Eq.~(\ref{f_fss}). The solid-line
segment correspond to the asymptotic small-$x$ behavior of the scaling
function $f(x)$ given in Eq.~(\ref{f_fss}).  }
\end{figure}

In Fig.~\ref{fig:softSWscaling}, for quenched plain SW networks, a similar
finite-size correction can be observed. From the perturbative calculations
$\Sigma \propto p^2$ but the constant factor and higher-order terms were
undetermined. In conclusion, $\xi \propto p^{-1}$. Using this knowledge and
doing the same scaled plot as for the hard network, a good data collapse is
observed, though due to our lack of knowledge about the accurate behavior
of $\xi$ the scaling function is not as precise as in the hard case.

\begin{figure}[tb]
\centering
\includegraphics[width=.45\textwidth]{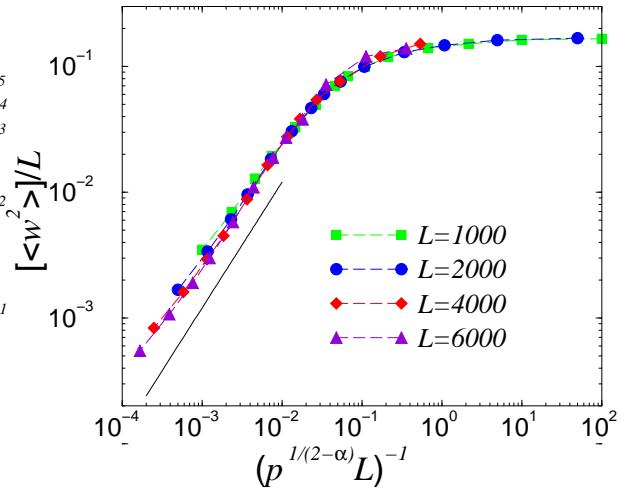}
\vspace*{-0.50cm}
\caption{
\label{fig:a1.4 scaled}
The scaling function Eq.~(\ref{eq:PL_f(x)}) for quenched PL-SW networks
with $\al=1.4$ obtained by exact numerical diagonalization of the diffusion
matrix. The straight line represents the slope predicted by the finite-size
scaling argument.}
\end{figure}

Strictly speaking, power-law networks do not have a correlation length like
plain SW networks ($\al=0$). Still, something can be said about the
finite-size behavior, as observed in Fig.~\ref{fig:a_1.4}. In the general
case, there are two competing terms determining the large scale behavior of
the Green's function, - the diffusion through the regular links, $\Delta$,
and through the random ones, $\DDr$. At small length scales, the number of
the random links is small, their contribution to the ensemble average is
negligible, and the behavior is close to that of a $d$-dimensional
unperturbed system. The effect of the random interactions takes over when
the self-energy generated by these random links start to dominate the
infrared behavior of $G(k)$, which is when $k_{\times}^2 \approx
\Sigma(k_{\times})$. Having these different regimes and crossovers, one can
construct a scaling function of $G(0)$ for finite systems similar to the
one of plain SW networks:
\begin{eqnarray}
[\langle w^2 \rangle ]=G(0)= L f(L_\times/L)
\label{eq:PL_f(x)}
\end{eqnarray}
where $L_\times = k^{-1}_\times \propto p^{\frac{-1}{2-\alpha}}$ in
transient/smooth phase II. The collapse of the scaled numerical data in
Fig.~\ref{fig:a1.4 scaled} supports the validity of the above construction,
though slight deviation can be observed from the predicted straight line
for large system sizes.

\subsection{Disorder-averaged two-point function}

\begin{figure}[tb]
\centering
\includegraphics[width=.45\textwidth]{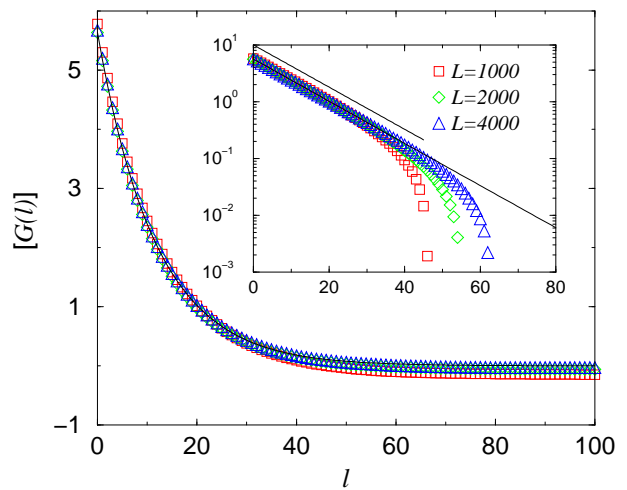}
\vspace*{-0.50cm}
\caption{
\label{fig:softGF}
Disorder-averaged two-point function as a function of the separation $l$ in
simple SW networks for $p=0.10$ and three system sizes. The solid line
corresponds to the exponential decay given by Eq.~(\ref{eq:G_l}).
The inset shows the same data in linear-log plot.}
\end{figure}

Finally, let us investigate the spatial behavior of the disorder-averaged GF,
$G(l)$ for quenched plain SW networks. From Eq.~(\ref{eq:s_1d_plain}),
$\Sigma(k) \propto p^2$, i.e. the GF is massive. In this case,
from Eq.~(\ref{eq:G(r) massive 1d}),
\begin{eqnarray}
[G(l)] \simeq  \frac{1}{2\sqrt{\Sigma}}e^{-\sqrt{\Sigma} l} \;,
\label{eq:G_l}
\end{eqnarray}
In Fig.~\ref{fig:softGF} we compared the above analytic result with ones
from exact numerical diagonalization and found good agreement up to
finite-size effects at large distances.

\section{SUMMARY}

The addition of random long-range links to a regular
$d$-dimensional network, producing a SW network, leads in many
cases to a crossover to mean-field-like behavior
\cite{HASTINGS03}, effectively becoming equivalent to averaging
over the long-range links in an annealed fashion. Here, we
investigated diffusion processes on distance-dependent SW
networks, where it is {\em not} the case (in low dimensions), and
the contrast between the quenched and annealed systems is strong.
The results are summarized in the phase diagrams of Figs.~\ref{fig:1d phases}
and \ref{fig:2d phases}, with dramatically different behavior
in the quenched and annealed case.  The
results for the quenched network were obtained by self-consistent
perturbation theory. Although the formalism was set up for the
propagator of the diffusion operator, it can be applied to other
processes where the mean-field behavior is violated or, if the
mean-field theory holds, to obtain higher-order corrections to it.
We also demonstrated the validity of the asymptotic theoretical
results by providing extensive numerical data based on the
spectral decomposition of the coupling matrix and numerical
integration of the stochastic EW process.

\acknowledgments
Discussions with H. Guclu, Z. Toroczkai, M.A. Novotny, L.A. Braunstein,
Z. R\'acz, G. Gy\"{o}rgyi, and T. Antal are gratefully acknowledged.
This research was supported in part by NSF Grant No. DMR-0426488,
the Research Corporation, and RPI's Seed Grant. M.B.H. was
supported by the U.S. DOE at LANL under Contract No. DE-AC52-06NA25396.
B.K. was partially supported by the EU under contract 001907 (DELIS).

\appendix

\section{The Fourier transform of the diffusion operator}
\label{app:[DDr](k)}

In a general case, the real space behavior of the averaged diffusion operator
(or self-energy) is
\begin{eqnarray}
[\Delta^{rnd}]_{\rr,\rr'}=
   \left\{
      \begin{array}{cl}
        p    & \mbox{if}  \ \rr = \rr' \\
        \frac{-p}{\NN}\frac{1}{ \left|\rr-\rr'\right|^{\alpha}}
            & \mbox{if}  \  \rr \ne \rr' \; . \\
      \end{array}
   \right.
\label{eq:ann_DDr}
\end{eqnarray}
Multiplying it with a Fourier mode, \(e^{i \kk \rr'}\),
\begin{eqnarray}
\sum_{\rr'}[\DDr_{\rr,\rr'}]\eeeP &=&
p \frac{\NN}{\NN} \eee -
 \frac{p}{\NN} \sum_{\rr' \ne \rr} \eeeP \frac{1}{|\rr-\rr'|^{\alpha}} \nn \\
&=& \frac{p}{\NN} \eee \sum_{\rr' \ne \rr}
   {\frac{1}{|\rr-\rr'|^\al}(1-e^{i\kk(\rr'-\rr)})} \nn \\
&=&\frac{p}{\NN} \eee \sum_{\rr'' \ne 0} {\frac{1}{|\rr''|^\al}(1-\eeePP)}.
\end{eqnarray}
The Fourier vectors are eigenvectors of the matrix with eigenvalues
\begin{eqnarray}
[\DDr](\kk)=\frac{p}{\NN} \sum_{\rr'' \ne 0}
{\frac{(\eeePP-1)}{|\rr''|^\al}}.
\end{eqnarray}
or, in the continuum limit,
\begin{eqnarray}
[\DDr](\kk)= \frac{p}{\NN} \int d^d r \frac{\eee-1}{r^\al}.
\label{eq:F_int}
\end{eqnarray}
Next, the scaling properties of $[\DDr](\kk)$ are calculated, when $a \ll
1/k \ll L$. For this reason $\NN$ is also approximated by its integral-form
limit
\begin{eqnarray}
\NN=\sum_{i=1}^{L^d} \frac{1}{r^\al} \propto \int^{L}_{a} r^{d-1} d r
\frac{1}{r^\al} \propto
   \left\{
      \begin{array}{ll}
        L^{d-\al} & \mbox{if $\al<d$} \\
        \ln (L/a) & \mbox{if $\al=d$} \\
        a^{d-\al} & \mbox{if $d<\al$}
      \end{array}
   \right. \nn \\
\end{eqnarray}
The approximations of the formulae below follow those of Appendix
\ref{app:G(0)-G(r)}, Eqs. (\ref{eq:1-cos(z)}) and (\ref{eq:x^2 d-dim}):
\begin{eqnarray}
\int^{L}_{a} d^d r \frac{1-\eee}{r^\al}&=&
\int_{a}^{L}  \frac{1-\cos (kx)}{x^{\al}}dx \nn \\
&\propto& k^2 \int_a^{1/k} \frac{dx x^2 }{x^\al}+
   \int_{1/k}^L \frac{dx}{x^\al} \nn  \\
&\propto&
   \left\{
      \begin{array}{ll}
        L^{1-\al} & \mbox{if $\al<1$} \\
        k^{\al-1} & \mbox{if $1<\al<3$}\\
        k^2 a^{3-\al} & \mbox{if $3<\al$}
      \end{array}
   \right. \nn \\
\end{eqnarray}
In higher dimensions, 
\begin{eqnarray}
\int^{L}_{a} d^d r \frac{1-\eee}{r^\al}&=&\int_{a}^{L} dr r^{d-1} \int d
\Omega \frac{1-e^{i k r \cos \theta}}{r^\al} \nn \\
&\propto& k^2 \int_a^{1/k} \frac{dr r^2 r^{d-1}}{r^\al}+
   \int_{1/k}^L \frac{dr r^{d-1}}{r^\al} \nn \\
&\propto&
   \left\{
      \begin{array}{ll}
        L^{d-\al}  & \mbox{if $\al<d$} \\
        \ln (kL)   & \mbox{if $\al=d$} \\
        k^{\al-d} & \mbox{if $d<\al<d+2$}\\
        k^2 |\ln(ka)| & \mbox{if $\al=d+2$}\\
        k^2 a^{d+2-\al} + ... & \mbox{if $d+2<\al$}
      \end{array}
   \right. \nn \\
\end{eqnarray}
Therefore, at leading order, assuming $a=1$ without loss of generality,
\begin{eqnarray}
[\DDr](\kk)\propto
   \left\{
      \begin{array}{ll}
        p & \mbox{if $\al<d$} \\
        p \LL(1+\frac{\ln k}{\ln L}\RR) & \mbox{if $\al=d$} \\
        p k^{\al-d}  & \mbox{if $d<\al<d+2$}\\
        p k^2 |\ln k| & \mbox{if $\al = d+2$}\\
        p k^2 + ... & \mbox{if $d+2<\al$}
      \end{array}
   \right.
\end{eqnarray}

In Subsection \ref{sec:2dquenched},
the behavior of the Fourier transform of $\Sigma(\rr)= \frac{s}{n \ r^4
  \ln^\mu (r/a)}$ is needed in two dimensions when $0>\mu>1$:
\begin{eqnarray}
\Sigma(k)&=&\frac{s \ 2 \pi}{n}\int_a^L \frac{(J_0(kr)-1)}
{r^4 \ln^\mu
  (r/a)}r dr \nn \\
&\propto& \frac{s}{n} k^2 \int_a^{1/k}\frac{1}{r
  \ln^\mu (r/a)} dr + ... \nonumber \\
&\propto& \frac{s c_1(\mu)}{n} k^2 \ln^{1-\mu}(ka) +
   \mathcal{O}\left( \frac{k^2}{\ln^\mu (ka)} \right) \;, \nn \\
\end{eqnarray}
where $c_1(\mu)$ is a constant, the result of the Fourier transformation.

For the present purposes of this work in most cases the scaling behavior of
$[\DDr](k)$ is sufficient to know. For more accurate calculations, one has to
cope with the divergences of the integrals at their limits (for example, by
introducing counterterms to the integrands) so as to obtain the constants in
the formulae. Since it is necessary in Subsection \ref{ss:1d quenched}, let us
calculate $\Sigma(k)$ (or equivalently $[\DD^{rnd}](k)$) with more scrutiny in one
dimension when $2<\al<3$:
\begin{eqnarray}
\Sigma(k)&=& 2 \frac{s}{n}\Re \int_{a}^{L}
   \frac{1-e^{ikr}}{r^\gamma}dr = 2 \frac{s}{n} \Re \ i
   \int_{a}^{L}\frac{1-e^{-kr}}{(ir)^\gamma}dr \nonumber \\ &=& 2
   \frac{s}{n} \sin \left(\frac{\pi}{2}\gamma\right)
   \int_{0}^{\infty}\frac{1-e^{-kr}}{r^\gamma}dr \nonumber \\ &=& 2
   \frac{s}{n} k^{\gamma-1} \sin \left(\frac{\pi}{2}\gamma\right)
   \Gamma(1-\gamma)
\end{eqnarray}
where $\Re$ is the real part of a complex function. In the above steps, we
transformed the complex contour integral from the real axis to the
imaginary one between $i a$ and $i L$, and then moved these limits
to $0$ and $\infty$ since the integrand was convergent at these limits and
the main contribution of the integral comes from $k$ values $\propto 1/r$ while
$a \ll r \ll L$.

\section{The scaling properties of $G(0)$ and $G(0,\omega=1/T)$ }
\label{app:G(0)}

Most of the results of this paper are derived for $G(r=0)$, or $G(0)$ for
short, and $G(r=0,\omega)$. Here, their scaling properties are obtained for
general $\Sigma(k)$-s when
\begin{eqnarray}
G(k)=\frac{1}{k^2+\Sigma(k)}
\label{eq:G(k)}
\end{eqnarray}
and similarly
\begin{eqnarray}
G(k,\omega)=\frac{1}{k^2+\Sigma(k)+\omega}
\end{eqnarray}

\begin{figure}[htb]
\centering
\includegraphics[width=.4\textwidth]{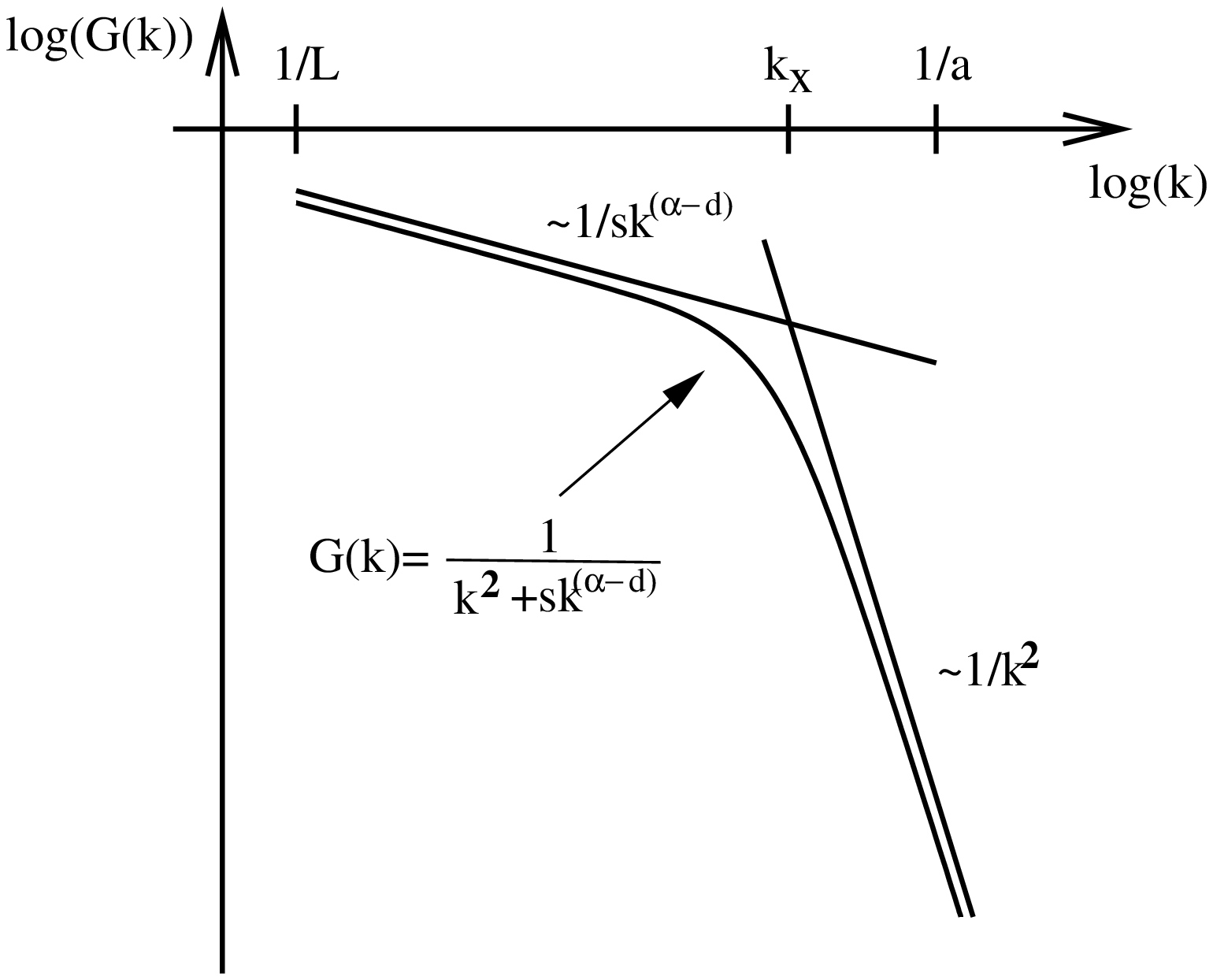}
\hspace*{0.2truecm}
\includegraphics[width=.4\textwidth]{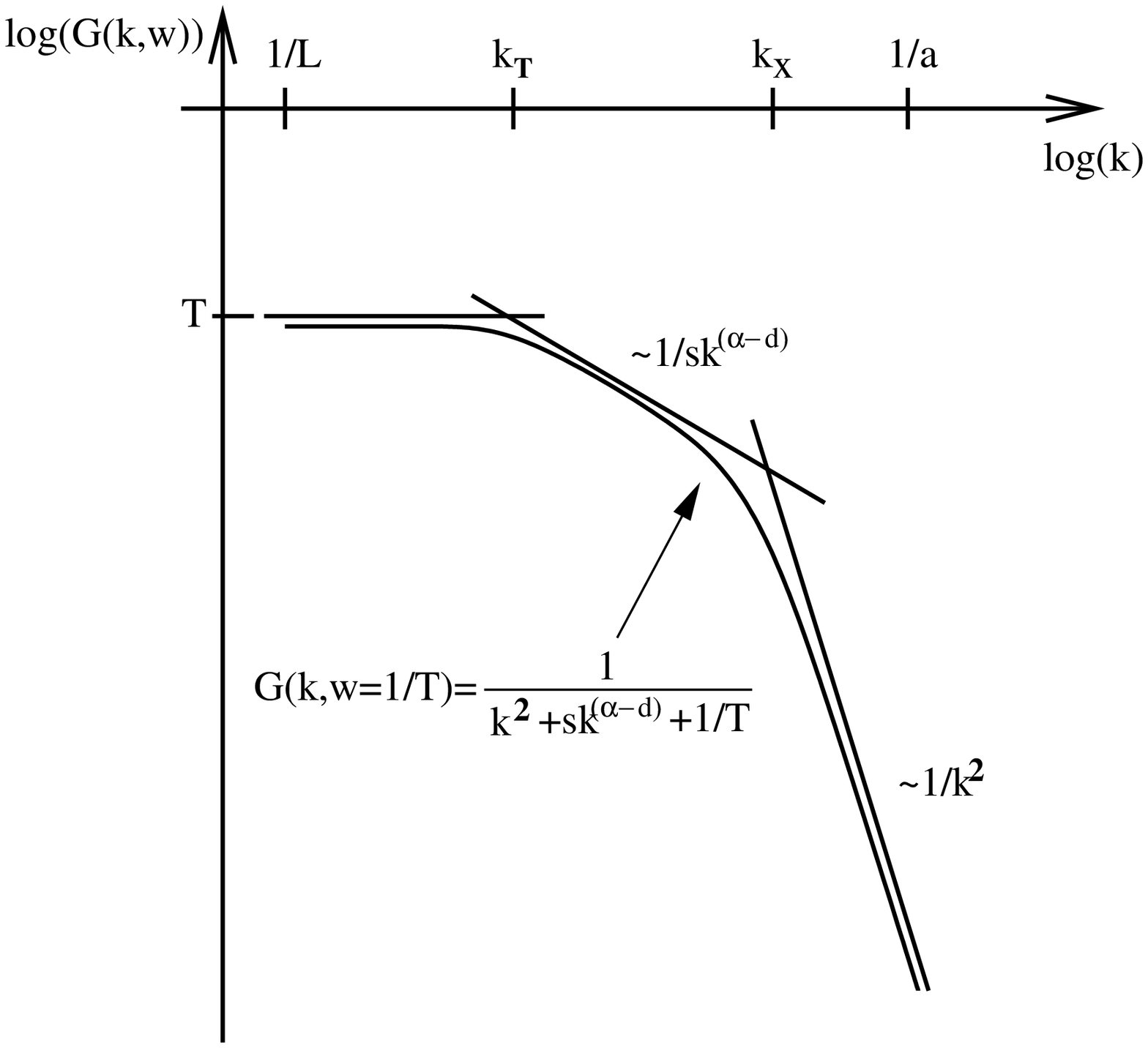}
\vspace*{0.50cm}
\caption{
\label{fig:k_scaling}
A sketch of the approximations used in Eq.-s (\ref{eq:approx}) and
(\ref{eq:approx_w}) to obtain the scaling of $G(0)$ and $G(0,\omega=1/T)$
for the specific case when $\Sigma(k)=sk^{\al-d}$. }
\end{figure}

First, let us investigate $G(0)$
\begin{eqnarray}
G(0) &=&  \int \frac{d^d k}{(2\pi)^d} \frac{1}{k^2+\Sigma(k)} \nn \\
&=& \frac{S_{d-1}}{(2\pi)^d} \int_{1/L}^{1/a} \frac{k^{d-1} dk}{k^2+\Sigma(k)}
\end{eqnarray}
Since, in the present work, our interest is only in the scaling properties
of $G(0)$, the following approximations are made: for different $k$ values
$G(k)$ is dominated by different terms in the denominator of
Eq.~(\ref{eq:G(k)}), as it is demonstrated in Fig.~\ref{fig:k_scaling} for the
specific case when $\Sigma(k)=s k^{\al-d}$. The crossover between the
different regimes is determined by \(k_\times\) defined by
\(k_\times^2=\Sigma(k_\times)\). From now on we assume that
\begin{eqnarray}
 \frac{\Sigma(k)}{k^2} \stackrel{k\to 0}{\longrightarrow} 0.
\end{eqnarray}
As can be seen in Fig.~\ref{fig:k_scaling},
\begin{eqnarray}
\frac{1}{k^2+\Sigma(k)} \approx
   \left\{
      \begin{array}{ll}
        \frac{1}{\Sigma(k)} & \mbox{if} \ k \ll k_\times \\ \\
        \frac{1}{k^2} & \mbox{if} \  k_\times \ll k.
      \end{array}
   \right.
\label{eq:approx}
\end{eqnarray}
Therefore,
\begin{eqnarray}
G(0) \propto \int_{1/L}^{k_\times} \frac{k^{d-1} dk}{\Sigma(k)} +
\int_{k_\times}^{1/a} \frac{k^{d-1} dk}{k^2}.
\end{eqnarray}

Next, $G(0)$ will be calculated for specific $\Sigma(k)$ as appear in the paper:\\
(i) If $\Sigma(k)=s$, $k_\times=\sqrt{s}$ and
\begin{eqnarray}
 G(0) &\propto& \int_{1/L}^{k_\times} \frac{k^{d-1} dk}{s}
   + \int_{k_\times}^{1/a}\frac{k^{d-1} dk}{k^2} \nn \\
  &=& \left\{
      \begin{array}{cl}
        s^{-1/2}                 & \mbox{if $d=1$}      \\
        \ln (s a^2)              & \mbox{if $d=2$}\\
        a^{2-d}          & \mbox{if $d \ge 3$}.
      \end{array}
      \right.
\end{eqnarray}
(ii) $\Sigma(k)=s k^{\al-d}$ where $d<\al<d+2$ \footnote{The exponent is
  expressed in this way because of the formulas appearing in the
  calculations of the perturbation expansion. Note that $0<\al-d<2$.},
$k_\times=s^{\frac{1}{2+d-\al}}$ and
\begin{eqnarray}
 G(0) &\propto& \int_{1/L}^{k_\times} \frac{k^{d-1} dk}{s k^{\al-d}}
   + \int_{k_\times}^{1/a}\frac{k^{d-1} dk}{k^2} \nn \\
   &=& \frac1s [k^{2d-\al}]_{1/L}^{k_\times} + [k^{d-2}]^{1/a}_{k_\times}
   \nn \\  \nn \\
  &=& \left\{
      \begin{array}{cl}
        s^{\frac{-1}{3-\al}}          & \mbox{if $d=1$ and $\al<2$}\\
        s^{-1} \ln(sL)                & \mbox{if $d=1$ and $\al=2$}\\
        \frac{1}{s} L^{\al-2}         & \mbox{if $d=1$ and $2<\al$}\\
        \ln (s a^{\al-4})             & \mbox{if $d=2$}\\
        a^{2-d}                       & \mbox{if $d \ge 3$}.
      \end{array}
      \right. \nn \\
\end{eqnarray}
(iii) $\Sigma(k)=s k^2 |\ln (ka)|$, $k_\times=\frac1a e^{1/s}$ and
\begin{eqnarray}
 G(0) &\propto&  \int_{1/L}^{k_\times} \frac{k^{d-1} dk}{s k^{2} |\ln (ka)|}
   + \int_{k_\times}^{1/a}\frac{k^{d-1} dk}{k^2} \nn \\ \nn  \\
   &\propto& \frac1s
        \LL[({d-2})^{-1}k^{d-2}|\ln (ka)|^{-1}\RR]_{1/L}^{k_\times}
       + [k^{d-2}]^{1/a}_{k_\times} \nn \\  \nn \\
  &\propto& \left\{
      \begin{array}{cl}
        \frac{L}{s \ln L}          & \mbox{if $d=1$}\\
        \frac1s \ln \ln L             & \mbox{if $d=2$}\\
        a^{2-d}                       & \mbox{if $d \ge 3$}.
      \end{array}
      \right. \nn \\
\end{eqnarray}
(iv) $\Sigma(k)=s k^2$, $\Sigma(k)$ is negligible to the $k^2$ term in the
integral and the calculation is straightforward
\begin{eqnarray}
 G(0) \propto
    \left\{
      \begin{array}{cl}
        L         & \mbox{if $d=1$}\\
        \ln (L/a)             & \mbox{if $d=2$}\\
        a^{2-d}                       & \mbox{if $d \ge 3$}.
      \end{array}
      \right.
\end{eqnarray}

Note that a similar approximation can be done as above for
$G(0,\omega=1/T)$ (see Fig.~\ref{fig:k_scaling}). Introducing $k_T$,
defined as $1/T=\Sigma(k_T)$,
\begin{eqnarray}
\frac{1}{k^2+\Sigma(k)+1/T} \approx
   \left\{
      \begin{array}{ll}
        T                   & \mbox{if} \ k \ll k_T \\
        \frac{1}{\Sigma(k)} & \mbox{if} \ k_T \ll k \ll k_\times \\
        \frac{1}{k^2} & \mbox{if} \  k_\times \ll k.
      \end{array}
   \right.
\label{eq:approx_w}
\end{eqnarray}
Though, one must be careful because, for some $\Sigma(k)$, the definition
of $k_T$ cannot be satisfied for any $k$-s (while $1/L<k<1/a$): for
example, if $\Sigma(k)=s$ and $T$ is large. This tells us that the $1/T$
term is irrelevant in the integrand while calculating the scaling
properties of $G(0,\omega=1/T)$, the low $k$ behavior is not affected by
$1/T$ (for calculational purposes one may interpret it as $k_T=1/L$).
Another case one must be careful with, is when $k_T>k_\times$: in this case
$\Sigma(k)$ becomes irrelevant in the determination of the integral since
this case corresponds to the scenario that $1/T$ cuts off the low $k$
behavior even before $\Sigma(k)$ would have contributed to the integral at
all (see Fig.~\ref{fig:k_scaling}).  Keeping these in mind the calculation
of the scaling of $G(0,\omega)$ is straightforward.

\vspace{0.3cm}

\section{The scaling properties of $(G(0)-G(r))$}
\label{app:G(0)-G(r)}

Since $(G(0)-G(r))$ appears frequently in the calculations, its
long-distance properties are calculated and summarized in this section.

\begin{eqnarray}
(G(\oo)-G(\rr))&=& \int \frac{1-e^{i\vec{k}\vec{r}}}{k^2+\Sigma(k)}
\frac{d^dk}{(2\pi)^d} \nn \\
&=& \int_{1/L}^{1/a} \frac{dk k^{d-1}}{(2\pi)^d} \int d
\Omega \frac{1-e^{i k r \cos \theta}}{k^2+\Sigma(k)} . \nn \\
\end{eqnarray}
In one dimension,
\begin{eqnarray}
(G(0)-G(x))= 2 \int_{1/L}^{1/a} \frac{1-\cos
  (kx)}{k^2+\Sigma(k)}\frac{dk}{2\pi}.
\end{eqnarray}
In order to extract the long-distance behavior of this integral, the
approximation
\begin{eqnarray}
(1-\cos(z)) \approx
   \left\{
      \begin{array}{ll}
        z^2 & \mbox{if} \ z<<1 \\
        const. & \mbox{if} \ z>>1
      \end{array}
   \right.
\label{eq:1-cos(z)}
\end{eqnarray}
is used\footnote{Note that, strictly speaking, the large-$z$ approximation
  is not really accurate: the function oscillates between $0$ and $1$ but,
  since it is multiplied with a slowly varying function in Eqn.
  (\ref{eq:G(0)-G(r) 1d}), the approximation is adequate to calculate
  scaling properties.}.\\
In higher dimensions, $d > 1$,
\begin{eqnarray}
 (G(\oo)-G(\rr)) &=& \frac{1}{(2\pi)^d} \int_a^L dk \int d \Omega
    \frac{1 - e^{i k r \cos \theta}}{k^2+\Sigma(k)}k^{d-1} \; . \nn \\
\label{eq:G(r)-G(0) d-dim}
\end{eqnarray}
From Appendix \ref{app:massiveGF},
\begin{widetext}
\begin{eqnarray}
\int_0^{\pi} e^{i z cos \theta} \sin^{d-2} \theta d \Omega
=J_\frac{d-2}{2}(z) \left(\frac{z}{2}\right)^{\frac{2-d}{2}}\pi^{1/2}\
\Gamma\left(\frac{d-1}{2}\right) S_{d-1}.
\end{eqnarray}
The asymptotic behavior of $J_\nu(z)$ is
\begin{eqnarray}
 J_\nu(z) \approx
   \left\{
      \begin{array}{ll}
        \frac{1}{\Gamma(\nu+1)} \left(\frac{z}{2} \right)^{\nu} \left( 1
           - \frac{\Gamma(\nu+1)}{2 \Gamma (\nu+2)}
         \left( \frac{z}{2} \right)^2 + ...\right)  & \mbox{if} \ z<<1 \\ \\
        \sqrt{\frac{2}{\pi z}}
             \cos\left(z-\frac{\pi}{4}-\frac{\nu\pi}{2}\right)
        & \mbox{if} \ z>>1
      \end{array}
   \right. 
\end{eqnarray}
and
\begin{eqnarray}
\int_0^{\pi} \sin^{d-2} \theta d \theta = \pi^{1/2} \frac{\Gamma
  \left(\frac{d-1}{2}\right)}{\Gamma \left(\frac{d}{2}\right)}
\end{eqnarray}
Therefore,
\begin{eqnarray}
\int_0^{\pi}(1- e^{i z \cos \theta}) \sin^{d-2} \theta \ d\theta \approx
   \left\{
      \begin{array}{ll}
        z^2 \sqrt{\pi} \frac{\Gamma\left(\frac{d-1}{2}\right)}{4 \Gamma\left(
            \frac{d+2}{2}\right)} + ... & \mbox{if} \ z<<1 \\ \\
         \sqrt{\pi} \frac{\Gamma
                 \left(\frac{d-1}{2}\right)}
          {\Gamma \left(\frac{d}{2}\right)} & \mbox{if} \ z>>1
      \end{array}
   \right.
\end{eqnarray}
\end{widetext}
Since, in the present work, we are only interested in the scaling
properties of the quantities investigated, it is sufficient to conclude
that after performing the solid angle integrations
\begin{eqnarray}
\int(e^{i k r \cos \theta}-1) \ d \Omega \propto
\left\{
      \begin{array}{cl}
        (kr)^2              & \mbox{if $ kr \ll 1$}\\
        const.          & \mbox{if $1 \ll kr$}.
      \end{array}
      \right.
\label{eq:x^2 d-dim}
\end{eqnarray}

Next, using the above approximations, let us calculate the scaling of
$(G(\oo)-G(\rr))$ when $\rr$ is large, for specific forms of $\Sigma(k)$,
appearing in the calculations. The crossover wavenumber $k_\times$ is
defined by $k^2_\times=\Sigma(k_\times)$.\\
(i) $\Sigma(k)=s$, $k_\times=\sqrt{s}$ and
\begin{eqnarray}
 (G(\oo)-G(\rr)) &\propto& r^2 \int_{1/L}^{1/r} \frac{k^2}{s}k^{d-1}dk \nn
 \\
&+& \int_{1/r}^{k_\times} \frac{k^{d-1} dk}{s}
   + \int_{k_\times}^{1/a}\frac{k^{d-1} dk}{k^2} \nn \\
  &=& \left\{
      \begin{array}{cl}
        s^{-1/2}                 & \mbox{if $d=1$}      \\
        \ln (s a^2)              & \mbox{if $d=2$}\\
        a^{2-d}          & \mbox{if $d \ge 3$}.
      \end{array}
      \right.
\end{eqnarray}
(ii) $\Sigma(k)=s k^{\al-d}$ where $d<\al<d+2$ \footnote{The exponent is
  expressed in this way because of the formulas appearing in the
  calculations of the perturbation expansion. Note that $0<\al-d<2$.},
$k_\times=s^{\frac{1}{2+d-\al}}$ and
\begin{eqnarray}
 (G(\oo)-G(\rr)) &\propto& r^2 \int_{1/L}^{1/r}
      \frac{k^2}{s k^{\al-d}}k^{d-1}dk \nn \\
&+& \int_{1/r}^{k_\times} \frac{k^{d-1} dk}{s k^{\al-d}}
   + \int_{k_\times}^{1/a}\frac{k^{d-1} dk}{k^2} \nn \\
   &=&r^2\frac1s [k^{2+2d-\al}]^{1/r}_{1/L} + \frac1s
        [k^{2d-\al}]_{1/r}^{k_\times} \nn \\
   &+& [k^{d-2}]^{1/a}_{k_\times} \nn \\
      \nn \\
  &=& \left\{
      \begin{array}{cl}
        s^{\frac{-1}{3-\al}}          & \mbox{if $d=1$ and $\al<2$}\\
        \frac1s \ln (rs)              & \mbox{if $d=1$ and $\al=2$}\\
        \frac{1}{s} r^{\al-2}         & \mbox{if $d=1$ and $2<\al$}\\
        \ln (s a^{\al-4})             & \mbox{if $d=2$}\\
        a^{2-d}                       & \mbox{if $d \ge 3$}.
      \end{array}
      \right. \nn \\
\end{eqnarray}
(iii) $\Sigma(k)=s k^2 |\ln (ka)|$, $k_\times=\frac1a e^{1/s}$ and
\begin{eqnarray}
 (G(\oo)-G(\rr)) &\propto& r^2 \int_{1/L}^{1/r}
      \frac{k^2}{s k^{2} |\ln (ka)| }k^{d-1}dk \nn \\
     &+& \int_{1/r}^{k_\times} \frac{k^{d-1} dk}{s k^{2} |\ln (ka)|}
   + \int_{k_\times}^{1/a}\frac{k^{d-1} dk}{k^2} \nn \\ \nn \\
   &\propto&r^2\frac1s [k^{d-2}|\ln (ka)|^{-1}]^{1/r}_{1/L} \nn \\
     &+& \frac1s \LL[({d-2})^{-1}k^{d-2}|\ln
      (ka)|^{-1}\RR]_{1/r}^{k_\times} \nn \\
       &+& [k^{d-2}]^{1/a}_{k_\times} \nn \\  \nn \\
  &\propto& \left\{
      \begin{array}{cl}
        \frac1s \frac{1}{r \ln(r/a)}    & \mbox{if $d=1$}\\
        \frac1s \ln(s\ln (r/a))       & \mbox{if $d=2$}\\
        a^{2-d}                       & \mbox{if $d \ge 3$}.
      \end{array}
      \right.
\end{eqnarray}
(iv) $\Sigma(k)=s k^2 \ln^{(1-\mu)}(ka)$ in two dimensions while $0<\mu<1$.
$k_\times=\frac1a e^{s^{1/(\mu-1)}}$ and
\begin{eqnarray}
  G(\oo)-G(\rr)&=&\int_{1/L}^{1/a} \frac{2 \pi (1-J_0(kr))}{k^2+s k^2
    \ln^{1-\mu}(ka) }\frac{k dk}{(2\pi)^2} \nn \\
  & \propto& \frac{1}{s}
    \int_{1/r}^{k_\times} \frac{dk}{k \ln^{1-\mu} (ka)} + ... \nonumber \\
     &\propto& \frac{c_2(\mu)}{s} \ln^\mu (r/a) +\mathcal{O}
        \left(\ln^{\mu-1}(r/a) \right) \; , \nn \\
\end{eqnarray}
where $c_2(\mu)$ is a constant, the result of the inverse-Fourier
transformation.

(v) $\Sigma(k)=s k^2$, $\Sigma(k)$ is negligible to the $k^2$ term in the
integral and the calculation is straightforward
\begin{eqnarray}
 (G(\oo)-G(\rr)) \propto
    \left\{
      \begin{array}{cl}
        r         & \mbox{if $d=1$}\\
        \ln (r/a)             & \mbox{if $d=2$}\\
        a^{2-d}                       & \mbox{if $d \ge 3$}.
      \end{array}
      \right.
\end{eqnarray}

\section{$G(r)$ for massive models}
\label{app:massiveGF}

In the literature massive GF are usually appear in their $k$-space
representation:
\begin{eqnarray}
G(k)=\frac{1}{k^2+m^2} \; .
\end{eqnarray}
In such models the correlation length, in terms of the mass, is
\begin{eqnarray}
\xi=m \; .
\end{eqnarray}
Here, the real space behavior of massive GF-s is reviewed.

In one dimension,
\begin{eqnarray}
G(x)=\int_{-\infty}^{\infty} \frac{d k}{2\pi} \frac{e^{i
    kx}}{k^2+m^2}=\frac{1}{2 m }e^{- m x},
\label{eq:G(r) massive 1d}
\end{eqnarray}
assuming $x \ge 0$ without loss of generality. The integration was performed
using contour integration by completing the integral contour with a
semi-circle in the upper half complex plane and using the residuum-theorem
with a pole at $k_0=i m $.

In higher dimensions, $d>1$,
\begin{eqnarray}
G(\rr)&=&\int \frac{d^d k}{(2\pi)^d} \frac{e^{i \kk \rr}}{k^2+m^2} \\ &=&
\int \frac{dk k^{d-1}}{(2 \pi)^d} \frac{1}{k^2+m^2} \int d \Omega e^{i k r
\cos \theta}.
\end{eqnarray}
The integration over the solid angles, $d \Omega$, can be performed
\begin{eqnarray}
\int d \Omega e^{i k r \cos \theta} = S_{d-1} \int^\pi_0 d \theta
\sin^{d-2} \theta e^{ikr \cos \theta},
\end{eqnarray}
where
\begin{eqnarray}
S_d=\int d \Omega = \frac{2 \pi^{d/2}}{\Gamma(d/2)} \; ,
\end{eqnarray} the surface
of a $d$-dimensional unit sphere. From one of the integral representations
of the Bessel functions \cite{GRADSTEIN},
\begin{eqnarray}
J_\nu(z)=\frac{\left(\frac{z}{2}\right)^{\nu}}{\pi^{1/2} \Gamma(\nu+1/2)}
\int_0^\pi  d \theta  \sin^{2\nu} \theta e^{iz \cos \theta}
\label{eq:J_nu}
\end{eqnarray}
and
\begin{eqnarray}
\int_0^\infty \frac{t^{\nu+1}}{(t^2+m^2)^{\mu+1}} J_\nu(at) dt = \frac{a^\mu
    m^{\nu-\mu}}{2^\mu \Gamma(\mu-1)} K_{\nu-\mu}(am) \; . \nn \\
\end{eqnarray}
As a result,
\begin{eqnarray}
G(r) &=& \int \frac{k^{d-1}dk}{(2 \pi)^d} \frac{S_{d-1}}{k^2+m^2}
  J_{\frac{d-2}{2}} (kr) \frac{\pi^{1/2} \Gamma \left(\frac{d-2}{2}
      +\frac12\right)}{\left(\frac{kr}{2}\right)^{\frac{d-2}{2}}} \nn  \\
 &=& \frac{S_{d-1} \pi^{1/2} \Gamma\left(\frac{d-1}{2}\right) 2^{\frac{d-2}{2}}}
     {(2\pi)^d} x^{\frac{2-d}{2}} \nn \\
  &\times& \int_0^{\infty} dk \frac{k^{d/2}}{k^2+m^2}
   J_{\frac{d}{2}-1} (kr) \nn \\
 &=& \frac{S_{d-1} \pi^{1/2} \Gamma\left(\frac{d-1}{2}\right) 2^{\frac{d-2}{2}}}
     {(2\pi)^d} r^{\frac{2-d}{2}} m^{\frac{d}{2}-1} K_{\frac{d}{2}-1}(mr)
  \nn \\
 &=& (2 \pi)^{-d/2} \frac{m^{\frac{d-2}{2}}}{r^{\frac{d-2}{2}}}
  K_{\frac{d-2}{2}}(mr) \; .
\end{eqnarray}

The asymptotic behavior of $K_\nu(x)$ is:\\
If $\nu=0$,
\begin{eqnarray}
  K_0(u)\approx
   \left\{
      \begin{array}{ll}
        -\gamma - \ln \LL( \frac{u}{2} \RR) +... & \mbox{if} \ u<<1 \\
        \sqrt{\frac{\pi}{2u}} e^{-u}(1+\mathcal{O}\LL(\frac{1}{u}\RR)) &
        \mbox{if} \ u>>1 \; .
      \end{array}
   \right.
\end{eqnarray}
If $\nu \ne 0$,
\begin{eqnarray}
  K_\nu(u)\approx
   \left\{
      \begin{array}{ll}
        \frac{\Gamma(|\nu|)}{2} \LL(\frac{2}{u} \RR)^{|\nu|} & \mbox{if} \
        u<<1 \\ \sqrt{\frac{\pi}{2u}}
        e^{-u}(1+\mathcal{O}\LL(\frac{1}{u}\RR)) & \mbox{if} \ u>>1 \;.
      \end{array}
   \right.
\end{eqnarray}

In some examples of the main text, the real space behavior of the
one-dimensional massless propagator is needed to know. In Fourier space,
$G^0(k)=k^{-2}$. The real space behavior of it can be obtained by
\begin{eqnarray}
G(x)&=& \frac{1}{2 \pi} \int_{2 \pi /L}^{2 \pi /a} \frac{e^{ikx} dk }{k^2}
 \approx \frac{1}{2 \pi} \int_{2 \pi /L}^{\infty} \frac{e^{ikx} dk }{k^2}
 \nonumber \\ &=& \frac{1}{2 \pi} \int_{2 \pi /L}^{\infty}
 \frac{(e^{ikx}-1) dk }{k^2} + \frac{1}{2 \pi} \int_{2 \pi /L}^{\infty}
 \frac{dk}{k^2} \; . \nn \\
\end{eqnarray}
First, since our interest is in the case $a \ll r \ll L$ and the integral
is convergent at the upper bound so $a \to 0$ can be taken.  Second, a
counter term is introduced to the integrand so as to make the integral
convergent at the lower bound too. Next, the $L \to \infty$ limit can be
taken
\begin{eqnarray}
G(x)&\approx& \frac{1}{2 \pi} \int_{0}^{\infty} \frac{(e^{ikx}-1) dk }{k^2}
   +  L \nn \\
&=& \frac{1}{2 \pi} x \int_{0}^{\infty}
   \frac{(e^{i\kappa}-1) d\kappa}{\kappa^2} + L \;.
\end{eqnarray}
In conclusion
\begin{eqnarray}
G(x) \approx L-x \gamma \;,
\end{eqnarray}
where \( \gamma =\frac{1}{2 \pi} \int_{0}^{\infty} \frac{(1-e^{i\kappa})
  d\kappa}{\kappa^2}\) is a constant.

\end{document}